\documentclass[twoside,openright,pdftex,a4paper]{report}

\usepackage[top=1in, bottom=1.25in, left=1.45in, right=1.05in,]{geometry}
\usepackage[T1]{fontenc}

\usepackage{lmodern}
\usepackage{siunitx}
\usepackage{amsmath}
\usepackage{bm}
\usepackage{amsfonts}
\usepackage{amssymb}
\usepackage{amsthm}
\usepackage{amstext}
\usepackage{amsbsy}
\usepackage{mathbbol}
\usepackage{multicol}
\usepackage{dsfont}
\usepackage{lscape}
\usepackage{physics}
\usepackage{amsopn}
\usepackage{braket}
\usepackage{amscd}
\usepackage{amsxtra}
\usepackage{enumerate}
\usepackage{caption}
\usepackage[ngerman]{babel}
\usepackage{graphicx}
\usepackage{upgreek}
\usepackage{placeins}
\usepackage{ifthen} 
\usepackage[utf8]{inputenc}
\usepackage{paralist} 
\usepackage{color}
\usepackage{cleveref}
\usepackage{titlesec}
\usepackage{mathrsfs}
\usepackage{comment}
\usepackage{marginnote}
\usepackage{multirow, bigstrut}
\usepackage{url}
\usepackage{appendix}
\usepackage{fancyhdr}

\usepackage{imakeidx}
\usepackage{xr-hyper}

\excludecomment{Loesung}
\DeclareMathOperator{\sgn}{sgn}

\usepackage{cleveref}
\usepackage{fancyhdr}
\usepackage{etoolbox}

\newcommand{\partfont}{\Huge\bfseries}
\makeatletter
\patchcmd{\@part}{\par\nobreak}{\enspace}{}{}
\patchcmd{\@part}{\Large\bfseries}{\partfont}{}{}
\patchcmd{\@part}{\huge\bfseries}{\partfont}{}{}
\patchcmd{\@spart}{\huge\bfseries}{\partfont}{}{}
\makeatother

\renewcommand{\d}[1]{\ensuremath{\mathrm{d}{#1}\;}}

\titleformat{\chapter}[display]{\normalfont\sffamily\Huge\bfseries}{\chaptertitlename\ \thechapter}{20pt}{\Huge}

\titleformat{\section}{\normalfont\sffamily\LARGE\bfseries}{\thesection}{1em}{\LARGE}

\titleformat{\subsection}{\normalfont\sffamily\Large\bfseries}{\thesubsection}{1em}{\Large}

\begin{document}\large
	
	\begin{titlepage}
		\thispagestyle{empty}
		\begin{center}
			\vspace*{\fill}
			\begin{huge}
				\textsf{\textbf{Modellierung der Messung von Flussqubits mittels Quanten-Fluss-Parametron}}
			\end{huge}\\
			\vspace{3em}
			
			\begin{figure}[h!]
				\centering
				\includegraphics[height=3cm]{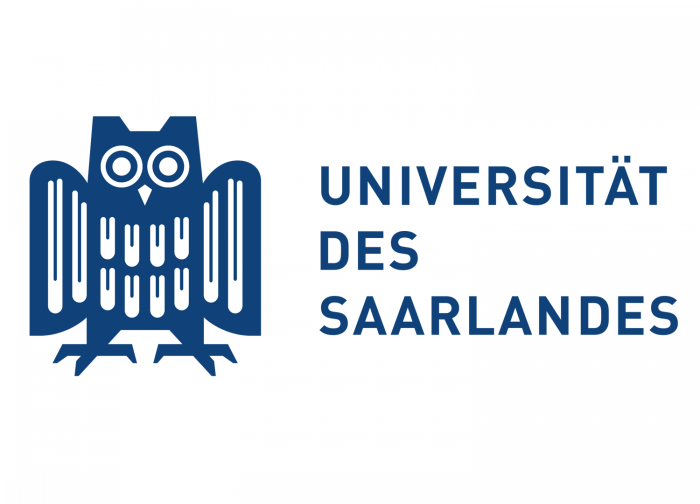}
			\end{figure}
			
			\vspace{1.5em}
			\begin{LARGE}
				\textsf{\textbf{Masterarbeit}}\\[0.3cm]
			\end{LARGE}
			
			\vspace{3.9em}
			
			\begin{large}
				\textsf{zur Erlangung des akademischen Grades\\
					Master of Science\\
					im Studiengang Physik\\
					der Naturwissenschaftlich-Technischen Fakultät\\
					der Universität des Saarlandes}\\[2.7cm]
			\end{large}
			
			\begin{large}
				\textsf{von}\\[0,3cm]
				\textsf{Yanjun Ji}\\[1.6cm]
			\end{large}

			\begin{large}
				\textsf{Saarbrücken}\\
				\textsf{16. Dezember 2020}\\
			\end{large}
			\vspace*{\fill}
			
		\end{center}
	\end{titlepage}

	\thispagestyle{empty}
	\cleardoublepage
	\vspace*{\fill}
	\pagenumbering{roman}
	\addcontentsline{toc}{chapter}{Eidesstattliche Erklärung}
	{\centering\bfseries\huge Eidesstattliche Erklärung\par}
	\vspace{1.8cm}
	\noindent Ich versichere hiermit, dass ich die vorliegende Arbeit selbständig verfasst und keine anderen als die angegebenen Quellen und Hilfsmittel benutzt habe.\par\vspace{6cm}
	\parbox{5cm}{\centering \dotfill \\
		\strut \centering \footnotesize Ort, Datum} \hfill\parbox{5cm}{\dotfill \\
		\strut \centering \footnotesize Unterschrift}
	\par\vspace{6cm}

	\vspace*{\fill}
	\thispagestyle{plain}
	\newpage
	\thispagestyle{empty}
	\cleardoublepage
	
	\addcontentsline{toc}{chapter}{Danksagungen}
	{\centering\bfseries\huge Danksagungen\par}
	\thispagestyle{plain}
	
	\vspace{3em}
	
	An dieser Stelle möchte ich mich bei all denjenigen bedanken, die mich während der Anfertigung dieser Masterarbeit unterstützt und motiviert haben.\\
	
	Mein Dank gilt zunächst Herrn Prof. Dr. Frank Wilhelm-Mauch, der meine Masterarbeit betreut und begutachtet hat. Auch danke ich der gesamten Arbeitsgruppe.\\
	
	Vielen Dank Prof. Dr. Uwe Hartmann für die Zweitkorrektur der Arbeit.\\
	
	Ganz besonders bedanke ich mich bei Susanna Kirchhoff, die mich sehr gut betreut hat.\\
	
	Zudem danke ich Dr. Marius Schöndorf für die gegebenen Materialien.\\
	
	Zuletzt möchte ich mich noch bei meiner Familie und meinen Freunden für die Unterstützung bedanken.
	
	\clearpage
	\thispagestyle{empty}
	
	\cleardoublepage

	\tableofcontents
	
	\thispagestyle{empty}
	\cleardoublepage
	
	\pagestyle{fancy}
	\pagenumbering{arabic}
	
	\part{Einleitung} 
	
	\chapter*{Einleitung}
	
	\vspace{3em}

	Ziel der Arbeit ist es, die Messung von Flussqubits mittels Quanten-Fluss-Parametron (QFP) zu modellieren und die Fidelity der Messung in verschiedenen Basen theoretisch zu untersuchen und numerisch zu simulieren.\\
	
	\vspace{3em}

	Ein Quantencomputer (QC) ist ein auf den Gesetzen der Quantenmechanik basierender Rechner. QC arbeiten mit Qubits. Im Unterschied zum klassischen Bit kann das Qubit einen Überlagerungszustand aus $0$ und $1$ einnehmen. Dieses Prinzip der Superposition und das Prinzip der Verschränkung ermöglichen, dass QC bestimmte Probleme effizienter lösen können, als klassische Algorithmen. QC können u. a. in folgenden Bereichen angewendet werden: Quantensimulation \cite{QS}, Quantenchemie \cite{QC} und Machine Learning \cite{ML}.\\
	
	Es gibt verschiedene physikalische Systeme, die als QC dienen können. Die folgenden Architekturen sind am weitesten verbreitet: Ionenfalle, Stickstoff-Fehlstellen-Zentren (NV-Zentren) in Diamant sowie supraleitender QC. Außerdem gibt es zwei Typen von Quantencomputern: Quantum-Annealing-Maschinen, die auf Optimierungsprobleme spezialisiert sind, und Quantengattercomputer, die beliebige Berechnungen ausführen können -- man nennt sie daher universell \cite{uni}. Dazu zählen zum Beispiel D-Waves adiabatischer QC \cite{dwave}, Googles supraleitender QC, IBMs universeller QC und Microsofts topologischer QC.\\
	
	D-Wave präsentierte im Jahr 2011 seinen ersten QC mit 128 Qubits. Im Jahr 2019 bot D-Wave zudem ein 2048-Qubit-Annealing-Quantencomputer an. Sie verwenden einen Prozess namens Quanten-Annealing, um nach Lösungen für bestimmte Optimierungsprobleme zu suchen. Dazu kann die Lösung der klassischen Optimierungsprobleme in den Grundzustand eines zeitabhängigen Hamiltonoperators encodiert werden. Quanten-Annealing beschreibt die Strategie, einen leicht zu implementierenden Grundzustand langsam in den gewünschten Endzustand zu entwickeln. Der Endzustand encodiert dabei die Lösung des betrachteten Problems \cite{Vinci2017}. Dabei verwendet man das supraleitende Qubit. Davon gibt es verschiedenen Arten, z. B. Flussqubit, Ladungsqubit und Phasenqubit.\\
	
	\thispagestyle{plain}
	
	Quantenmessung beschreibt die Strategie zum Auslesen des Zustands eines Qubits. Das Ergebnis der Messung hängt von der Wahl der Basis, in welcher der Zustand gemessen wird, ab.
	Die Basiswahl ist für das Auslesen des Qubits wichtig, da die Fidelity, mit der die Güte einer einzelnen Messung quantifiziert werden kann, von der Basis abhängig ist \cite{Pommerening2020}. Wir wollen daher untersuchen, welche Unterschiede zwischen Messungen in verschiedenen Basen für unsere betrachtete Architektur gibt. Dadurch kann das Messpostulat richtig bleiben. Man kann auch den Widerspruch zwischen Messbasis und Eigenbasis besser kennenlernen.
	Es ist auch wichtig für die Messung während der Rechnung, sodass der Messfehler reduziert werden kann.\\

	\vspace{3em}
	
	Die Arbeit besteht aus drei Teilen. Im Teil 1 werden zuerst das Flussqubit und das Quanten-Fluss-Parametron (QFP) eingeführt. Flussbasis und Energiebasis werden dann vorgestellt. Um das Qubit auszulesen, wird ein Resonator benutzt. Die Wechselwirkung eines Qubits mit einem monochromatischen, resonanten Lichtfeld wird durch das Jaynes-Cummings-Modell (JCM) beschrieben. Die Methode der Messung von zwei gekoppelten Qubits wird in Kapitel drei dargestellt.\\

	Mit den im Teil 1 gelegten Grundlagen wird im Teil 2 ein Modell der Messung eines Flussqubits mittels QFP erstellt. Es wird zuerst die Verschränkung zwischen Flussqubit und QFP untersucht. Der Überlapp zwischen Bare- und Dressed-Zustand wird dabei berechnet und simuliert. Das verschränkte System lässt sich durch einen effektiven Flussqubit-Hamiltonoperator beschreiben. Dies wird durch einen Resonator ausgelesen. Der Messprozess wird mit dem JCM analysiert. Die Fidelity wird in der Flussbasis und Energiebasis theoretisch und numerisch untersucht.\\
	
	Im dritten Teil wird die Messung von zwei gekoppelten Flussqubits modelliert. Jedes Flussqubit ist mit einem entsprechenden QFP verbunden. Dabei wird das System einmal mit und einmal ohne QFP1-Annealing betrachtet. Die Fidelity wird in verschiedenen Basen theoretisch analysiert und dann numerisch simuliert.\\
	
	\thispagestyle{plain}
	
	\vspace{3em}
	
	Durch die Modellierung der Messung von Flussqubits mittels QFP und die Berechnung sowie die numerische Simulation wurden die folgenden Ergebnissen von der Autorin gefunden:
	\begin{itemize}
		
		\item Sowohl in der Flussbasis als auch in der Energiebasis wird das Flussqubit-Signal durch eine adiabatische Durchführung vom QFP-Annealing im QFP erfolgreich gespeichert.
		
		\item Für ein einzelnen Flussqubit liefert eine Messung in der Energiebasis immer höhere Messgüte als eine Messung in der Flussbasis.
		
		\item Bei der Messung von FQ2 ohne QFP1-Annealing ist die Fidelity zur festgelegten Messzeit sowohl in der FQ2-Energiebasis als auch in der FQ1-FQ2-Energiebasis  höher als in der Flussbasis, wobei die Messung näher an der Dressed-Basis als an der Bare-Basis ist.
		
		\item Eine hohe Fidelity wird auch bei einer kurzen Messzeit bei der Messung von FQ2 mit QFP1-Annealing  erreicht.
		
		\item Für die verschiedenen Setups wurden Bedingungen dafür formuliert, welche Basis die effektivere ist. Dabei bedeutet "\text{effektiver}", dass eine Messung in dieser Basis den Qubitzustand genauer wiedergibt, als eine Messung in der anderen Basis.\\
		
	\end{itemize}
	
	\clearpage
	\thispagestyle{empty}
	
	\part{Grundlagen}
	\thispagestyle{empty}
	
	\chapter{Supraleitende Quantenbauteile}\label{SQED}

	In diesem Kapitel werden zuerst der Josephson-Kontakt und das darauf basierende Flussqubit erklärt. Danach kommen wir zu dem wichtigen Element der Masterarbeit, dem Quanten-Fluss-Parametron (QFP). Das QFP ist ein supraleitendes Bauteil. Je nach Regime kann das QFP als Qubit oder Resonator fungieren. Zum Schluss werden dann die in der Arbeit häufig diskutierten Basen, Flussbasis und Energiebasis, vorgestellt.\\

	\section{Josephson-Kontakt und Flussqubit (FQ)}
	
	Supraleiter sind Materialien, deren elektrischer Widerstand beim Unterschreiten der sogenannten Sprungtemperatur auf Null fällt. Die Supraleitung wurde zuerst 1911 von Heike Kamerlingh Onnes beobachtet \cite{kittel}.\\
	
	Ein mikroskopisches Modell der Supraleitung liefert die BCS-Theorie, die 1957 von John Bardeen, Leon Neil Cooper und John Robert Schrieffer entwickelt wurde \cite{BCS}. Cooper stellte fest, dass in einem Supraleiter zwei Elektronen durch Gitterschwingungen miteinander wechselwirken können. Dadurch können sie sogenannte Cooper-Paare bilden. Die BCS-Theorie postuliert nun einen Grundzustand der Supraleitung, in dem alle beteiligten Elektronen in Cooper-Paaren gebunden sind. Da die Elektronen zu Paaren gebunden sind, verhalten sich die Paare wie Bosonen und können alle den gleichen Grundzustand besetzen.
	
	Daher können sich supraleitende Qubits -- auch wenn sie aus Milliarden Atomen bestehen -- wie ein einzelnes quantenmechanisches System verhalten. Ein supraleitender LC-Schaltkreis ist ein einfaches Beispiel für ein solches System. Er verhält sich wie ein quantenmechanischer harmonischer Oszillator (HO). Allerdings hat ein HO äquidistante Energieniveaus, sodass man den Qubitzustand nicht identifizieren kann, weswegen sich ein LC-Schaltkreis nicht für die Realisierung eines Qubits eignet. Ein nichtlineares Element, hier ein Josephson-Kontakt, wird dafür benötigt, die Äquidistanz der Energielevel aufzuheben.\\
	
\begin{figure}[ht]
		\centering
		\includegraphics[width=0.7\textwidth]{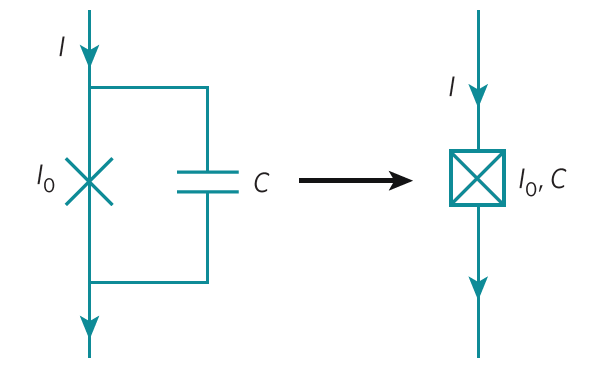}
		\caption{Zwei Darstellungen eines Josephson-Kontakts mit kritischem Strom $I_0$ durch das Josephson-Element und einer parallelgeschalteten Kapazität $C$. (Quelle: \cite{Clarke2008})}
		\label{fig:jj}
	\end{figure}

	Ein Josephson-Kontakt besteht aus zwei Supraleitern, die durch eine isolierende Barriere geeigneter Dicke, typischerweise 2-3 $\si{\nano \meter}$, getrennt sind, durch die Cooper-Paare tunneln können. Die entsprechende elektronische Schaltung ist wie in Abbildung \ref{fig:jj} dargestellt \cite{Clarke2008}.\\
	
\begin{figure}[ht]
		\centering
		\includegraphics[width=0.7\textwidth]{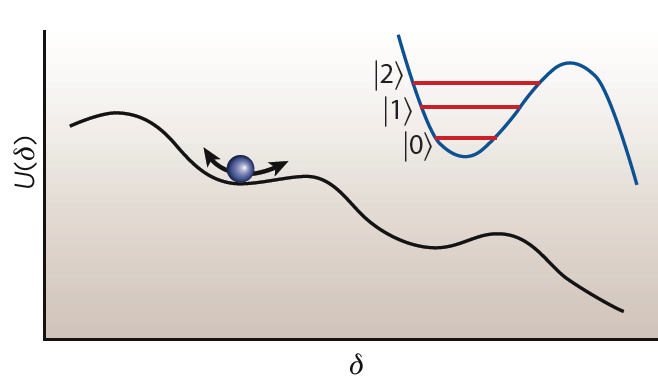}
		\caption{Die entsprechende potenzielle Energie $U(\delta)$ von Josephson-Kontakt. Wobei $\delta$ die Phasendifferenz zwischen beiden Supraleitern ist. (Quelle:  \cite{Clarke2008})}
		\label{fig:jjpotential}
	\end{figure}

	Der Suprastrom $I$ durch die Barriere hängt mit der Phasendifferenz $\delta(t)$ zwischen beiden Supraleitern durch die Strom-Phasen-Beziehung zusammen:\\
	\begin{align}
		I=I_0 \sin(\delta).
	\end{align}\\
	$I_0$ ist der materialabhängige kritische Strom. Dabei gilt für die Potenzialdifferenz\\
	\begin{align}
		U=\frac{\hbar}{2e}\frac{d\delta}{dt}=\frac{\Phi_0}{2 \pi} \dot{\delta},
		\label{equ:jju}
	\end{align}
	\\
	wobei $\Phi _{0}\equiv h/2e$ das magnetische Flussquant ist. Die potenzielle Energie wird in Abb. \ref{fig:jjpotential} dargestellt. Die Energiedifferenz zwischen benachbarten Niveaus ist unterschiedlich, sodass es zur Realisierung eines Qubits verwendet werden kann.\\
	
	Der Josephson-Kontakt verhält sich quantenmechanisch. Um das System quantenmechanisch zu beschreiben, kann man Operatoren $\hat{\delta}$ und $\hat{Q}=-i \hbar \partial/\partial \hat{\delta}$ mit $[\hat{\delta},\hat{Q}]=i\hbar$ definieren. Die beiden relevanten Operatoren sind der für $\hat{\delta}$, der mit der Josephson-Kopplungsenergie $E_J \equiv \hbar I_0 /2e=I_0 \Phi_{0}/2\pi$ zusammenhängt und der für die Cooper-Paar-Differenz $N$ durch die Kapazität $C$, die mit der Ladeenergie $E_c \equiv \left(2e\right)^2/2C$ zusammenhängt.\\
	
	Es gibt verschiedene Realisierungen von supraleitenden Qubits, wobei alle auf den drei Grundrealisierungen basieren: Flussqubit, Ladungsqubit und Phasenqubit. Einen Überblick über diese verschiedenen Typen findet man z. B. in \cite{Clarke2008} und \cite{Gu2017}.
	Diese Arbeit konzentriert sich auf das Flussqubit. Das wird daher im Folgenden vorgestellt.\\
	
	Flussqubits bestehen aus einer supraleitenden Schleife, die durch eine oder mehrere Josephson-Kontakte unterbrochen wird. Durch Stromkreisquantisierung \cite{Bishop2010} \cite{Wendin2005} kann man die entsprechende Energie des Flussqubits untersuchen. Als Beispiel wollen wir hier zwei Arten des Flussqubits, rf-SQUID und steuerbares rf-SQUID, im Detail anschauen.\\

	\subsection*{rf-SQUID}
\begin{figure}[ht]
		\centering
		\includegraphics[width=0.9\textwidth]{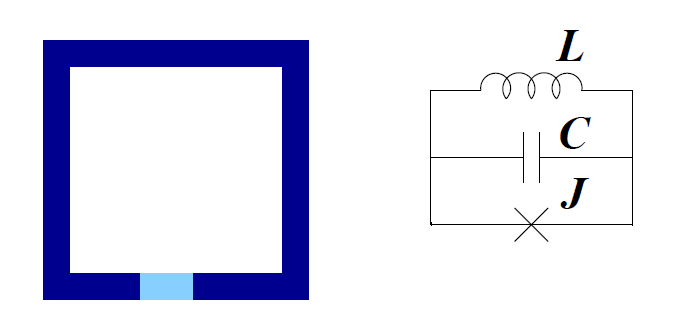}
		\caption{rf-SQUID und die entsprechende elektronische Schaltung. rf-SQUID, ein supraleitender Ring (dunkelblau), ist durch isolierendes Material (hellblau) unterbrochen. Der elektronische Schaltung besteht aus einer Induktivität $L$, einer Kapazität $C$, und einem Josephson-Element $J$. (Quelle:  \cite{Wendin2005})}
		\label{fig:rfsquid}
	\end{figure}

	Ein rf-SQUID (\textit{radio frequency-Superconducting quantum interference device}) besteht aus einem supraleitenden Ring, der an einer Stelle durch ein normal leitendes oder elektrisch isolierendes Material unterbrochen wird (siehe Abb. \ref{fig:rfsquid}).\\
	
	Zur Stromkreisquantisierung schreiben wir zuerst die Lagrangefunktion als kinetische Energie $T$ minus potenzielle Energie $U$:\\
	\begin{align}
		\mathcal{L}=T-U.
	\end{align}
	\\
	Die kinetische Energie ist im Kondensator gespeichert, während die potenzielle Energie in Spule und Josephson-Element stecken. Es ergibt sich dann\\
	\begin{align}
		\mathcal{L}=\frac{Q^2}{2C}-\left(-E_J \cos(2\pi\frac{\Phi_q}{\Phi_{0}})+\frac{\left(\Phi_q-\Phi_q^x\right)^2}{2L}\right),
	\end{align}
	\\
	wobei $Q$ der Ladung, $C$ der Kapazität, $E_J$ der Josephson-Energie, $\Phi_q$ dem magnetischen Fluss, $\Phi_q^x$ dem äußeren magnetischen Fluss und $L$ der Induktivität entsprechen. Der Phasenunterschied und der magnetische Fluss hat die Beziehung $\phi =2e\Phi/\hbar=2\pi \Phi/\Phi_{0}$. Man kann auch die Lagrangefunktion dadurch bekommen, dass man Kirchhoffsche Regeln mit Euler-Lagrange-Gleichungen vergleicht (für mehr Details siehe \cite{Wendin2005}).\\
	
	Stellt man die Ladung $Q=C U= C \dot{\phi}_q \Phi_0/\left(2\pi\right)$ durch $\dot{\phi}_q$ gemäß der Gl. \ref{equ:jju} dar, dann lässt sich die Hamiltonfunktion $H$ durch die Lagrangefunktion berechnen:\\
	\begin{align}
		H\left(Q, \phi_q\right)=\dot{\phi_q} \partial \mathcal{L}/\partial \dot{\phi}_q-\mathcal{L}.
	\end{align}\\	
	Für einen magnetischen Fluss, der gleich einer ganzzahligen Anzahl von Flussquanten ist, hat die potenzielle Energie des SQUIDs ein absolutes Minimum bei $\Phi_q=\Phi_q^x$. Für halbzahlige Flussquanten hat die potenzielle Energie zwei entartete Minima, die den beiden in der SQUID-Schleife in entgegengesetzter Richtung zirkulierenden Dauerstromzuständen entsprechen \cite{Wendin2005}. Die mit äußerem Magnetfeld verbundene zeitabhängige Spannung kann für ein digitales Magnetometer genutzt werden.\\
	
	\subsection*{Steuerbares rf-SQUID}
	
\begin{figure}[ht]
		\centering
		\includegraphics[width=0.9\textwidth]{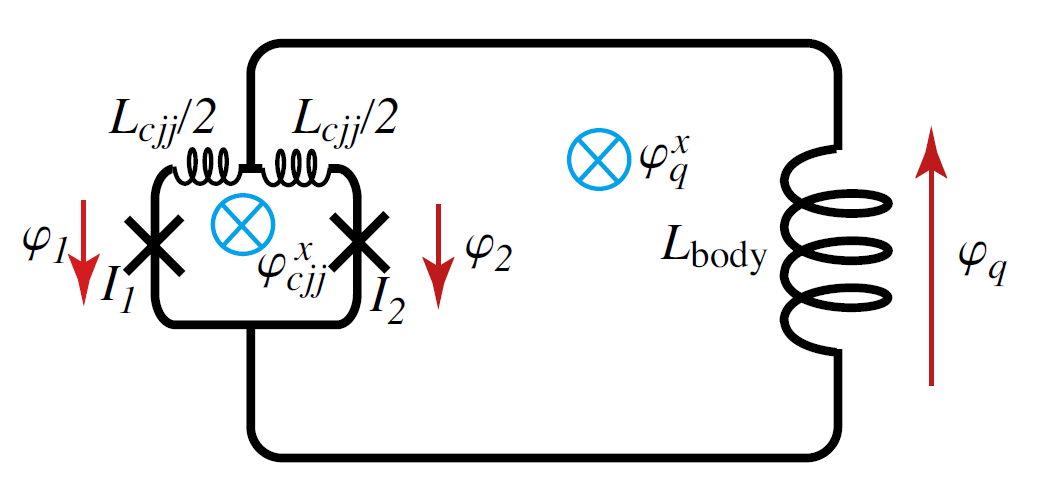}
		\caption{Der elektrische Stromkreis von steuerbarem rf-SQUID. Es besteht aus zwei Schleifen: einer großen q-Schleife und einer kleinen cjj-Schleife. In der kleinen Schleife gibt es zwei Josephson-Kontakten mit den kritischen Strömen $I_1$ und $I_2$. Die äußeren magnetischen Flüsse in beiden Schleifen werden mit $\varphi_q^x$ und $\varphi_{\text{cjj}}^x$ bezeichnet. (Quelle:  \cite{Harris2010})}
		\label{fig:cjjrfsquid}
	\end{figure}

	Ein steuerbares rf-SQUID (rf-SQUID mit zwei Josephson-Kontakten) hat eine mit Josephson-Kontakten zusammengesetzte (\textit{Compound Josephson Junction} (CJJ)) Schleife (siehe Abb. \ref{fig:cjjrfsquid}). Die beiden Josephson-Kontakte verhalten sich wie ein einziger Kontakt mit regelbarem kritischen Strom. Die kleine Schleife bezeichnen wir mit cjj und die Hauptschleife mit q.\\
	
	Die äußeren magnetischen Flüsse, die mit einer Spule von außen auf das Flussqubit aufgebracht werden können, sind $\Phi_{\text{cjj}}^{x}=\Phi_0 \varphi_{\text{cjj}}^x /2 \pi$ für die cjj-Schleife und $\Phi_{q}^x=\Phi_0 \varphi_{q}^{x} /2 \pi$ für die q-Schleife. $\varphi_{\text{cjj}}$ und $\varphi_{q}$ sind die Phasendifferenzen der cjj-Schleife und der q-Schleife.\\
	
	Bei der Stromkreisquantisierung spielt die Flussquantisierung \cite{Ferber2005} eine Rolle, die sagt, dass der gesamte magnetische Fluss durch einen geschlossenen Stromkreis nur ganzzahlige Vielfache des Flussquantums betragen kann. Dadurch kann man den gesamten magnetischen Fluss der Spule durch die magnetischen Flüsse durch andere Bauelemente darstellen. Damit haben wir für unser System\\
	\begin{align}
		\begin{split}
			\varphi_{\Sigma_q}+\varphi_1 - \varphi_q^x=0,~\varphi_{\Sigma_q}+\varphi_2 - \varphi_q^x=0,
		\end{split}
	\end{align}\\
	sowie\\
	\begin{align}
		\varphi_{\Sigma_{\text{cjj}}}+\varphi_1 - \varphi_2 - \varphi_{\text{cjj}}^x=0.
	\end{align}
	So bekommen wir\\
	\begin{align}
		\varphi_{\Sigma_q}&=\varphi_q^x-\left(\varphi_1 + \varphi_2\right)/2 \equiv \varphi_q^x-\varphi_q,\\
		\varphi_{\Sigma_{\text{cjj}}}&=\varphi_{\text{cjj}}^x -\left(\varphi_1 - \varphi_2\right)\equiv\varphi_{\text{cjj}}^x - \varphi_{\text{cjj}}.
	\end{align}\\
	Hier sind $\varphi_{\Sigma_q}$ ($\varphi_{\Sigma_{\text{cjj}}}$) die Phasendifferenz der gesamten Spulen in der q-Schleife (respektive cjj-Schleife).\\
	
	Die Born-Oppenheimer-Näherung ist eine Näherung zur Vereinfachung der Schrödingergleichung von Systemen aus mehreren Teilchen. Sie nutzt aus, dass schwere und leichte Teilchen in einem System ihre Bewegungsrichtung auf sehr unterschiedlichen Zeitskalen ändern, und dass die Bewegungsgleichungen der schnellen, leichten daher ohne Berücksichtigung der Bewegung der langsamen, schweren sinnvoll gelöst werden kann. Unter der Born-Oppenheimer-Näherung können wir die effektive potenzielle Energie für identischen Josephson-Kontakt wie folgt schreiben \cite{Harris2010}\cite{Ozfidan2019}\cite{Berkley2010}:\\
	\begin{align}
		U_{\text{eff}}\left(\Phi_q\right)=- E_J\left(\Phi_{\text{cjj}}^{x}\right) \cos(\frac{2 \pi \Phi_{q}}{\Phi_0})+\frac{\left(\Phi_{q}-\Phi_{q}^x\right)^2}{2 L_{q}},
		\label{equ:qfppotentiell}
	\end{align}
	mit\\
	\begin{align}
		E_J\left(\Phi_{\text{cjj}}^{x}\right)=\frac{\Phi_0 I_{q}^c}{2 \pi} \cos(\frac{\pi \Phi_{\text{cjj}}^{x}}{\Phi_0}),
	\end{align}	\\
	wobei $I_{q}^c=2 I_c$. $I_c$ entspricht dem kritischen Strom von einem Kontakt. Mithilfe des Lagrangeformalismus bekommen wir die Hamiltonfunktion\\
	\begin{align}
		H=\frac{Q_q^2}{2C_q}+U_{\text{eff}}\left(\Phi_q\right),
		\label{equ:hqfp}
	\end{align}\\
	mit $C_q=C_1+C_2$ der Kapazität der q-Schleife.\\
	
	So haben wir das nichtlineares Element Josephson-Kontakt erklärt und die Hamiltonfunktion von Flussqubits rf-SQUID sowie steuerbarem rf-SQUID hergeleitet. Im Folgenden wird das Quanten-Fluss-Parametron (QFP) vorgestellt.\\
	
	\section{Quanten-Fluss-Parametron (QFP)}
	\label{sec:qfp}
	
	Das Quanten-Fluss-Parametron (QFP) \cite{Hosoya91} ist ein steuerbares rf-SQUID (siehe Abb. \ref{fig:cjjrfsquid}). Das QFP enthält eine kleine Induktivität, eine große Kapazität und einen sehr großen kritischen Strom. Folglich besitzt das QFP eine wesentliche größere Tunnelbarriere im Vergleich zum Flussqubit, was seinen Zustand stabil gegenüber hochfrequenter Strahlung beim Auslesen macht \cite{Berkley2010}.\\
	
	Durch Einfügen eines QFPs zwischen Flussqubit und Resonator wird nicht nur das Stromsignal verstärkt, sondern auch das Qubit vom Resonator isoliert. Das Übersprechen zwischen Qubit und Resonator kann dadurch verbessert werden \cite{Przybysz2019}.\\
	
	Aus der Gl. \ref{equ:hqfp} bekommen wir die Hamiltonfunktion des QFPs\\	
	\begin{align}
		H_{\text{qfp}}&=\frac{Q^2}{2C}+\frac{\left(\Phi-\Phi_{z}^{\text{qfp}}\right)^2}{2L}+\beta\left(\Phi_{x}^{\text{qfp}}\right)\cos(\frac{2 \pi \Phi}{\Phi_0})
	\end{align}\\
	mit\\
	\begin{align}
		\beta\left(\Phi_{x}^{\text{qfp}}\right)=\frac{\Phi_0 I_{c}^{\text{qfp}}}{2\pi}\cos\left(\frac{\pi\Phi_{x}^{\text{qfp}}}{\Phi_0}\right)=E_J \cos\left(\frac{\pi\Phi_{x}^{\text{qfp}}}{\Phi_0}\right),
	\end{align}\\
	wobei $I_{c}^{\text{qfp}}$ dem kritischen Strom von einem Kontakt, $L$ der lineare Induktivität der Schleife, $C$ der Summe von beiden Kapazitäten und $E_J$ der Summe von beiden Josephson-Energien entsprechen. $\Phi_{z}^{\text{qfp}}$ und $\Phi_{x}^{\text{qfp}}$ sind der äußere magnetische Fluss von der großen und der kleinen Schleife respektive.\\
	
	Das QFP besteht aus einer kleinen cjj-Schleife, die die Energiebarriere des Potenzials steuert, und einer großen Hauptschleife, in der der Dauerstrom erzeugt wird und die die Symmetrie des Potenzials kontrolliert. Die Josephson-Kontakte des QFPs sind zehnmal so groß wie von dem Flussqubit, wodurch der Dauerstrom entsprechend zunimmt ist.\\
	
	\begin{figure}[ht]
		\centering
		\includegraphics[width=0.7\textwidth]{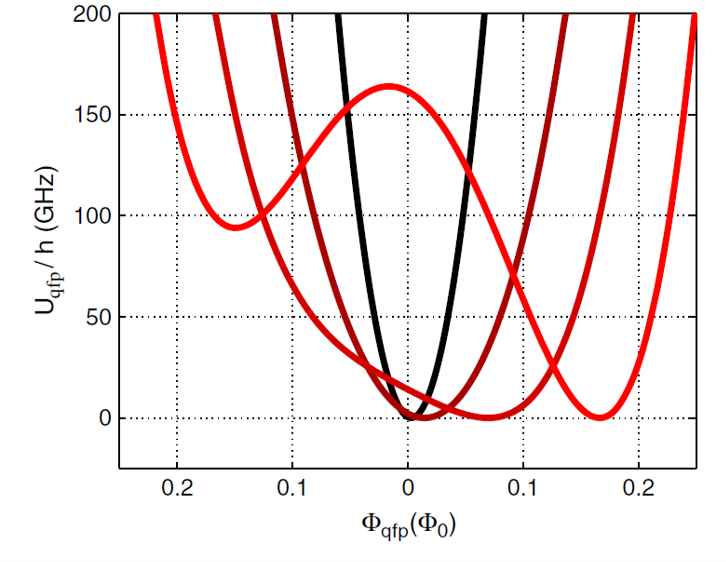}
		\caption{Potenzielle Energie beim QFP-Annealing. Durch Einschaltung des äußeren Magnetfeldes in der kleinen cjj-Schleife wird das Potenzial von monostabil (schwarz) zu bistabil (rot) umgewandelt. (Quelle:  \cite{Berkley2010})}
		\label{fig:qfp_potential}
	\end{figure}
	
	Durch Anlegen eines Magnetfeldes erzeugt man einen Dauerstrom in der größeren Schleife, der links oder rechtsherum zirkulieren kann. Die Richtung des Dauerstroms entscheidet dann den Qubitzustand, der $\ket{0}$ oder $\ket{1}$ ist. Dabei wird das Potenzial asymmetrisch sein (siehe Abb. \ref{fig:qfp_potential}). Durch die Kontrolle des magnetischen Flusses in der kleinen cjj-Schleife des QFP, wird die Barriere angehoben und dessen potenzielle Energie langsam von monostabil zu bistabil (schwarz zu rot) umgewandelt (siehe Abb. \ref{fig:qfp_potential}). Dieser Prozess wird auch als Annealing bezeichnet. Durch QFP-Annealing geht der Qubitzustand in den $\ket{0}$ oder $\ket{1}$. Im monostabilen Zustand kann man das QFP als einen Resonator und im bistabilen Zustand als Qubit betrachten.\\
	
	\section{Flussbasis und Energiebasis}

	Ein (reiner) quantenmechanischer Zustand kann als Vektor im sogenannten Hilbertraum beschrieben werden \cite{QM}. Die Messung eines quantenmechanischen Zustands entspricht einer Projektion des Zustands auf die Basisvektoren.
	Die Basis eines Vektorraums ist eine Menge von Vektoren in diesem Raum, die als Koordinaten für diesen Raum verwendet werden können. Jedes Paar von Vektoren, die linear unabhängig sind, könnte als Basis dienen.  Bei einem Quantensystem hängt die Wahl der Messbasis vom Messverfahren oder -gerät ab. Um Messfehler zu minimieren, ist es wichtig, die geeignete Basis zu benutzen \cite{Pommerening2020}. Hier werden zwei wichtige Basen für das Flussqubit vorgestellt.\\
	
	Zuerst ist die Fluss- oder Dauerstrombasis $\{\ket{\circlearrowright}, \ket{\circlearrowleft}\}$. Dabei sind $\ket{\circlearrowleft}$ und $\ket{\circlearrowright}$ die Zustände mit den links- und rechts zirkulierenden Dauerströme $I_p$. Die beiden Zustände können als Basisvektoren des Paulioperators $\hat{\sigma}_z$ ausgedrückt werden:\\
	\begin{align*}
		\ket{\circlearrowright}=\begin{pmatrix}
			1\\
			0
		\end{pmatrix}\quad \text{und}\quad \ket{\circlearrowleft}=\begin{pmatrix}
			0\\
			1
		\end{pmatrix}.
	\end{align*}\\
	Die Paulioperatoren lässt sich dann in der Flussbasis darstellen:\\
	\begin{align*}
		\hat{\sigma}_z=\ket{\circlearrowright}\bra{\circlearrowright}-\ket{\circlearrowleft}\bra{\circlearrowleft},\\ \hat{\sigma}_x=\ket{\circlearrowright}\bra{\circlearrowleft}+\ket{\circlearrowleft}\bra{\circlearrowright}.
	\end{align*}\\
	Die Physik der beiden tiefstgelegenen Zustände eines Flussqubits kann mit einem effektiven Hamiltonoperator in der Flussbasis $\{\ket{\circlearrowright}, \ket{\circlearrowleft}\}$ erfasst werden:\\	
\begin{align*}
	\hat{H}_{q}=-\frac{\hbar}{2}\left(\epsilon\hat{\sigma}_z  +\Delta\hat{\sigma}_x \right),
\end{align*}\\
wobei $\epsilon=2I_p\left(\Phi_{\text{cjj}}^{x}-\Phi_0/2\right)$ die Qubit-Magnetenergie ist.
Dabei kontrolliert der diagonale Term $\epsilon$ durch die Einstellung von $\Phi_{\text{cjj}}^{x}$ die Asymmetrie des  Doppelmuldenpotenzials. Das Außerdiagonalelement $\Delta$ beschreibt die Tunnelrate, die von der Höhe der Barriere und damit von der Josephson-Energie $E_J$ abhängig ist. Das heißt, durch Einstellung des äußeren magnetischen Flusses $\Phi_{\text{cjj}}^{x}$ von der cjj-Schleife (siehe Abb. \ref{fig:cjjrfsquid}) und $\Phi_{q}^x$ von der q-Schleife werden die Energiebarriere und die Symmetrie des Doppelmuldenpotenzials entsprechend verändert (siehe Abb. \ref{fig:qfp_potential}).\\

Durch Anwenden einer unitären Transformation\\
\begin{align*}
	\hat{U}=\begin{pmatrix}
		\phantom{0}\cos(\theta/2) & \sin(\theta/2)\\
		-\sin(\theta/2) & \cos(\theta/2)
	\end{pmatrix},
\end{align*}\\
wird der Hamiltonoperator des Flussqubits\\
\begin{align*}
	\hat{H}_{q}=-\frac{\hbar}{2}\left(\epsilon\hat{\sigma}_z  +\Delta\hat{\sigma}_x \right)=-\frac{\hbar}{2} \begin{pmatrix}
		\epsilon & \Delta \\ \Delta & -\epsilon
	\end{pmatrix},
\end{align*}\\
diagonalisiert und
in der Energiebasis darstellt, wobei $\theta$ durch $\tan(\theta)=\Delta/\epsilon$ definiert ist. So ergibt sich\\
\begin{align*}
	\tilde{\hat{H}}_{q}=\hat{U} \hat{H}_{q} \hat{U}^\dagger=-\frac{\hbar\sqrt{\Delta^2+\epsilon^2}}{2}\begin{pmatrix}
		1 & \phantom{0}0\\
		0 & -1
	\end{pmatrix}=-\frac{\hbar\omega_{q}}{2}\hat{\tilde{\sigma}}_z,
\end{align*}\\
mit $\omega_{q}$ der Qubitfrequenz. Dabei wird $\hat{\tilde{\sigma}}_z=\ket{g}\bra{g}-\ket{e}\bra{e}$ in der Energiebasis $\left\{\ket{g}, \ket{e}\right\}$ dargestellt. Die Beziehung zwischen Flussbasis und Energiebasis lässt sich wie folgt darstellen:\\
\begin{align*}
	\ket{g}&=\hat{U}\ket{\circlearrowright}=\begin{pmatrix}
		\phantom{0}\cos(\theta/2)\\
		-\sin(\theta/2)
	\end{pmatrix}=\cos(\theta/2)\ket{\circlearrowright}-\sin(\theta/2)\ket{\circlearrowleft},\\ \ket{e}&=\hat{U}\ket{\circlearrowleft}=\begin{pmatrix}
		\sin(\theta/2)\\
		\cos(\theta/2)
	\end{pmatrix}=\sin(\theta/2)\ket{\circlearrowright}+\cos(\theta/2)\ket{\circlearrowleft}.
\end{align*}

\clearpage
\thispagestyle{empty}

	\chapter{Jaynes-Cummings-Modell (JCM)}
	
	Die Grundlage der meisten Messschemata für supraleitende Qubits ist die Wechselwirkung des Qubits mit einem Mikrowellenresonator. Diese Wechselwirkung lässt sich durch den Rabi-Hamiltonoperator beschreiben. Wenn die Kopplung zwischen Qubit und Resonator klein genug ist, sodass die Drehwellennäherung (\textit{Rotating Wave Approximation}) (RWA) gültig ist, führt es zum Jaynes-Cummings-Hamiltonoperator \cite{Jaynes1963}. In diesem Kapitel wird verdeutlicht, wie man ein Qubit durch einen Resonator auslesen kann. Der Bare- und Dressed-Zustand werden dabei vorgestellt.\\
	
	\section{JCM}
	
	Die Wechselwirkung zwischen Resonator und Qubit lässt sich durch den Rabi-Hamiltonoperator beschreiben\\
	\begin{align}
		\hat{H}_{\text{Rabi}}=\underbrace{\frac{\hbar}{2}\omega_q \hat{\sigma}_z}_{\hat{H}_{q}}+\underbrace{\hbar\omega_r \left(\hat{a}^{\dagger}\hat{a}+\frac{1}{2}\right)}_{\hat{H}_{\text{r}}}+\underbrace{\hbar g \left(\hat{a}+\hat{a}^{\dagger}\right)\hat{\sigma}_x}_{\hat{H}_{\text{int}}},
		\label{equ:H_rabi}
	\end{align}\\
	wobei $\hbar=h/2\pi$, $h$ der Planck-Konstante, $g=\Omega_0 / 2$ der  Qubit-Resonator-Kopplungs-stärke entsprechen. $\Omega_0$ ist dabei die Vakuum-Rabifrequenz. $\omega_q$ und $\omega_r$ sind die Qubit- und Resonatorfrequenz respektive. $\hat{a}^{\dagger}$ und $\hat{a}$ sind die Erzeugungs- und Vernichtungsoperatoren der Resonatormode und erfüllen die  bosonischen Kommutatorrelationen: $[\hat{a}, \hat{a}^{\dagger}]=1, [\hat{a}, \hat{a}]=[\hat{a}^{\dagger}, \hat{a}^{\dagger}]=0$. $\hat{\sigma}_x=\hat{\sigma}_{+}+\hat{\sigma}_{-}$ ist die erste Pauli-Matrix. $\hat{\sigma}_{+}, \hat{\sigma}_{-}$ sind die Leiteroperatoren des Qubits und erfüllen $[\hat{\sigma}_-,\hat{\sigma}_+]=-\hat{\sigma}_z$. Hier wird angenommen, dass zwischen Qubit und Resonator eine Dipolkopplung besteht, d. h., die Wechselwirkung vollständig außerdiagonal ist. \\
	
	Transformieren wir $\hat{H}_{\text{Rabi}}$ in einem rotierenden Bezugssystem durch Anwendung der unitären Transformation\\
	\begin{align*}
		\hat{U}_r=\exp(i\omega_{r} t \hat{a}^\dagger \hat{a}-i\omega_q t \hat{\sigma}_z /2),
	\end{align*}\\
	ergibt sich\\
	\begin{align*}
		\tilde{\hat{H}}_{\text{Rabi}}=\hat{U}\hat{H}_{\text{Rabi}}\hat{U}^\dagger+i\hat{\dot{U}}\hat{U}^\dagger.
	\end{align*}\\
	Mithilfe der Baker-Campbell-Hausdorff-Formel (BCHF)\\	
	\begin{align}
		e^{\hat{S}}\hat{H}e^{-\hat{S}}=\hat{H}+\left[\hat{S}, \hat{H}\right]+\frac{1}{2!}\left[\hat{S},\left[\hat{S},\hat{H}\right]\right]+\cdots
		\label{equ:bchf}
	\end{align}\\
	erhalten wir\\
	\begin{align*}
		\tilde{\hat{H}}_{\text{Rabi}}=\hbar g\left(\hat{a}\hat{\sigma}_{+}e^{i\left(\omega_q-\omega_r\right)}+\hat{a}^\dagger\hat{\sigma}_{-}e^{i\left(\omega_r-\omega_q\right)}+\hat{a}\hat{\sigma}_{-}e^{-i\left(\omega_q+\omega_r\right)}+\hat{a}^\dagger\hat{\sigma}_{+}e^{i\left(\omega_q+\omega_r\right)}\right).
	\end{align*}\\
	Unter der Bedingung $\left(\omega_q + \omega_r\right) \gg \left\{g, \abs{\omega_q - \omega_r}\right\}$, also die Qubit- und Resonatorfrequenz befinden sich nahe der Resonanz, ist die Drehwellennäherung gültig. Das heißt, die beiden schnell oszillierenden Terme, mit $\left(\omega_q+\omega_r\right)$ im Exponenten der $e$-Funktion, können vernachlässigt werden. So lässt sich der Hamiltonoperator $\hat{H}_{\text{Rabi}}$ (Gl. \ref{equ:H_rabi}) durch den Jaynes-Cummings-Hamiltonoperator zurück im Schrödingerbild darstellen:\\
	\begin{align}
		\hat{H}_{JC}=\underbrace{\frac{\hbar}{2}\omega_q \hat{\sigma}_z}_{\hat{H}_q}+\underbrace{\hbar\omega_r \left(\hat{a}^{\dagger}\hat{a}+\frac{1}{2}\right)}_{\hat{H}_r}+\underbrace{\hbar g \left(\hat{\sigma}_{+}\hat{a}+ \hat{\sigma}_{-}\hat{a}^{\dagger}\right)}_{\hat{H}_{\text{int}}}.
		\label{equ:H_JC}
	\end{align}\\
	Durch Diagonalisierung erhalten wir die Eigenwerte im Unterraum $\ket{g,n+1}, \ket{e,n}$:\\
	\begin{align}
		E_{\pm,n}=\hbar \omega_r \left(n+\frac{1}{2}\right)\pm \frac{\hbar}{2}\sqrt{\delta^2+\Omega_0^2\left(n+1\right)},
	\end{align}\\
	wobei $\delta=\omega_q-\omega_r$ die Verstimmung zwischen Qubit- und Resonatorfrequenz ist. Die entsprechenden Eigenvektoren sind\\
	\begin{align}
		\begin{split}
			\ket{+,n}&=\cos(\frac{\theta_{n}}{2})\ket{e,n}+\sin(\frac{\theta_n}{2})\ket{g,n+1},\\
			\ket{-,n}&=-\sin(\frac{\theta_{n}}{2})\ket{e,n}+\cos(\frac{\theta_n}{2})\ket{g,n+1},
		\end{split}
		\label{equ:dressedstate}
	\end{align}\\
	wobei $\theta_n$ durch $\tan(\theta_n)=\Omega_0 \sqrt{n+1}/\delta$ definiert ist.\\
	
	Die beiden in Gl. \ref{equ:dressedstate} angegebenen Zustände werden in der Quantenoptik und Atomphysik als Dressed-Zustände (\textit{dressed states})  bezeichnet, während $\ket{g,n+1}$ und $\ket{e,n}$ Bare-Zustände (\textit{bare states}) genannt werden.\\
	
	\section{Dispersive Messung}
	
	Unter den Bedingungen $\omega_r \neq \omega_q$ und $g \ll \abs{\delta}$ befinden sich das Qubit und das Resonatorfeld im dispersiven Regime. In diesem Fall gibt es keine resonante Photonenabsorption oder -emission.\\
	
	Um den Jaynes-Cummings-Hamiltonoperator (Gl. \ref{equ:H_JC}) im dispersiven Regime darzustellen, können wir die Methode der Schrieffer-Wolff-Transformation (SWT) \cite{Blais2004} \cite{Gu2017} benutzen. Das heißt, wir wählen eine unitäre Transformation $\hat{U}=\exp(S)$, sodass $\left[\hat{S}, \hat{H}_0\right]=-\hat{H}_{\text{int}}$. Diese Wahl führt zum Vereinfachen von BCHF\\
	\begin{align*}
		e^{\hat{S}}\hat{H}e^{-\hat{S}}=\hat{H}+\left[\hat{S}, \hat{H}\right]+\frac{1}{2!}\left[\hat{S},\left[\hat{S},\hat{H}\right]\right]+\cdots=\hat{H}_0+\frac{1}{2}\left[\hat{S},\hat{H}_{\text{int}}\right]+\cdots
	\end{align*}\\
	Mithilfe der Kommutatorrelationen\\
	\begin{align*}
		\begin{array}{ll}
			\left[\hat{a},\hat{a}^\dagger\right]=1, & \left[\hat{\sigma}_+, \hat{\sigma}_{-}\right]=\hat{\sigma}_z,\\
			\left[\hat{a}, \hat{a}^\dagger\hat{a}\right]=\hat{a}, & \left[\hat{a}^\dagger, \hat{a}^\dagger\hat{a}\right]=-\hat{a}^\dagger,\\
			\left[\hat{\sigma}_+, \hat{\sigma}_z\right]=-2\hat{\sigma}_+, & \left[\hat{\sigma}_-, \hat{\sigma}_z\right]=2\hat{\sigma}_-,
		\end{array}
	\end{align*}\\
	erhalten wir\\
	\begin{align*}
		&\left[\hat{a}\hat{\sigma}_+ -\hat{a}^\dagger\hat{\sigma}_-, \hat{a}^\dagger\hat{a}\right]=\hat{a}\hat{\sigma}_+ +\hat{a}^\dagger\hat{\sigma}_-,\\
		&\left[\hat{a}\hat{\sigma}_+ -\hat{a}^\dagger\hat{\sigma
		}_-, \hat{\sigma}_z\right]=-2\left(\hat{a}\hat{\sigma}_+ +\hat{a}^\dagger\hat{\sigma}_-\right),
	\end{align*}\\
	und\\
	\begin{align*}
		\left[\hat{a}\hat{\sigma}_+ -\hat{a}^\dagger\hat{\sigma}_-, \hat{a}\hat{\sigma}_+ -\hat{a}^\dagger\hat{\sigma}_-\right]&=2\left(\mathbb{1}+\hat{\sigma}_z+\hat{a}^\dagger\hat{a}\hat{\sigma}_z\right)\\
		&=2\left(\hat{a}^\dagger\hat{a}+1\right)\hat{\sigma}_z+2.
	\end{align*}\\
	Daraus bekommen wir $\hat{S}=\lambda\left(\hat{a}\hat{\sigma}_+ -\hat{a}^\dagger\hat{\sigma}_-\right)$ mit $\lambda=g/\delta$ und $\delta=\omega_q-\omega_r$. Durch Anwenden der Transformation $\hat{U}=\exp(\lambda\left(\hat{a}\hat{\sigma}_+ -\hat{a}^\dagger\hat{\sigma}_-\right))$ auf verschiedene Operatoren erhalten wir\\
	\begin{align*}
		&U \hat{\sigma}_z \hat{U}^\dagger=\hat{\sigma}_z-2\lambda\left(\hat{\sigma}_+ \hat{a}+\hat{\sigma}_- \hat{a}^\dagger\right)-2\lambda^2\left(\hat{a}^\dagger\hat{a}+\frac{1}{2}\right)\hat{\sigma}_z+\mathcal{O}(\lambda^3),\\
		&U \left(\hat{a}^\dagger\hat{a} + \frac{1}{2}\right) \hat{U}^\dagger=\hat{a}^\dagger\hat{a}+\frac{1}{2}+\lambda\left(\hat{\sigma}_+ \hat{a}+\hat{\sigma}_- \hat{a}^\dagger\right)+\lambda^2\left(\hat{a}^\dagger\hat{a}+\frac{1}{2}\right)\hat{\sigma}_z+\mathcal{O}(\lambda^3),\\
		&U \left(\hat{\sigma}_+ \hat{a}+\hat{\sigma}_- \hat{a}^\dagger\right) \hat{U}^\dagger=\hat{\sigma}_+ \hat{a}+\hat{\sigma}_- \hat{a}^\dagger+2\lambda\left(\hat{a}^\dagger\hat{a}+\frac{1}{2}\right)\hat{\sigma}_z+\mathcal{O}(\lambda^2).
	\end{align*}\\
	So wird der Jaynes-Cummings-Hamiltonoperator im dispersiven Regime dargestellt:\\
	\begin{align*}
		\hat{H}_{\text{disp}}
		&= \frac{\hbar}{2}\omega_q U \hat{\sigma}_z \hat{U}^\dagger+\hbar\omega_r U \left(\hat{a}^\dagger\hat{a} + \frac{1}{2}\right) \hat{U}^\dagger+\hbar g U \left(\hat{\sigma}_+ \hat{a}+\hat{\sigma}_- \hat{a}^\dagger\right) \hat{U}^\dagger\\
		&=\hbar\left(\omega_r+\chi\hat{\sigma}_z\right)\left(\hat{a}^\dagger\hat{a}+\frac{1}{2}\right)+\frac{\hbar}{2}\omega_q \hat{\sigma}_z+\mathcal{O}(\lambda^2).
	\end{align*}\\
	$\chi = g^2/\delta$ ist dabei die dispersive Kopplungsstärke des Resonators an das Qubit.
	Dann bekommen wir den Jaynes-Cummings-Hamiltonoperator bis zur ersten Ordnung in $\lambda$ im dispersiven Regime:\\
	\begin{align}
		\hat{H}_{\text{disp}}\approx\hbar\left(\omega_r +\chi \hat{\sigma}_z\right)\hat{a}^{\dag}\hat{a}+\frac{\hbar}{2}\left(\omega_q+\chi\right)\hat{\sigma}_z.
		\label{equ:H_disp}
	\end{align}\\	
	Aus dem ersten Term können wir schließen, dass die effektive Resonatorfrequenz eine vom Qubitzustand abhängige Verschiebung hat. Es ist daher möglich, eine zerstörungsfreie Quantenmessung (\textit{Quantum Nondemolition Measurement}) (QND-Messung) des Qubits durchzuführen, indem man die Mikrowellenübertragung in der Nähe der Resonatorfrequenz überwacht.\\
	
	Bei QND-Messungen ist sowohl der Qubit-Hamilton- als auch der Wechselwirkungsoperator der Kopplung zwischen System und Messapparatur mit der zu messenden Größe vertauschbar. Das bedeutet, die Messung führt bei wiederholter Ausführung zum gleichen Resultat.
	Die Gl. \ref{equ:H_disp} kann umgeschrieben werden zu\\
	\begin{align}
		\hat{H}_{\text{disp}}\approx\hbar\omega_r\left(\hat{a}^{\dagger}\hat{a}+\frac{1}{2}\right)+\hbar\left[\frac{\omega_q}{2}+\left(\hat{a}^{\dag}\hat{a}+\frac{1}{2}\right)\chi\right]\hat{\sigma}_z.
		\label{equ:H_disp_2}
	\end{align}\\
	Obwohl
	\begin{align*}
		\left[\hat\sigma_z,~ \hat{H}_q\right]=\frac{\hbar}{2}\omega_q \left[\hat\sigma_z,~ \hat{\sigma}_z\right]=0,
	\end{align*}\\
	erfüllt ist, ist die Messung nicht prinzipiell QND, da:\\
	\begin{align*}
		\left[\hat\sigma_z,~ \hat{H}_{\text{int}}\right]=\hbar g \left[\hat\sigma_z,~ \left(\hat{\sigma}_{+}\hat{a}+ \hat{\sigma}_{-}\hat{a}^{\dagger}\right)\right]=2\hbar g\left(\hat \sigma_+ \hat{a}-\hat{\sigma}_-\hat{a}^\dagger\right)\neq 0.
	\end{align*}

	\subsection*{Notation und Rechnungsregeln}
	
	Hier werden die in der Masterarbeit benutzte Notation und Rechnungsregeln gegeben.
	Die Pauli Matrizen sind\\
	\begin{align}
		\hat{\sigma}_{x}=\begin{bmatrix}
			0&1\\
			1&0\\
		\end{bmatrix},~
		\hat{\sigma}_{y}=\begin{bmatrix}
			0&-i\\
			i&0\\
		\end{bmatrix},~
		\hat{\sigma}_{z}=\begin{bmatrix}
			1&0\\
			0&-1\\
		\end{bmatrix}.
	\end{align}\\
	Dadurch definieren wir die Leiteroperatoren (Erzeugungs- und Vernichtungsoperator):\\
	\begin{align}
		\hat{\sigma}_{+}&=\frac{1}{2}\left(\hat{\sigma}_{x}+i \hat{\sigma}_{y}\right)=\begin{bmatrix}
			0 & 1\\
			0 & 0
		\end{bmatrix},\\
		\hat{\sigma}_{-}&=\frac{1}{2}\left(\hat{\sigma}_{x}-i \hat{\sigma}_{y}\right)=\begin{bmatrix}
			0 & 0\\
			1 & 0
		\end{bmatrix}.
	\end{align}\\
	Die Rechnungsregeln zwischen zwei Qubits 1 und 2 lautet:\\
	\begin{align}
		\hat{\sigma}_{1}^i&=\hat{\sigma}_{1}^i \otimes \mathbb{1}\\
		\hat{\sigma}_{2}^{i}&= \mathbb{1} \otimes \hat{\sigma}_{2}^{i}\\
		\hat{\sigma}_{1}^{i}\hat{\sigma}_{2}^{i}&=\hat{\sigma}_{1}^{i}\otimes\hat{\sigma}_{2}^{i},
	\end{align}\\
	wobei $i\in \left\{x,y,z\right\}$. So erhalten wir\\
	\begin{align}
		\begin{split}
			\hat{\sigma}_{1}^z &=\hat{\sigma}_{1}^z \otimes \mathbb{1} = \begin{bmatrix}
				1 & 0 \\
				0 & -1 \\
			\end{bmatrix} \otimes \begin{bmatrix}
				1 & 0 \\
				0 & 1 \\
			\end{bmatrix} = \begin{bmatrix}
				1 & 0 & 0 & 0\\
				0 & 1 & 0 & 0\\
				0 & 0 & -1 & 0\\
				0 & 0 & 0 & -1\\
			\end{bmatrix} \\
			&=\ket{00}\bra{00}+\ket{01}\bra{01}-\ket{10}\bra{10}-\ket{11}\bra{11}.
		\end{split}
	\end{align}\\

	Analog dazu bekommen wir auch\\
	\begin{align}
		\begin{split}
			\hat{\sigma}_{2}^z&=\mathbb{1} \otimes \hat{\sigma}_{2}^z=\ket{00}\bra{00}-\ket{01}\bra{01}+\ket{10}\bra{10}-\ket{11}\bra{11},
		\end{split}
	\end{align}
	\begin{align}
		\begin{split}
			\hat{\sigma}_{1}^x \hat{\sigma}_{2}^x&=\hat{\sigma}_{1}^x \otimes \hat{\sigma}_{2}^x =\ket{00}\bra{11}+ \ket{01}\bra{10} +\ket{10}\bra{01}+\ket{11}\bra{00},
		\end{split}
	\end{align}
	\begin{align}
		\begin{split}
			\hat{\sigma}_{1}^{y} \hat{\sigma}_{2}^{y}&=\hat{\sigma}_{1}^{y} \otimes \hat{\sigma}_{2}^{y}=-\ket{00}\bra{11}+ \ket{01}\bra{10} +\ket{10}\bra{01}-\ket{11}\bra{00}.
		\end{split}
	\end{align}\\
	
	\clearpage
	\thispagestyle{empty}

	\chapter{Auslesen von zwei gekoppelten Qubits}
	\label{messmethode}
	
	In diesem Kapitel wird die in \cite{Pommerening2020} benutzte Methode zum Auslesen von zwei gekoppelten Qubits vorgestellt, sodass man die Messung in verschiedenen Basen vergleichen kann. Dabei wird die Berechnung aus \cite{Pommerening2020} ergänzt. Teile der Methode werden dann in den folgenden Kapiteln benutzt.\\

\section{Modell}

In \cite{Pommerening2020} wird ein Protokoll zum Auslesen von zwei durch einen Resonator gekoppelten Qubits vorgestellt. Es wird verdeutlicht, wie Messgeschwindigkeit und -intensität auf das System einwirken und in welcher Basis zu messen vorteilhaft ist, um niedrigere Messfehler zu bekommen.\\

\begin{figure}[ht]
	\centering
	\includegraphics[width=1\textwidth]{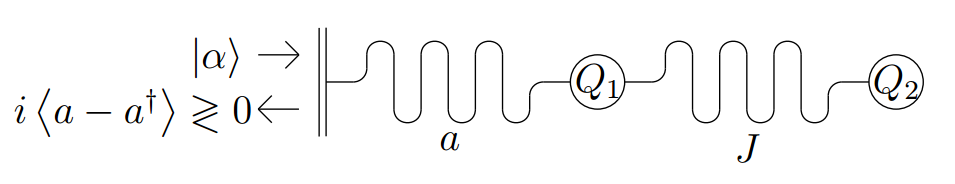}
	\caption{Auslesen von zwei durch einen Resonator gekoppelten Qubits mit $J$ der Kopplungsstärke zwischen beiden Qubits. Qubit 1 wird mittels Resonator ausgelesen. (Quelle: \cite{Pommerening2020})}
	\label{fig:auslesenschema}
\end{figure}

Das Ausleseschema wird in Abb. \ref{fig:auslesenschema} dargestellt.\\

Zwei durch einen Resonator (Bus) gekoppelte Qubits (siehe Abb. \ref{fig:auslesenschema}) haben eine durch virtuelle Photonen vermittelte Austauschwechselwirkung\\
\begin{align}
	\hat{H}_{\text{int}}=J\left(\hat{\sigma}^{+}_{1}\hat{\sigma}^{-}_{2}+\hat{\sigma}^{-}_{1}\hat{\sigma}^{+}_{2}\right)=\frac{J}{2} \left(\hat{\sigma}_1^x \hat{\sigma}_2^x + \hat{\sigma}_1^y \hat{\sigma}_2^y\right).
\end{align}\\
Wenn man den Ausleseresonator in seinem rotierenden Bezugssystem bewegt und die Lamb-Verschiebung vernachlässigt \cite{Pommerening2020}, bleibt sich dann nur die Qubit ac-Stark-Verschiebung $\chi$. Der Hamiltonoperator des Systems in der Bare-Basis ist daher\\
\begin{align}
	\hat{H}=-\frac{\omega_{1}}{2} \hat{\sigma}_1^z -\frac{\omega_{2}}{2} \hat{\sigma}_2^z +\frac{J}{2} \left(\hat{\sigma}_1^x \hat{\sigma}_2^x + \hat{\sigma}_1^y \hat{\sigma}_2^y\right) + \chi \hat{\sigma}_1^z \hat{a}^{\dag}\hat{a}.
	\label{equ.hamilton1}
\end{align}\\	
Man kann den Hamiltonoperator in der Dressed-Basis (im Folgenden mit Tilde indiziert) diagonalisieren. Bare- und Dressed-Basis haben folgende Beziehung\\	
\begin{align}
	\begin{split}
		\ket{\tilde{0} \tilde{0}} &= \ket{00},\\
		\ket{\tilde{0} \tilde{1}} &=\cos(\frac{\gamma_0}{2}) \ket{0 1}+\sin(\frac{\gamma_0}{2}) \ket{1 0},\\
		\ket{\tilde{1} \tilde{0}}&= -\sin(\frac{\gamma_0}{2})\ket{0 1}+\cos(\frac{\gamma_0}{2})\ket{1 0},\\
		\ket{\tilde{1} \tilde{1}} &= \ket{11}.
	\end{split}
	\label{equ:bdb1}
\end{align}\\	
$\tilde{\hat{\sigma}}_1^z=\ket{\tilde{0} \tilde{0}}\bra{\tilde 0 \tilde 0}+\ket{\tilde 0 \tilde 1}\bra{\tilde 0 \tilde 1}-\ket{\tilde 1 \tilde 0}\bra{\tilde 1 \tilde 0}-\ket{\tilde 1 \tilde 1}\bra{\tilde 1 \tilde 1}$ wird in der Dressed-Basis dargestellt.\\

Um den Hamiltonoperator in der Dressed-Basis darzustellen, finden wir zuerst die Beziehung zwischen $\hat{\sigma}_i$ und $\tilde{\hat{\sigma}}_i$, wobei $\hat{\sigma}_i$ ein beliebiger Paulioperator in der Bare-Basis ist. Durch die Beziehung zwischen Bare- und Dressed-Basis (\ref{equ:bdb1}) lässt sich $\tilde{\hat{\sigma}}_1^z$, $\tilde{\hat{\sigma}}_2^z$, $\tilde{\hat{\sigma}}_1^x \tilde{\hat{\sigma}}_2^x$ und $\tilde{\hat{\sigma}}_1^y \tilde{\hat{\sigma}}_2^y$ durch $\hat{\sigma}_1^z$, $\hat{\sigma}_2^z$, $\hat{\sigma}_1^x \hat{\sigma}_2^x$ und $\hat{\sigma}_1^y \hat{\sigma}_2^y$ darstellen:\\
\begin{align*}
		\tilde{\hat{\sigma}}_1^z&
		= \frac{1}{2} \left(\hat{\sigma}_1^z +\hat{\sigma}_2^z\right)+\frac{\cos(\gamma_0)}{2}  \left(\hat{\sigma}_1^z -\hat{\sigma}_2^z\right)+\frac{\sin(\gamma_0)}{2} \left(\hat{\sigma}_1^x  \hat{\sigma}_2^x + \hat{\sigma}_1^y \hat{\sigma}_2^y\right),\\
		\tilde{\hat{\sigma}}_2^z
		&=\frac{1}{2} \left(\hat{\sigma}_1^z +\hat{\sigma}_2^z\right)-\frac{\cos(\gamma_0)}{2}  \left(\hat{\sigma}_1^z -\hat{\sigma}_2^z\right)-\frac{\sin(\gamma_0)}{2} \left(\hat{\sigma}_1^x  \hat{\sigma}_2^x + \hat{\sigma}_1^y \hat{\sigma}_2^y\right),\\
		\tilde{\hat{\sigma}}_1^x \tilde{\hat{\sigma}}_2^x
		&=-\frac{\sin(\gamma_0)}{2}  \left(\hat{\sigma}_1^z -\hat{\sigma}_2^z\right)+\frac{\cos(\gamma_0)}{2} \left(\hat{\sigma}_1^x  \hat{\sigma}_2^x + \hat{\sigma}_1^y \hat{\sigma}_2^y\right)+\frac{1}{2} \left(\hat{\sigma}_1^x  \hat{\sigma}_2^x - \hat{\sigma}_1^y \hat{\sigma}_2^y\right),\\
		\tilde{\hat{\sigma}}_1^y \tilde{\hat{\sigma}}_2^y
		&=-\frac{\sin(\gamma_0)}{2}  \left(\hat{\sigma}_1^z -\hat{\sigma}_2^z\right)+\frac{\cos(\gamma_0)}{2} \left(\hat{\sigma}_1^x  \hat{\sigma}_2^x + \hat{\sigma}_1^y \hat{\sigma}_2^y\right)-\frac{1}{2} \left(\hat{\sigma}_1^x  \hat{\sigma}_2^x - \hat{\sigma}_1^y \hat{\sigma}_2^y\right).
\end{align*}\\
So ergibt sich\\
\begin{align}
	\begin{split}
		\begin{bmatrix}
			\tilde{\hat{\sigma}}_1^z+\tilde{\hat{\sigma}}_2^z \\
			\tilde{\hat{\sigma}}_1^z-\tilde{\hat{\sigma}}_2^z \\
			\tilde{\hat{\sigma}}_1^x\tilde{\hat{\sigma}}_2^x + \tilde{\hat{\sigma}}_1^y\tilde{\hat{\sigma}}_2^y\\
		\end{bmatrix}=\begin{bmatrix}
			1&0&0\\
			0&\cos(\gamma_0)&\sin(\gamma_0)\\
			0&-\sin(\gamma_0)&\cos(\gamma_0)\\
		\end{bmatrix} \begin{bmatrix}
			\hat{\sigma}_1^z+\hat{\sigma}_2^z \\
			\hat{\sigma}_1^z-\hat{\sigma}_2^z\\
			\hat{\sigma}_1^x \hat{\sigma}_2^x+\hat{\sigma}_1^y \hat{\sigma}_2^y
		\end{bmatrix},
	\end{split}
\end{align}\\
oder\\
\begin{align}
	\begin{split}
		\begin{bmatrix}
			\hat{\sigma}_1^z+\hat{\sigma}_2^z \\
			\hat{\sigma}_1^z-\hat{\sigma}_2^z\\
			\hat{\sigma}_1^x \hat{\sigma}_2^x+\hat{\sigma}_1^y \hat{\sigma}_2^y\\
		\end{bmatrix}=\begin{bmatrix}
			1 & 0 & 0\\
			0 & \cos(\gamma_0) & -\sin(\gamma_0)\\
			0 & \sin(\gamma_0) & \cos(\gamma_0)\\
		\end{bmatrix} \begin{bmatrix}
			\tilde{\hat{\sigma}}_1^z+\tilde{\hat{\sigma}}_2^z \\
			\tilde{\hat{\sigma}}_1^z-\tilde{\hat{\sigma}}_2^z \\
			\tilde{\hat{\sigma}}_1^x\tilde{\hat{\sigma}}_2^x + \tilde{\hat{\sigma}}_1^y\tilde{\hat{\sigma}}_2^y\\
		\end{bmatrix}.\\
	\end{split}
\end{align}\\
Der Hamiltonoperator (\ref{equ.hamilton1}) lässt sich in der Dressed-Basis mit $\delta_0=\frac{\omega_{2}-\omega_{1}}{2}$,~$\sin(\gamma_0)=\frac{\sgn\left(\delta_{0}\right)J}{\sqrt{\delta_{0}^{2}+J^2}}$ und $\cos(\gamma_0)=\frac{\abs{\delta_{0}}}{\sqrt{\delta_{0}^{2}+J^2}}$ schreiben:\\
\begin{align*}
	\begin{split}
		\hat{H}&=-\frac{\omega_{1}}{2} \hat{\sigma}_1^z -\frac{\omega_{2}}{2} \hat{\sigma}_2^z +\frac{J}{2} \left(\hat{\sigma}_1^x \hat{\sigma}_2^x + \hat{\sigma}_1^y \hat{\sigma}_2^y\right) + \chi \hat{\sigma}_1^z \hat{a}^{\dag}\hat{a}\\
		&=\frac{1}{2}\left(-\frac{\omega_{1}+\omega_{2}}{2}+\chi \hat{a}^{\dag}\hat{a}\right)\left(\tilde{\hat{\sigma}}_1^z+\tilde{\hat{\sigma}}_2^z\right)+\frac{1}{2}\left(\left(\delta_0+\chi \hat{a}^{\dag}\hat{a}\right)\cos(\gamma_0)+J\sin(\gamma_0)\right)\left(\tilde{\hat{\sigma}}_1^z-\tilde{\hat{\sigma}}_2^z\right)\\
		&~~~+\frac{1}{2}\left(\left(-\delta_{0}-\chi \hat{a}^{\dag}\hat{a}\right)\sin(\gamma_0)+J\cos(\gamma_0)\right)\left(\tilde{\hat{\sigma}}_1^x\tilde{\hat{\sigma}}_2^x + \tilde{\hat{\sigma}}_1^y\tilde{\hat{\sigma}}_2^y\right)\\
		&=\frac{1}{2}\left(-\frac{\omega_{1}+\omega_{2}}{2}+\chi \hat{a}^{\dag}\hat{a}\right)\left(\tilde{\hat{\sigma}}_1^z+\tilde{\hat{\sigma}}_2^z\right)+\frac{1}{2}\left(\sgn\left(\delta_{0}\right)\sqrt{\delta_{0}^{2}+J^2}+ \frac{\chi \abs{\delta_{0}}}{\sqrt{\delta_{0}^{2}+J^2}}\hat{a}^{\dag}\hat{a}\right)\left(\tilde{\hat{\sigma}}_1^z-\tilde{\hat{\sigma}}_2^z\right)\\
		&~~~- \frac{ J \chi \sgn\left(\delta_{0}\right)}{2\sqrt{\delta_{0}^{2}+J^2}}\hat{a}^{\dag}\hat{a}\left(\tilde{\hat{\sigma}}_1^x\tilde{\hat{\sigma}}_2^x + \tilde{\hat{\sigma}}_1^y\tilde{\hat{\sigma}}_2^y\right).
	\end{split}
\end{align*}\\
So haben wir den Hamiltonoperator in den Bare- und Dressed-Basen dargestellt. Im Folgenden stellen wir das Messprinzip und die Bedingung zum Vergleichen der Basen vor.\\

\section{Messprinzip}

Eine Messung der Qubitzustände kann wie folgt durchgeführt werden \cite{Pommerening2020}:

\begin{enumerate}
	\item Initialisiere den Resonator in einem kohärenten Zustand $\ket{\alpha}$ (sei $\alpha$ reell und positiv).
	\item Lass den Resonator mit dem/den Qubit(s) für das Zeitintervall $t_m = \pi/\left(2\abs{\chi}\right)$ wechselwirken, was durch den Zeitentwicklungsoperator $U\left(t \right) = \exp(-iHt)$ beschrieben wird.
	\item Auslesen des Resonators durch Positive Operator Valued (Probability) Measure (POVM) mit Elementen\\
	\begin{align}
		E_{\pm}=\frac{1}{\pi} \int_{\Omega_{\pm}}d^{2}\beta \ket{\beta}\bra{\beta}, ~E_{+}+E_{-}=\mathbb{1}.
	\end{align}\\
	Die Integrale sind über kohärente Zustände in der unteren $\left(\Omega_{\sgn \chi }\right)$ und oberen Halbebene $\left(\Omega_{-\sgn \chi }\right)$.
	\item  Miss den Resonator.\\
\end{enumerate}

Dann ist der zugehörige Superoperator, der die Wirkung der Messung mit dem Ergebnis $ x \in \{\pm \}$ auf einen anfänglichen Zwei-Qubit-Zustand $\rho$ beschreibt \cite{Pommerening2020}\\
\begin{align}
	\begin{split}
		\mathcal{E}_{x}\left(\rho\right)&=\tr_{r}E_{x}U\left(t_m\right) \rho \otimes \ket{\alpha}\bra{\alpha}\hat{U}^{\dagger}\left(t_m\right)\\
		&=\sum_{n,m=0}^{\infty}g_{x}\left(m,n\right)\bra{n} U\left(t_m\right)\ket{n}\rho \bra{m} \hat{U}^{\dagger}\left(t_m\right)\ket{m},
	\end{split}
	\label{equ:E_x}
\end{align}\\
mit\\
\begin{align}
	\begin{split}
		g_{x}\left(m,n\right)
		=\mathrm{e}^{-\alpha^2} \left(\frac{\alpha^{2n}}{2n!}-\frac{ix}{\pi}\frac{ \alpha^{n+m} \Gamma\left(\frac{m+n}{2}+1\right)}{ m! n! \left(m-n\right)} \text{odd}\left(m-n\right)\right).
	\end{split}
	\label{equ:g_x_form}
\end{align}
Wobei\\
\[ \text{odd}\left(n\right) = \left\{ \begin{array}{ll}
	1, & \mbox{$n$ ist eine ungerade ganze Zahl}\\
	0, & \mbox{sonst}.\end{array} \right. \]\\

\section{Bedingung zum Vergleich der Basen}

Nach \cite{Pommerening2020} ermöglicht es die Drehwellennäherung, die Außerdiagonalelemente des Hamiltonoperators zu vernachlässigen, und ist dies nur dann eine gute Näherung wenn folgende Bedingung erfüllt ist:\\
\begin{align}
	\left\Vert \frac{- \frac{ J \chi \sgn\left(\delta_{0}\right)}{\sqrt{\delta_{0}^{2}+J^2}}\hat{a}^{\dag}\hat{a}}{\sgn(\delta_{0})\sqrt{\delta_{0}^{2}+J^2}+ \frac{\chi \abs{\delta_{0}}}{\sqrt{\delta_{0}^{2}+J^2}}\hat{a}^{\dag}\hat{a}}\right\Vert \ll 1,
	\label{equ:gutebedingung}
\end{align}\\	
also wenn:\\
\begin{align}
	\left\Vert\frac{J \chi \hat{a}^{\dag}\hat{a}}{\delta_{0}^{2}+J^2+\chi \delta_{0}\hat{a}^{\dag}\hat{a}} \right\Vert=\left\Vert \frac{J}{\delta_{0}+\chi \hat{a}^{\dag}\hat{a}}\right\Vert \left\Vert \frac{\chi \hat{a}^{\dag}\hat{a}}{\delta_{0}+\frac{J^2}{\delta_{0}+\chi \hat{a}^{\dag}\hat{a}}} \right\Vert \ll 1.
\end{align}\\
Der Hamiltonoperator (Gl. \ref{equ.hamilton1}) in der Bare-Basis ist\\
\begin{align}
	\hat{H}=\begin{bmatrix}
		-\frac{\omega_{1}+\omega_{2}}{2}+\chi \hat{a}^{\dag}\hat{a} & 0 & 0 & 0\\
		0 & \delta_{0}+\chi \hat{a}^{\dag}\hat{a} & J & 0\\
		0 & J & \delta_{0}+\chi \hat{a}^{\dag}\hat{a} & 0\\
		0 & 0 & 0 & \frac{\omega_{1}+\omega_{2}}{2}-\chi \hat{a}^{\dag}\hat{a}\\
	\end{bmatrix},
\end{align}\\
daher ist der erste Term $\left\Vert \frac{J}{\delta_{0}+\chi \hat{a}^{\dag}\hat{a}}\right\Vert$ die gleiche Voraussetzung in der Bare-Basis. Ob der zweite Term $\left\Vert \frac{\chi \hat{a}^{\dag}\hat{a}}{\delta_{0}+\frac{J^2}{\delta_{0}+\chi \hat{a}^{\dag}\hat{a}}} \right\Vert $ kleiner oder größer als $1$ ist, bestimmt daher ob es besser ist in der Bare- oder der Dressed-Basis zu messen.  Wenn $\left\Vert \frac{\chi \hat{a}^{\dag}\hat{a}}{\delta_{0}+\frac{J^2}{\delta_{0}+\chi \hat{a}^{\dag}\hat{a}}} \right\Vert =1 $, also $\left\Vert \chi \hat{a}^{\dag}\hat{a} \right\Vert =\sqrt{\delta_{0}^2+J^2}$ gilt, macht es keinen Unterschied.

\clearpage
\thispagestyle{empty}

	\part{Modellierung der Messung eines Flussqubits mittels QFP}
	
	\clearpage
	\thispagestyle{plain}
	
	\vspace*{\fill}
	
	In diesem Teil wird zuerst das gekoppelte Flussqubit-QFP-System untersucht. Dabei wird die Fidelity der Messung in verschiedenen Basen analysiert. Die Verschränkung zwischen Flussqubit und QFP wird mittels QuTip \cite{qutip1}\cite{qutip2} numerisch untersucht.
	Der direkte Unterschied zwischen Bare- und Dressed-Zustand bei dem gekoppelten Flussqubit-QFP-System wird auch diskutiert.\\
	
	Nach dem QFP-Annealing lässt sich das gekoppelte Flussqubit-QFP-System durch einen effektiven Flussqubit-Hamiltonoperator beschreiben, dessen Tunnel-Amplitude verkleinert, während die Magnetenergie vergrößert ist. Dann wird das gekoppelte System durch einen Resonator ausgelesen. Dies wird durch das JCM für Flussqubit im dispersiven Regime beschrieben. Die Gültigkeit der Drehwellennäherung wird dabei durch die in \cite{Pommerening2020} beschriebene Methode untersucht. Die Fidelity in verschiedenen Basen wird zuerst theoretisch untersucht und dann numerisch simuliert.\\
	
	\vspace{15em}
	\vspace*{\fill}
	
	\clearpage
	\thispagestyle{empty}
	
	\chapter{Das gekoppelte Flussqubit-QFP-System}
	
	In diesem Kapitel wird zuerst das gekoppelte Flussqubit-QFP-System analysiert. Die Verschränkung dazwischen wird zuerst theoretisch untersucht. QuTip \cite{qutip1}\cite{qutip2} wird danach zur numerischen Simulation der Verschränkung benutzt.
	Zum Schluss werden Bare- und Dressed-Zustand des Systems verglichen.\\
	
	\section{Flussqubit gekoppelt an QFP}
	 
	Der Flussqubit-Hamiltonoperator lautet\\
	\begin{align}
		\hat{H}_{q}  =-\frac{1}{2}\left(\epsilon \hat{\sigma}_z + \Delta \hat{\sigma}_x\right),
	\end{align}\\
	wobei $\epsilon$ der Qubit-Magnetenergie und $\Delta$ der Tunnel-Amplitude entsprechen.\\
	
	Zusammen mit den Grundlagen zum QFP im Abschnitt \ref{sec:qfp} bekommen wir die Hamiltonfunktion für das QFP mit Kopplung an das Flussqubit\\
	\begin{align}
		H_0&=\frac{Q^2}{2C}+\frac{\left(\Phi-M_q \abs{I_p^{q}} \sigma_z \right)^2}{2L}-\beta\left(\Phi_{x}^{\text{qfp}}\right)\cos(\frac{2 \pi \Phi}{\Phi_0}),
	\end{align}\\
	mit $\sigma_z=\expval{\hat{\sigma}_z}=\pm 1$. Der durch das Flussqubit induziert magnetische Fluss lautet $M_q I_p^{q}=M_{q}\abs{I_p^{q}} \sigma_z$, wobei $\abs{I_p^{q}}$ der Größe des Qubit-Dauerstroms und $M_q$ der Gegeninduktivität (auch als Koppelinduktivität bezeichnet) zwischen Flussqubit und QFP entsprechen.\\
	
	Vernachlässigen wir die Konstante und drücken wir die Hamiltonfunktion erneut in Form von dimensionslosen Standardparametern aus \cite{Kafri2017}\cite{Schoendorf2020}\\
	\begin{align}
		H_0&=E_L\left(4\xi^2 \frac{q^2}{2}+\frac{\varphi^2}{2}+\beta_{\text{qfp}}\left(\varphi_x^{\text{qfp}}\right)\cos\left(\varphi\right)-\lambda\varphi\sigma_z\right),
		\label{equ:hint}
	\end{align}\\
	wobei\\
	\begin{align*}
		\begin{array}[t]{l l}
			E_L=\left(\Phi_0/2\pi\right)^2/L, &\quad  q=Q/(2e),\\
			\xi=2\pi e/\Phi_0\sqrt{L/C}=4\pi Z/R, &\quad Z=\sqrt{L/C},\\
			R=h/e^2, &\quad \lambda=2\pi M_q\abs{I_p^{q}}/\Phi_0,\\
			\varphi=2\pi\Phi/\Phi_0+\pi,&\quad \varphi_{x}^{\text{qfp}}=2\pi\Phi_{x}^{\text{qfp}}/\Phi_0+\pi,
		\end{array}
	\end{align*}\\
	und\\
	\begin{align*}
		\beta_{\text{qfp}}\left(\varphi_x^{\text{qfp}}\right)&=\frac{E_J}{E_L}\cos(\frac{\pi \Phi_{x}^{\text{qfp}}}{\Phi_0})=\frac{4\pi I_c^{\text{qfp}}L}{\Phi_0}\cos(\frac{\varphi_{x}^{\text{qfp}}}{2}-\frac{\pi}{2})=\frac{4\pi I_c^{\text{qfp}}L}{\Phi_0}\sin(\frac{\varphi_{x}^{\text{qfp}}}{2}).
	\end{align*}\\
	Beim QFP-Annealing wird der äußere magnetische Fluss $\Phi_{x}^{\text{qfp}}$ von $\Phi_0 /2$ auf sein Maximum $\Phi_0$ und die entsprechende Phase $\varphi_{x}^{\text{qfp}}$ von $2\pi$ auf $3\pi$ erhöht. Sei die gesamte Zeitdauer vom QFP-Annealing $ t_{\text{qfp}} = \pi/2\omega$, wobei $\omega$ die Kreisfrequenz ist. Es wird auch angenommen, dass die Phase linear gestiegen ist.
	Dann können wir $\beta_{\text{qfp}}\left(\varphi_x^{\text{qfp}}\right)$ umschreiben:\\
	\begin{align*}
		\beta_{\text{qfp}}\left(\omega, t\right)=
		\left[\Theta\left(t\right)-\Theta\left(t-t_{\text{qfp}}\right)\right]\beta_{\text{max}}\sin(\omega t) +\Theta\left(t-t_{\text{qfp}}\right)\beta_{\text{max}},
	\end{align*}\\
	wobei $\beta_{\text{max}}=4\pi I_c^{\text{qfp}}L/\Phi_0$ und $\Theta(t)$ die Heaviside-Funktion ist.\\

	Um die Hamiltonfunktion zu quantisieren, definieren wir zwei Operatoren\\
	\begin{align*}
		\hat\varphi&=\frac{1}{\sqrt{2m\Omega}}\left(\hat{a}^\dagger+\hat{a}\right),\\
		\hat q&=i\sqrt{\frac{m\Omega}{2}}\left(\hat{a}^\dagger-\hat{a}\right),
	\end{align*}\\
	mit $m=1/(2\xi)^2$ der effektive Masse, $\Omega= 2\xi$, $\hbar\equiv1$, $\left[\hat{a}, \hat{a}^\dagger\right]=1$ und $\left[\hat{\varphi}, \hat{q}\right]=i$.\\
	
	So bekommen wir\\	
	\begin{align}
		\hat{H}_0=E_L\left(\Omega\hat{a}^\dagger \hat{a}+m\Omega^2 \beta_{\text{qfp}}(\omega, t)\cos\left(\frac{1}{\sqrt{2m\Omega}}\hat{a}^\dagger+\hat{a}\right) -\lambda(\hat{a}^{\dag}+\hat{a}) \hat{\sigma}_z\right).
	\end{align}\\
	Der dimensionslos potenzielle Teil\\
	\begin{align*}
		U\left(\varphi\right)=\frac{\varphi^2}{2}+\beta_{\text{qfp}}\left(\omega, t\right)\cos\left(\varphi\right)-\lambda\varphi\sigma_z
	\end{align*}\\
	lässt sich um sein Minimum entwickeln. Zuerst finden wir das Minimum mit der Bedingung\\
	\begin{align}
		0\equiv\pdv{U}{\varphi}=\varphi-\beta_{\text{qfp}}\left(\omega, t\right) \sin(\varphi)-\lambda\sigma_z.
		\label{equ:minimum_bd}
	\end{align}\\
	Wegen $\sigma_z=\expval{\hat{\sigma}_z}=\pm 1$ erhalten wir die Lösung der Gl. \ref{equ:minimum_bd}\\
	\begin{align*}
		\varphi=\pm \varphi_p =\sigma_z \varphi_p,
	\end{align*}\\
	wobei $\varphi_p$ den positiven Positionswert bezeichnet.\\
	
	Mit Hilfe der Gl. \ref{equ:minimum_bd} bekommt man die minimale Energie
	
	\begin{align*}
		U_{\text{min}}=&\frac{\left(\varphi_p \sigma_z\right)^2}{2}+\beta_{\text{qfp}}\left(\omega, t\right)\cos\left(\varphi_p \sigma_z\right)-\lambda\left(\varphi_p \sigma_z\right)\sigma_z\\
		=&\frac{1}{2}\left[\beta_{\text{qfp}}\left(\omega, t\right) \sin(\varphi_p \sigma_z)+\lambda\sigma_z\right]^2+\beta_{\text{qfp}}\left(\omega, t\right)\cos\left(\varphi_p \sigma_z\right)\\
		&-\lambda\left[\beta_{\text{qfp}}\left(\omega, t\right) \sin(\varphi_p \sigma_z)+\lambda\sigma_z\right]\sigma_z\\
		=&\frac{1}{2}\left[\beta_{\text{qfp}}^2(t)\sin[2](\varphi_p \sigma_z)-\lambda^2\right]+\beta_{\text{qfp}}\left(\omega, t\right)\cos\left(\varphi_p \sigma_z\right).
	\end{align*}\\
	Daraus schließen wir, dass $U_{\text{min}}$ eine gerade Funktion und daher unabhängig von $\sigma_z$ ist. Die wird dann als eine Konstante vernachlässigt.\\
	
	So wird $U$ an der Stelle $\sigma_z \varphi_p(t)$ in eine Taylorreihe bis zur zweiten Ordnung entwickelt:\\
	\begin{align}
		\begin{split}
		U&=U_{\text{min}}+\frac{1}{2}\left[1-\beta_{\text{qfp}}\left(\omega, t\right)\cos(\varphi_p)\right] \left(\varphi-\varphi_p \sigma_z\right)^2+\mathcal{O}\left(\left(\varphi-\varphi_p(t) \sigma_z\right)^3\right)\\
		&\approx \frac{1}{2}\left[1-\beta_{\text{qfp}}\left(\omega, t\right)\cos(\varphi_p(t))\right] \left[\varphi-\varphi_p(t) \sigma_z\right]^2.
		\end{split}
		\label{equ:potential_a}
	\end{align}\\

	Durch die Quantisierung bekommen wir den Hamiltonoperator\\
	\begin{align}
		\hat{H}_0/E_L=\Omega(t)\hat{a}^\dagger \hat{a}+\Omega(t)\sqrt{m\Omega(t)/2}\varphi_p(t)\left(\hat{a}^\dagger+\hat{a}\right) \hat{\sigma}_z,
		\label{equ:h_0quant}
	\end{align}\\
	mit $\Omega(t)= 2\xi\sqrt{1-\beta_{\text{qfp}}\left(\omega, t\right)\cos(\varphi_p(t))}$. Dabei werden $\Omega(t)/2$ und $m\Omega^2(t)\left[\varphi_p(t)\hat{\sigma}_z\right]^2/2$ vernachlässigt, weil die beiden Terme unabhängig von dem Qubitzustand sind. Die Gl. \ref{equ:h_0quant} beschreibt einen verschobenen harmonischen Oszillator.\\
	
	Der Zustand des gekoppelten Flussqubit-QFP-Systems entwickelt sich durch eine adiabatische Durchführung vom QFP-Annealing schließlich zu einem verschränkten Zustand \cite{Schoendorf2020}\\
	\begin{align*}
		\left(\alpha\ket{0}+\beta\ket{1}\right)\otimes \ket{0}\overset{U}{\longrightarrow}	\left(\alpha_{\text{eff}}\ket{0,L}+\beta_{\text{eff}}\ket{1,R}\right).
	\end{align*}
	Das heißt, das $\ket{0}$ oder $\ket{1}$ im Flussqubit Zustand ist mit dem links  $\ket{L}$ oder rechts $\ket{R}$ verschobenen harmonischen Oszillator im QFP verschränkt. Die Bedingung zur adiabatischen Durchführung findet man in \cite{Schoendorf2020}.\\
	
	\section{Fidelity in verschiedenen Basen}
	
	Die Fidelity in der Flussbasis lässt sich schreiben als \cite{Schoendorf2020}\\
	\begin{align*}
		\mathcal{F}(\omega, t_m)=\Phi\left(\frac{\varphi_p(t_m)}{\hat{\sigma}(\omega, t_m)}\right),
	\end{align*}\\
	wobei $\hat{\sigma}(\omega, t_m)=\left[\sqrt{1-\beta_{\text{qfp}}\left(\omega, t_m\right)\cos(\varphi_p(t_m))}/\xi\right]^{-1/2}$ die Standardabweichung des gaußschen Wellenpakets und $t_m$ die beliebige Messzeit ist. $\Phi(x)=\frac{1}{\sqrt{2\pi}}\int_{-\infty}^{x}e^{-t^2/2}\d{t}$ beschreibt dabei die Normalverteilung.\\
	
	In der Energiebasis wird die Position des minimalen Potenzials durch eine Rotation entsprechend zu $\pm \left[\cos(\theta_{\text{qb}})-i\sin(\theta_{\text{qb}})\right]\varphi_p(t)$ kommen und die Fidelity lautet dann\\
	\begin{align*}
		\tilde{\mathcal{F}}(\omega, t_m)=\Phi\left(\frac{\tilde{\varphi}_p(t_m)}{\tilde{\hat{\sigma}}(\omega, t_m)}\right),
	\end{align*}\\
	wobei $\tilde{\varphi}_p(t_m)=\left[\cos(\theta_{\text{qb}})-i\sin(\theta_{\text{qb}})\right]\varphi_p(t_m)$.\\
	
	So können wir die Fidelity in beiden Basen numerisch untersuchen.\\

\begin{figure}[ht]
		\centering
		\includegraphics[width=0.7\textwidth]{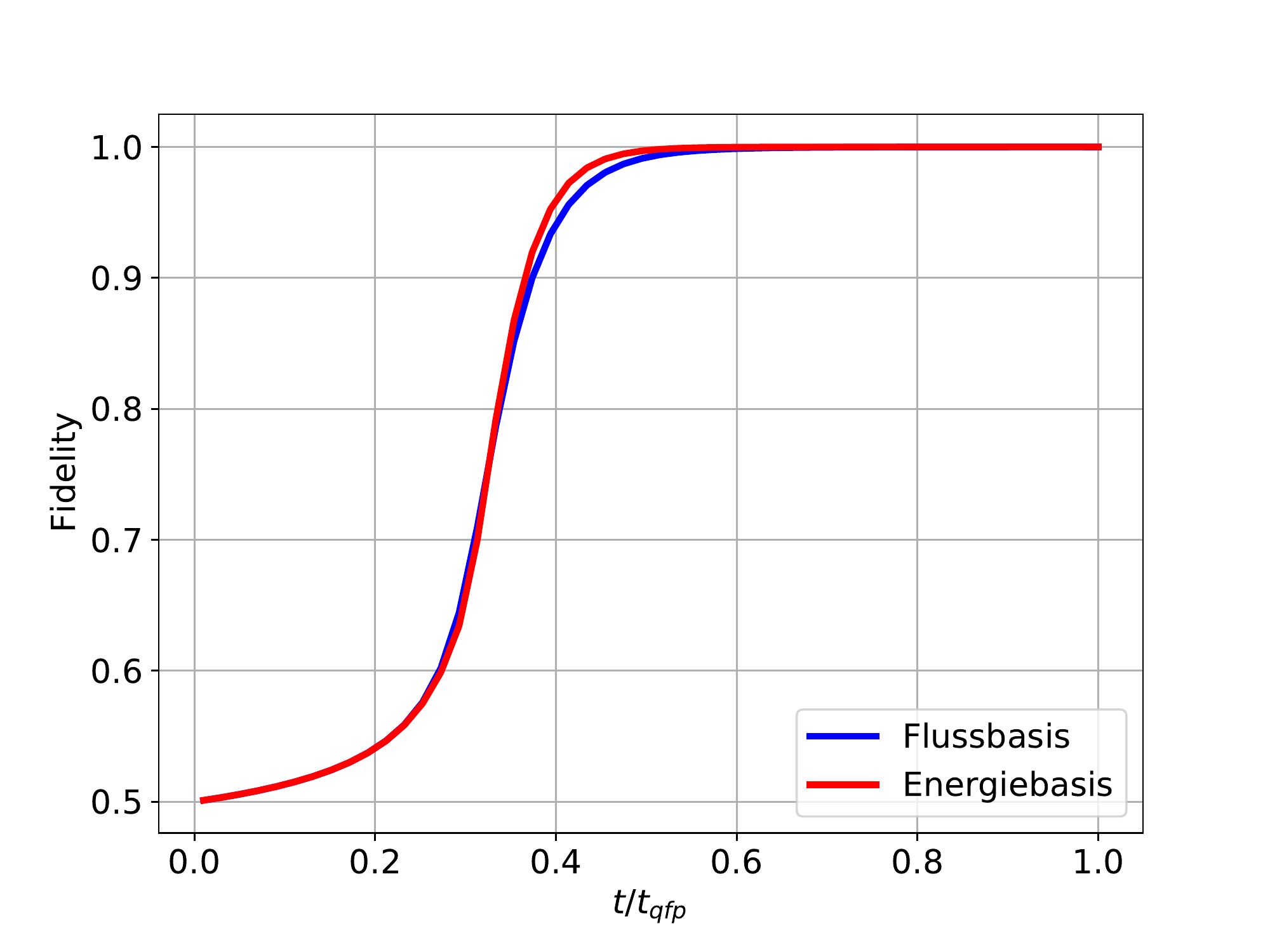}
		\caption{Fidelity in Abhängigkeit von der Messzeit $t/t_{\text{qfp}}$ in der Flussbasis (blau) und Energiebasis (rot) mit $\Delta_{q}/\epsilon_{q}=1,~ \xi=0.4,~ \beta_{\text{max}}=2.5$, wobei $t_{\text{qfp}}$ die gesamte Zeitdauer vom QFP-Annealing ist. Die Fidelity wird mit steigender Zeit höher und erreicht einen Sättigungswert.}
		\label{fig:an_t}
	\end{figure}

	In Abb. \ref{fig:an_t} wird die Fidelity in Abhängigkeit von $t/t_{\text{qfp}}$ dargestellt, wobei $t_{\text{qfp}}$ die gesamte Zeitdauer des QFP-Annealing ist. Kurz vor dem Ende des QFP-Annealing-Prozesses, also $t\approx0.6 t_{\text{qfp}}$, geht die Fidelity in beiden Basen schon gegen $1$. Am Ende von QFP-Annealing ($t= t_{\text{qfp}}$) wird das Flussqubit-Signal sowohl in der Flussbasis als auch in der Energiebasis im QFP effizient verschränkt und gespeichert. Für $t\gtrsim 0.3t_{\text{qfp}}$ ist der Speicherprozess bei fester Fidelity in der Energiebasis schneller.
	\begin{figure}[ht]
		\centering
		\includegraphics[width=0.7\textwidth]{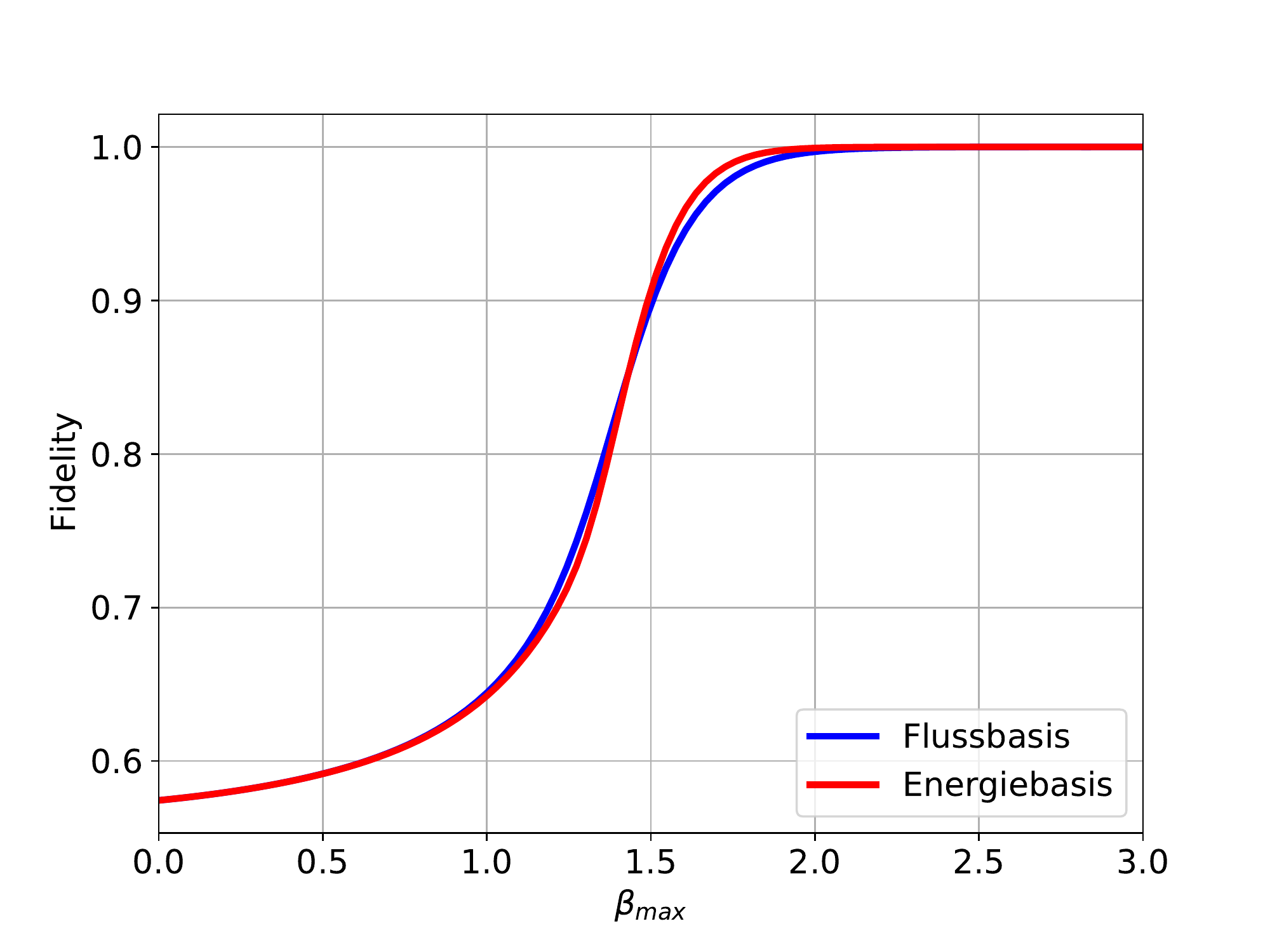}
		\caption{Fidelity in Abhängigkeit von $\beta_{\text{max}}$ in der Flussbasis (blau) und Energiebasis(rot) mit $\Delta_{q}/\epsilon_{q}=1/2$, $\xi = 0.4$, der Messzeit $t_m=t_{\text{qfp}}$. Die Fidelity wird also mit steigendem $\beta_{\text{max}}$ höher und erreicht einen Sättigungswert.}
		\label{fig:an_betamax}
	\end{figure}

	\clearpage
		
	Außerdem wird die Beziehung zwischen Fidelity und $\beta_{\text{max}}$ untersucht (siehe Abb. \ref{fig:an_betamax}). Für die angegebenen Parameter wird die Fidelity bei $\beta_{\text{max}}\gtrsim2$ in beiden Basen am besten. Ein kleiner Wert von $\beta_{\text{max}}$ bedeutet, dass die Barriere des Potenzials beim QFP-Annealing nicht hoch genug, die Zustände deshalb weniger gut unterscheidbar, die Fidelity daher niedrig ist.\\

	Durch eine adiabatische Durchführung vom QFP-Annealing wird das Flussqubit-Signal sowohl in der Flussbasis als auch in der Energiebasis erfolgreich gespeichert. Bei diesem Speicherprozess wird eine gleichermaßen hohe Fidelity in der Energiebasis mit kürzeren Messzeiten erreicht als in der Flussbasis.\\
	
	\section{Vergleich der Bare- und Dressed-Zustände}
	
	Um den Bare- und Dressed-Zustand des Systems zu vergleichen, schreiben wir den gesamten Hamiltonoperator als Summe von dem verschobenen harmonischen Oszillator (QFP) und dem Flussqubit:\\	
	\begin{align}
		\hat{H}(t)  = \omega_r(t) \hat{a}^{\dag}\hat{a}+ g\left(t\right) \hat{\sigma}_z\left(\hat{a}^{\dag}+\hat{a}\right)-\frac{1}{2}\left(\epsilon \hat{\sigma}_z + \Delta \hat{\sigma}_x\right),
		\label{eqn:Htotal}
	\end{align}	\\	
	wobei $\omega_r(t)=E_L \Omega(t)$ der Resonatorfrequenz und $g(t)=E_L\Omega(t)\sqrt{m\Omega(t)/2}\varphi_p(t)$ der zeitabhängigen QFP-Flussqubit-Kopplungsstärke entsprechen.\\
	
	Die Gl. \ref{eqn:Htotal} kann man nicht analytisch lösen, deshalb ist eine Näherung erforderlich.  Der häufigste benutzte Ansatz ist die Annahme, dass die Resonatorfrequenz $\omega _{r}$ nahe der Qubitfrequenz  $\omega_{q}$ liegt und die Kopplung dazwischen schwach ist. Die schneller oszillierenden Terme werden durch die Drehwellennäherung vernachlässigt.\\
	
	In unserem Fall ist die Qubitfrequenz viel kleiner als die Resonatorfrequenz. Außerdem ist die Kopplungsstärke dazwischen in der Größenordnung oder größer als die Resonatorfrequenz. Die Drehwellennäherung ist in diesem Regime nicht gültig, aber dafür die adiabatische Näherung \cite{Irish2005}. Das heißt, die Qubitfrequenz ist so klein, dass es nicht genug Energie aufbringen kann, um den Resonantor anzuregen.\\
	
	Wir stellen den Hamiltonoperator in der Basis des verschobenen Oszillators \cite{Irish2005} dar, indem wir zuerst die Qubit-Energie vernachlässigen. Setzen wir $\hat{\sigma}_z$ entsprechend auf seinen Eigenwert $\pm 1$, ergibt sich\\	
	\begin{align}
		\left[\pm g\left(\hat{a}^{\dag}+\hat{a}\right)+\omega_r \hat{a}^{\dag}\hat{a}\right] \ket{\phi_{\pm}}=E\ket{\phi_{\pm}}.
	\end{align}	\\	
	Es lässt sich schreiben als\\	
	\begin{align}
		\left[\left(\hat{a}^{\dag} \pm \frac{g}{\omega_r}\right) \left(\hat{a} \pm \frac{g}{\omega_r}\right) \right] \ket{\phi_{\pm}}=\left(\frac{E}{\omega_r}+\frac{g^2}{\omega_r^2}\right)\ket{\phi_{\pm}}.
		\label{equ:vhood}
	\end{align}	\\	
	Jetzt zeigen wir, dass die Eigenzustände und entsprechende Eigenenergie wie folgt sind:\\	
	\begin{align}
		\begin{split}
			\ket{\phi_{\pm}}&=\hat{D}\left(\mp\frac{g}{\omega_r}\right)\ket{N}\\
			&=\mathrm{e}^{\mp \frac{g}{\omega_{r}}\left(\hat{a}^{\dag}-\hat{a}\right)}\ket{N}\\
			&\equiv\ket{N_{\pm}}
		\end{split}
		\label{equ:egv}
	\end{align}	\\
	und\\
	\begin{align}
		\begin{split}
		E&=E_N=\omega_{r}\left(N-\frac{g^2}{\omega_{r}^2}\right),
		\label{equ:eig}
	\end{split}
	\end{align}\\
	wobei $\ket{N}$ mit $N=0,1,2,...$ der Fock-Zustand ist.\\
	
	Der Verschiebungsoperator lautet\\
	\begin{align}
		\hat{D}\left(\nu\right)=\exp[\nu \hat{a}^{\dag}-\nu ^*\hat{a}].
	\end{align}
	Es gelten
	\begin{align}
		\hat{D}^{\dag}\left(\alpha\right)\hat{a}\hat{D}\left(\alpha \right)=\hat{a}+\alpha,\\ \hat{D}\left(\alpha\right)\hat{a}\hat{D}^{\dag}\left(\alpha\right)=\hat{a}-\alpha,
	\end{align}
	und\\
	\begin{align}
		\hat{D}^\dagger\left(\nu\right)=\hat{D}^{-1}\left(\nu \right)=\hat{D}\left(-\nu\right).
	\end{align}\\
	So lässt sich die linke Seite der Gl. \ref{equ:vhood} mit Verschiebungsoperatoren schreiben \cite{Irish2005}\\
	\begin{align}
		\left[\left
		(\hat{a}^{\dag} \pm \frac{g}{\omega_r}\right) \left(\hat{a} \pm \frac{g}{\omega_r}\right) \right]=\hat{D}\left(\mp \frac{g}{\omega_{r}}\right)\hat{a}^{\dag}\hat{a}\hat{D}^{\dag}\left(\mp \frac{g}{\omega_{r}}\right).
		\label{equ:vhold}
	\end{align}	\\	
	Aus den Gl. \ref{equ:vhood} -- \ref{equ:eig} sowie Gl. \ref{equ:vhold} haben wir\\	
	\begin{align}
		\begin{split}
		\hat{D}\left(\mp \frac{g}{\omega_{r}}\right)\hat{a}^{\dag}\hat{a}\hat{D}^{\dag}\left(\mp \frac{g}{\omega_{r}}\right) \ket{\phi_{\pm}}
		&=\hat{D}\left(\mp \frac{g}{\omega_{r}}\right)\hat{a}^{\dag}\hat{a}\hat{D}^{\dag}\left(\mp \frac{g}{\omega_{r}}\right) \hat{D}\left(\mp\frac{g}{\omega_r}\right)\ket{N}\\
		&=\hat{D}\left(\mp\frac{g}{\omega_r}\right)N\ket{N}\\
		&=N\hat{D}\left(\mp\frac{g}{\omega_r}\right)\ket{N}\\
		&=N\ket{\phi_{\pm}}
	\end{split}
	\end{align}\\
	Vergleicht man mit der Gl. \ref{equ:vhood}, ergibt sich dann die Eigenenergie (Gl. \ref{equ:eig}).\\
	
	Unter adiabatischer Näherung bekommen wir schließlich (für mehr Details siehe \cite{Irish2005})\\	
	\begin{align}
		\hat{H}_N \approx \begin{bmatrix} E_N-\epsilon/2 & -\Delta\braket{N_{-}|N_{+}}/2\\ -\Delta\braket{N_{-}|N_{+}}/2 & E_N+\epsilon/2 \end{bmatrix}.
	\end{align}\\
	Durch Diagonalisierung erhalten wir die Eigenenergie\\
	\begin{align}
		E_{\pm,N}=E_N \pm \frac{1}{2} \sqrt{\epsilon^2+\Delta^2\braket{N_{-}|N_{+}}},
	\end{align}
	und die Eigenvektoren\\	
	\begin{align}
		\begin{bmatrix} \ket{\phi_{+,N}} \\ \ket{\phi_{-,N}} \end{bmatrix} =\begin{bmatrix} \cos(\frac{\theta}{2}) & \sin(\frac{\theta}{2})\\ -\sin(\frac{\theta}{2}) & \cos(\frac{\theta}{2})  \end{bmatrix} \begin{bmatrix} \ket{+,N_+} \\ \ket{-,N_-} \end{bmatrix},
	\end{align}\\
	wobei $\theta$ durch $\tan (\theta)=\Delta \braket{N_{-}|N_{+}}/\epsilon$ definiert ist.\\
	
	Um Bare- und Dressed-Basis zu vergleichen, berechnen wir zuerst die Eigenenergien und Eigenvektoren vom Qubit. Dafür diagonalisieren wir den Hamiltonoperator des Flussqubits $\hat{H}_{q}$. So erhalten wir die Eigenzustände\\	
	\begin{align}
		\begin{bmatrix} \ket{\varphi_+} \\ \ket{\varphi_-} \end{bmatrix} =\begin{bmatrix} \cos(\frac{\theta_q}{2}) & \sin(\frac{\theta_q}{2})\\ -\sin(\frac{\theta_q}{2}) & \cos(\frac{\theta_q}{2})  \end{bmatrix} \begin{bmatrix} \ket{+} \\ \ket{-} \end{bmatrix},
	\end{align}	\\	
	wobei $\theta_q$ durch $\tan (\theta_q)=\Delta/\epsilon$ definiert ist.
	Die Eigenvektoren des Oszillators sind einfach Fock-Zustände $\ket{N}$. So bekommen wir die Bare-Zustände\\
	\begin{align}
		\begin{bmatrix} \ket{\phi_{+,N}} \\ \ket{\phi_{-,N}} \end{bmatrix} \equiv \begin{bmatrix} \ket{\phi_+} \\ \ket{\phi_-} \end{bmatrix} \otimes \ket{N} =\begin{bmatrix} \cos(\frac{\theta_q}{2}) & \sin(\frac{\theta_q}{2})\\ -\sin(\frac{\theta_q}{2}) & \cos(\frac{\theta_q}{2})  \end{bmatrix} \begin{bmatrix} \ket{+,N} \\ \ket{-, N} \end{bmatrix}.
	\end{align}	\\
	Dann lässt sich der Überlapp zwischen Bare- und Dressed-Zustand berechnen\\	
	\begin{align*}
		\begin{split}
			\braket{\varphi_{+,N}|\phi_{+,N}}&=\Big[\cos(\frac{\theta_q}{2})\cos(\frac{\theta}{2})+\sin(\frac{\theta_q}{2}) \sin(\frac{\theta}{2})\Big] \braket{N|N_+} = \cos(\frac{\theta-\theta_q}{2})\braket{N|N_+},\\
			\braket{\varphi_{-,N}|\phi_{-,N}}&=\Big[\sin(\frac{\theta_q}{2})\sin(\frac{\theta}{2})+\cos(\frac{\theta_q}{2}) \cos(\frac{\theta}{2})\Big] \braket{N|N_-} = \cos(\frac{\theta-\theta_q}{2})\braket{N|N_-},
		\end{split}
	\end{align*}\\
	also\\
	\begin{align}
		\braket{\varphi_{\pm,N}|\phi_{\pm,N}}=\cos(\frac{\theta-\theta_q}{2})\braket{N|\hat{D}(\mp\frac{g}{\omega_r})|N}.
		\label{gl.bare_dressed}
	\end{align}\\
	Der Überlapp zwischen Bare- und Dressed-Zustand des gekoppelten Systems lässt sich durch einen Erwartungswert der Verschiebungsoperatoren in den Fock-Zuständen darstellen.\\
	
	\begin{figure}[ht]
		\centering
		\includegraphics[width=0.7\textwidth]{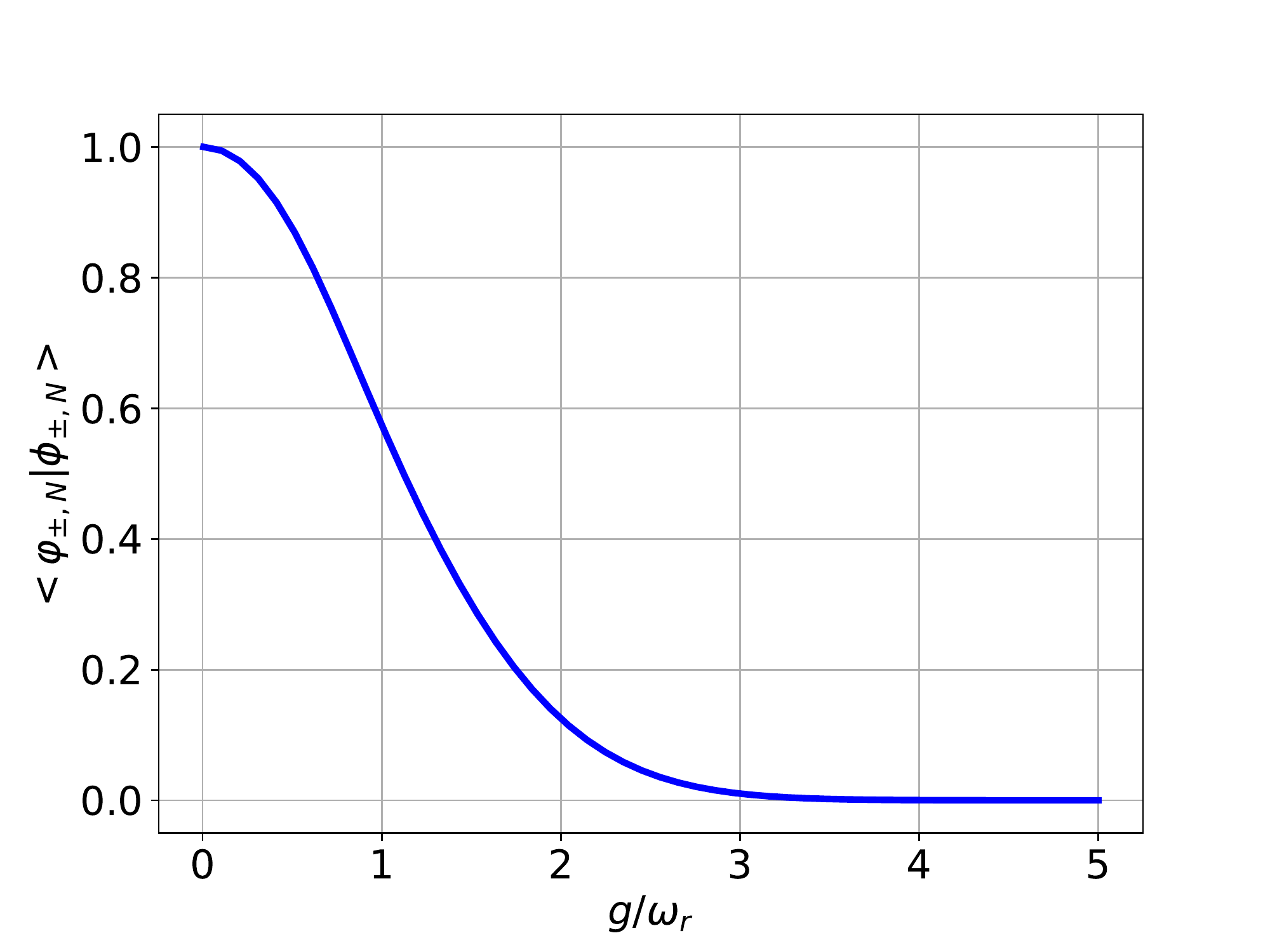}
		\caption{Der Überlapp zwischen Bare- und Dressed-Zustand (Gl. \ref{gl.bare_dressed}) in Abhängigkeit von $g/\omega_r$ mit $\theta_q=\pi/4$ und $N=49$. Der Überlapp nimmt mit steigendem $g/\omega_r$ ab.}
		\label{fig:bare_dressed}
	\end{figure}
	
	Die numerische Simulation der Gl. \ref{gl.bare_dressed} wird in Abb. \ref{fig:bare_dressed} dargestellt. Mit steigender QFP-Flussqubit-Kopplungsstärke unterscheiden sich die Bare- und Dressed-Zustände zunehmend. Bei $g\approx 3 \omega_r$ sind die beiden Basen ungefähr orthogonal.\\
	
	\clearpage
	\thispagestyle{empty}

	\chapter{Messung eines Flussqubits}
		
\begin{figure}[ht]
		\centering
		\includegraphics[width=0.7\textwidth]{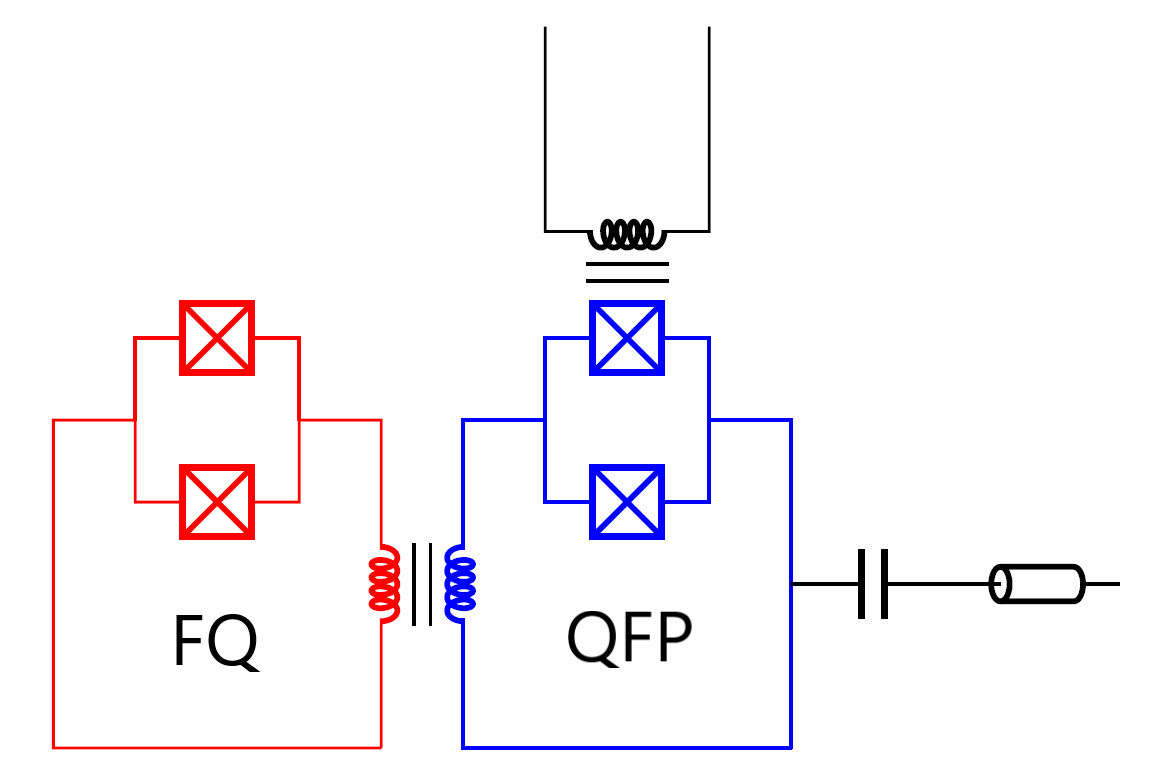}
		\caption{Modellierung der Messung eines Flussqubits. Ein Flussqubit (rot) ist an ein QFP (blau) gekoppelt. Nach dem QFP-Annealing wird das Signal von Flussqubit im QFP gespeichert. Dann wird das QFP durch einen Resonator ausgelesen.}
		\label{fig:modell1}
	\end{figure}

	Nach dem QFP-Annealing wird das Flussqubit-Signal im QFP gespeichert. Dies wird durch einen effektiven Flussqubit-Hamiltonoperator beschrieben. Sobald das Flussqubit-Signal im QFP verschränkt ist, wird das QFP durch einen Resonator ausgelesen (siehe Abb. \ref{fig:modell1}). Dieser Messprozess wird durch das JCM für Flussqubit im dispersiven Regime beschrieben. Dabei betrachten wir zwei Situationen, eine ohne treibendes Feld, andere mit treibendem Feld. Wir werden zeigen, dass die Anwesenheit einer externen resonanten Ansteuerung keinen Einfluss auf die Dynamik des Messprozesses gibt. Die Fidelity der Messung ohne externe resonante Ansteuerung wird in verschiedenen Basen numerisch simuliert.
	
	\section{Ohne treibendes Feld}
	
	Sowohl in der Flussbasis als auch in der Energiebasis kann das Flussqubit-Signal durch eine adiabatische Durchführung vom QFP-Annealing im QFP gespeichert werden. Danach wird das gekoppelte Flussqubit-QFP-System durch einen effektiven Flussqubit-Hamiltonoperator \cite{Wilhelm2003} beschrieben:\\
	\begin{align*}
		\hat{H}_{\text{eff}}=-\frac{1}{2}\left(\epsilon_{\text{eff}}\hat{\sigma}_z+\Delta_{\text{eff}}\hat{\sigma}_x\right),
	\end{align*}\\
	wobei $\hbar = 1$ gesetzt und $\epsilon_{\text{eff}}$ wegen $I_p^{\text{qfp}}\approx 10I_{p}^{q}$ vergrößert wird. Dabei entsprechen $I_p^{\text{qfp}}$ dem Dauerstrom von QFP und $I_{p}^{q}$ dem von Flussqubit. $\Delta_{\text{eff}} = \Delta_{\text{q}}~ e^{-\eta}$ wird entsprechend verkleinert. $\eta$ ist von QFP abhängig (für mehrere Details siehe \cite{Wilhelm2003}).\\
	
	Dann wird der Messprozess durchgeführt. Dies lässt sich mit dem folgenden Hamiltonoperator beschreiben \cite{FD10}:\\
	\begin{align}
		\hat{H}&=-\frac{1}{2}\left(\epsilon_{\text{eff}}\hat{\sigma}_z+\Delta_{\text{eff}}\hat{\sigma}_x\right)+\omega_r \hat{a}^\dagger\hat{a}+ g \hat{\sigma}_z  \left(\hat{a} +\hat{a}^\dagger\right),
	\end{align}\\
	wobei $g$ die QFP-Resonator-Kopplungsstärke ist. In der Energiebasis haben wir\\
	\begin{align}
		\tilde{\hat{H}}=-\frac{ \omega_{\text{q,eff}}}{2}\tilde{\hat{\sigma}}_z+ \omega_r \hat{a}^\dagger\hat{a}+ g\left[\cos(\theta_{\text{eff}}) \tilde{\hat{\sigma}}_z- \sin(\theta_{\text{eff}}) \tilde{\hat{\sigma}}_x\right] \left(\hat{a}^\dagger+\hat{a}\right),
	\end{align}\\
	mit $\tan (\theta_{\text{eff}}) =\Delta_{\text{eff}}/\epsilon_{\text{eff}}$ und $\omega_{\text{q,eff}}=\sqrt{\epsilon_{\text{eff}}^2+\Delta^2_{\text{eff}}}$.\\
	
	Unter der Drehwellennäherung ergibt sich\\
	\begin{align}
		\tilde{\hat{H}}=-\frac{\omega_{\text{q,eff}}}{2} \tilde{\hat{\sigma}}_z+ \omega_r \hat{a}^\dagger\hat{a}+ g  \cos(\theta_{\text{eff}})  \tilde{\hat{\sigma}}_z \left(\hat{a}^\dagger+\hat{a}\right)- g  \sin(\theta_{\text{eff}})  \left(\hat{a}\tilde{\hat{\sigma}}_+ +\hat{a}^\dagger\tilde{\hat{\sigma}}_-\right).
		\label{equ:H_t}
	\end{align}\\
	Der Term $g  \sin(\theta_{\text{eff}})  \left(\hat{a}\tilde{\hat{\sigma}}_+ +\hat{a}^\dagger\tilde{\hat{\sigma}}_-\right)$ oszilliert mit $e^{\pm i \left(\omega_{\text{q,eff}}-\omega_r\right) t}$ im rotierenden Bezugssystem langsamer als der Term $g  \cos(\theta_{\text{eff}})  \tilde{\hat{\sigma}}_z \left(\hat{a}^\dagger+\hat{a}\right)$ mit $e^{\pm i \omega_r t}$, der für eine kleine Werte $g \cos(\theta_{\text{eff}})$ vernachlässigt werden kann. Außerdem lässt sich der Term $\tilde{\hat{\sigma}}_z\left(\hat{a}^\dagger+\hat{a}\right)$ durch einen Verschiebungsoperator \cite{CG69}\\
	\begin{align}
		\hat{D}(\alpha)=\exp(\alpha \hat{a}^{\dag} -\alpha^* \hat{a}),
	\end{align}\\
	eliminieren und führt nur zu einer globalen Verschiebung der Energie \cite{Irish2005}. Durch Anwenden des Verschiebungsoperators
	auf der Gl. \ref{equ:H_t} ergibt\\
	\begin{align*}
		\tilde{\hat{H}}^{(1)}&=\hat{D}^\dagger(\alpha)\tilde{\hat{H}}	\hat{D}(\alpha)\\
		&=-\frac{\omega_{\text{q,eff}}}{2} \tilde{\hat{\sigma}}_z+   \omega_r \left(\hat{a}^\dagger+\alpha^*\right)(\hat{a}+\alpha)\\
		&\quad- g  \sin(\theta_{\text{eff}})  \left[\left(\hat{a}+\alpha\right)\tilde{\hat{\sigma}}_+ +(\hat{a}^\dagger+\alpha^*)\tilde{\hat{\sigma}}_-\right]\\
		&\quad+ g  \cos(\theta_{\text{eff}})  \tilde{\hat{\sigma}}_z \left(\hat{a}^\dagger+\alpha^*+\hat{a}+\alpha\right)\\
		&=-\frac{\omega_{\text{q,eff}}}{2} \tilde{\hat{\sigma}}_z+\omega_r \hat{a}^{\dag}\hat{a}- g  \sin(\theta_{\text{eff}})\left(a \tilde{\hat{\sigma}}_+ + \hat{a}^{\dag} \tilde{\hat{\sigma}}_-\right)\\
		&\quad + \omega_r\left( \hat{a}^\dagger\alpha+\alpha^*\hat{a}+\abs{\alpha}^2\right)- g  \sin(\theta_{\text{eff}})  \left(\alpha\tilde{\hat{\sigma}}_+ +\alpha^*\tilde{\hat{\sigma}}_-\right)\\
		&\quad+ g  \cos(\theta_{\text{eff}})  \tilde{\hat{\sigma}}_z \left(\hat{a}^\dagger+\hat{a}+\alpha^*+\alpha\right).
	\end{align*}\\
	Wählen wir $\alpha=-g\cos(\theta_{\text{eff}})\tilde{\hat{\sigma}}_z /\omega_r$ und vernachlässigen alle Transienten in $\alpha$ \cite{Blais2007}, folgt\\
	\begin{align}
		\tilde{\hat{H}}^{(1)}=-\frac{\omega_{\text{q,eff}}}{2} \tilde{\hat{\sigma}}_z+\omega_r \hat{a}^{\dag}\hat{a}- g  \sin(\theta_{\text{eff}})\left(a \tilde{\hat{\sigma}}_+ + \hat{a}^{\dag} \tilde{\hat{\sigma}}_-\right).
		\label{equ:Htilde1}
	\end{align}\\
	Wenn wir uns zu einem Bezugssystem bewegen, das sich sowohl für das Qubit als auch für die Feldoperatoren mit der Frequenz $\omega_r$ dreht, erhalten wir den Hamiltonoperator\\
	\begin{align}
		\tilde{\hat{H}}^{(2)}=-\frac{\delta}{2} \tilde{\hat{\sigma}}_z- g  \sin(\theta_{\text{eff}})\left(a \tilde{\hat{\sigma}}_+ + \hat{a}^{\dag} \tilde{\hat{\sigma}}_-\right),
		\label{equ:H2}
	\end{align}\\
	im dispersiven Regime mit $g\sin(\theta_{\text{eff}}) \ll \abs{\delta}$ und $\delta = \omega_{\text{q,eff}} - \omega_r  \neq 0$ der Verstimmung zwischen effektivem Flussqubit- und Resonatorfrequenz.\\
	
	Durch Anwenden einer unitären Transformation\\
	\begin{align}
		\hat{U}_3=\exp(\lambda\left(a\tilde{\hat{\sigma}}_+ - \hat{a}^{\dag}\tilde{\hat{\sigma}}_- \right)),
		\label{equ:U_3}
	\end{align}\\
	mit $\lambda=g\sin(\theta_{\text{eff}})/\delta$ auf der Gl. \ref{equ:H2}, ergibt sich der Hamiltonoperator in der Energiebasis bis zur ersten Ordnung in $\lambda$\\
	\begin{align}
		\tilde{\hat{H}}^{(3)}=-\left[\frac{\delta}{2}+\chi\left(\hat{a}^{\dag}\hat{a}+\frac{1}{2}\right)\right]\tilde{\hat{\sigma}}_z=-\frac{\delta+\chi}{2}\tilde{\hat{\sigma}}_z-\chi \hat{a}^{\dag}\hat{a}\tilde{\hat{\sigma}}_z.
		\label{equ:H_3}
	\end{align}\\
	Zurück zur Flussbasis ergibt sich\\
	\begin{align}
		\hat{H}^{(3)}=-\left[\frac{\delta}{2}+\chi\left(\hat{a}^{\dag}\hat{a}+\frac{1}{2}\right)\right]\left[\cos(\theta_{\text{eff}})\hat{\sigma}_z+\sin(\theta_{\text{eff}})\hat{\sigma}_x\right],
		\label{equ:H_3_b}
	\end{align}\\
	
	\begin{figure}[ht]
		\centering
		\includegraphics[width=0.7\textwidth]{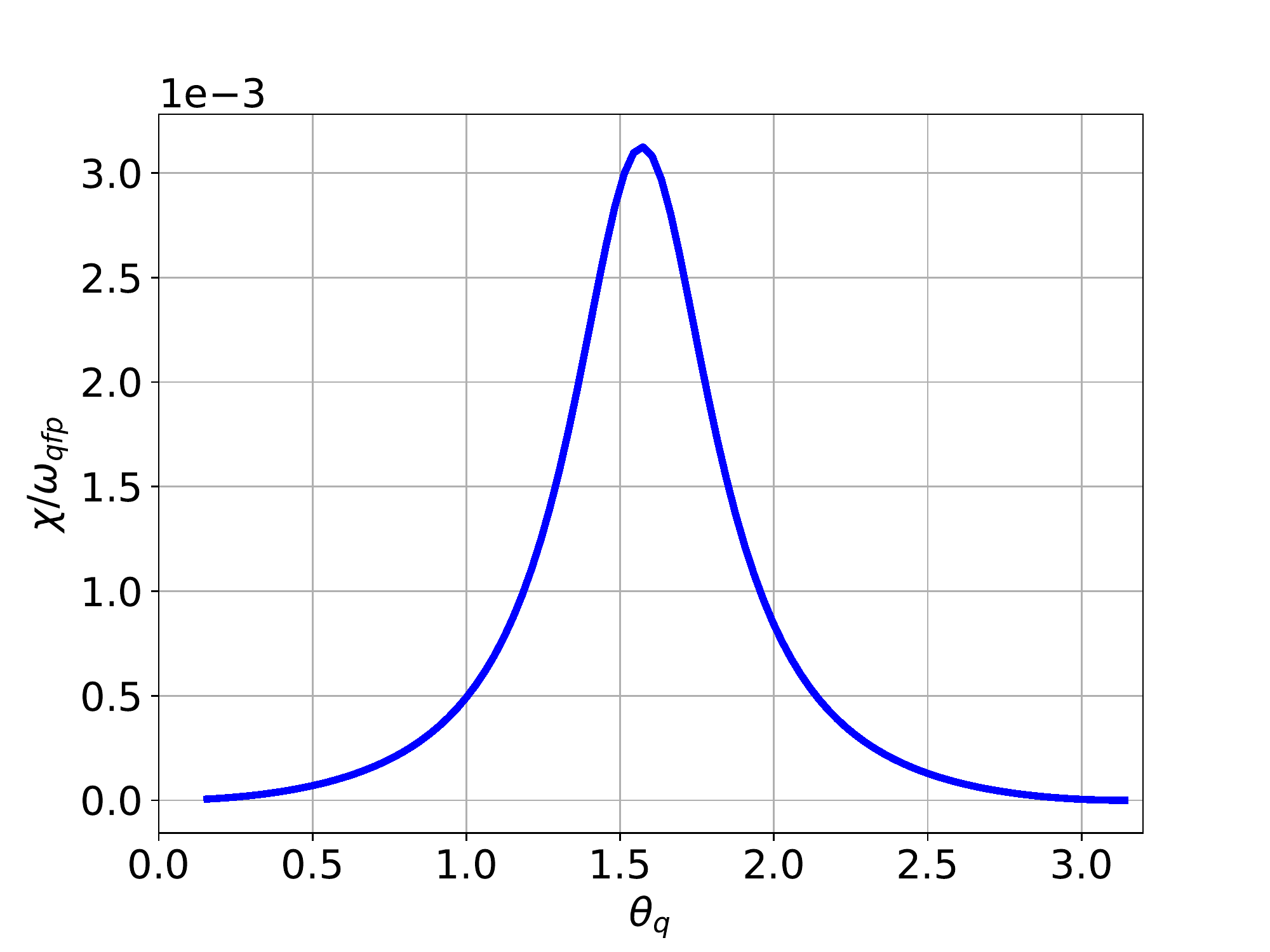}
		\caption{$\chi/\omega_{\text{qfp}}$ in Abhängigkeit von $\theta_q$ mit $\delta/g=8$ und $\eta=1.25$.}
		\label{fig:1q_chi_theta}
	\end{figure}
\begin{figure}[ht]
		\centering
		\includegraphics[width=0.7\textwidth]{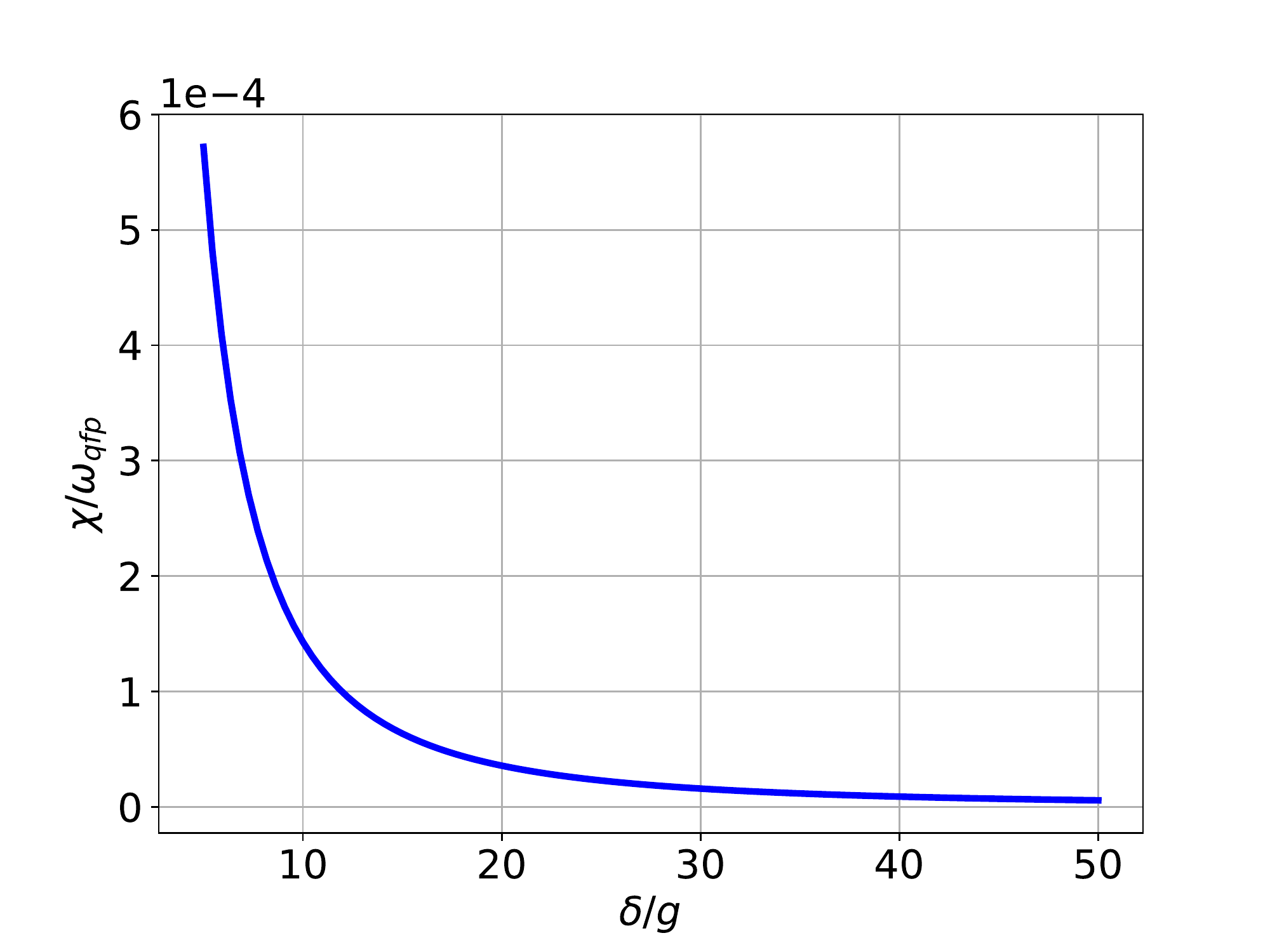}
		\caption{$\chi/\omega_{\text{qfp}}$ in Abhängigkeit von $\delta/g$ mit $\theta_{q}=\pi/4$ und $\eta=1.25$.  Mit steigendem $g$ wird die AC-Stark-Verschiebung größer.}
		\label{fig:1q_chi_delta}
	\end{figure}


	mit $\chi = g^2\sin[2](\theta_{\text{eff}})/\delta$. $\chi/\omega_{\text{qfp}}$ wird in Abhängigkeit von $\theta_q$ in Abb. \ref{fig:1q_chi_theta} und von $\delta/g$ in Abb. \ref{fig:1q_chi_delta} mit $\eta=1.25$ respektive dargestellt.\\
	
	Der Resonator wird benutzt, um den Qubitzustand auszulesen. Die Wechselwirkung zwischen Qubit und Resonator wird durch das Jaynes-Cummings-Modell beschrieben. So bekommen wir mit der Methode aus \cite{Pommerening2020} die Bedingung für eine gute Drehwellennäherung in der Flussbasis\\
	\begin{align}
		\abs{\frac{\sin(\theta_{\text{eff}})}{\cos(\theta_{\text{eff}})}} = \abs{\frac{\Delta_{\text{eff}}}{\epsilon_{\text{eff}}}}\ll 1.
		\label{equ:bed_1}
	\end{align}\\
	Im Vergleich dazu ist der Hamiltonoperator in der Energiebasis (Gl. \ref{equ:H_3}) selbst schon in einer diagonalisierten Form, daher ist die Bedingung bereits erfüllt. Das heißt, in der Energiebasis ist die Bedingung immer besser als in der Flussbasis erfüllt. Die Fidelity wird also in der Energiebasis immer höher sein als in der Flussbasis.\\
	
	\section{Mit treibendem Feld}
	
	Wir betrachten nun eine externe Ansteuerung des Resonators, die durch den folgenden Hamiltonoperator beschrieben wird \cite{Blais2004}\\
	\begin{align}
		\hat{H}_d=\epsilon_d(t) \hat{a}^{\dag} e^{-i\omega_d t}+\epsilon_d^*(t) \hat{a} e^{i\omega_d t}.
		\label{equ:H_d}
	\end{align}\\
	Mit Hilfe von Gleichungen \cite{CG69}\\
	\begin{align}
		-\pdv{\hat{D}(\alpha)}{\alpha^*}&=\left(\hat{a}-\frac{\alpha}{2}\right)\hat{D}(\alpha)=\hat{D}(\alpha)\left(\hat{a}+\frac{\alpha}{2}\right),\\
		\pdv{\hat{D}(\alpha)}{\alpha}&=\left(\hat{a}^{\dag}-\frac{\alpha^*}{2}\right)\hat{D}(\alpha)=\hat{D}(\alpha)\left(\hat{a}^{\dag}+\frac{\alpha^*}{2}\right),
	\end{align}\\
	können wir die Feldoperatoren mit dem zeitabhängigen Verschiebungsoperator $\hat{D}(\alpha)$ verschieben\\
	\begin{align}
		\tilde{\hat{H}}^{(1)}=\hat{D}^\dagger(\alpha)\left(\tilde{\hat{H}}+\hat{H}_d\right)\hat{D}(\alpha)-i\hat{D}^\dagger(\alpha)\dot{D}(\alpha).
	\end{align}\\
	Es folgt\\
	\begin{align}
		\begin{split}
			\tilde{\hat{H}}^{(1)}&=\frac{\omega_{\text{q,eff}}}{2}\tilde{\hat{\sigma}}_z+\omega_r \hat{a}^{\dag} a -g\sin(\theta_{\text{eff}})\left(a\tilde \hat{\sigma}_++\tilde{\hat{\sigma}}_{-} \hat{a}^{\dag}\right) +\omega_r\left(\hat{a}^{\dag} \alpha+\alpha^* \hat{a} +\abs{\alpha}^2\right)\\
			&\quad +g\cos(\theta_{\text{eff}})\tilde{\hat{\sigma}}_z\left(\hat{a}^{\dag}+\hat{a}+\alpha^*+\alpha\right)+\epsilon_d(t)\left(\hat{a}^{\dag}+\alpha^*\right)e^{-i\omega_d t}\\
			&\quad+\epsilon_d^*(t)\left(\hat{a}+\alpha\right)e^{i\omega_d t} - i \left[\dot{\alpha}\left(\hat{a}^{\dag}+\frac{\alpha^*}{2}\right)-\dot{\alpha}^*\left(\hat{a}+\frac{\alpha}{2}\right)\right].
		\end{split}
	\end{align}\\
	Um den Direktantrieb auf den Resonator (Gl. \ref{equ:H_d}) und den Term mit $\tilde{\hat{\sigma}}_z\left(\hat{a}^{\dag}+a\right)$ vom Hamiltonoperator zu eliminieren, wählen wir \cite{Blais2007}\\
	\begin{align}
		\dot{\alpha}=-i\left(\omega_r \alpha +g\cos(\theta_{\text{eff}})\tilde{\hat{\sigma}}_z+\epsilon_d (t) e^{-i\omega_d t}\right).
	\end{align}
	Vernachlässigen wir alle Transienten in $\alpha(t)$, ergibt sich\\
	\begin{align}
		\tilde{\hat{H}}^{(1)}=\frac{\omega_{\text{q,eff}}}{2}\tilde{\hat{\sigma}}_z+\omega_r \hat{a}^{\dag} a -g\sin(\theta_{\text{eff}})\left(a\tilde \hat{\sigma}_++\tilde{\hat{\sigma}}_{-} \hat{a}^{\dag}\right).
		\label{equ:H_d_1}
	\end{align}\\
	In dem Fall, dass die Antriebsamplitude $\epsilon_d$ zeitunabhängig ist, und indem wir uns sowohl für das Flussqubit als auch für den Feldoperator zu einem mit der Frequenz $\omega_d$ rotierenden Bezugssystem bewegen, 
	erhalten wir\\
	\begin{align}
		\tilde{\hat{H}}^{(2)}=\Delta_r \hat{a}^{\dag} a +\frac{\Delta_{\text{q,eff}}}{2}\tilde{\hat{\sigma}}_z-g\sin(\theta_{\text{eff}})\left(a\tilde \hat{\sigma}_++\tilde{\hat{\sigma}}_{-} \hat{a}^{\dag}\right),
		\label{equ:tildeH_d}
	\end{align}\\
	wobei $\Delta_r=\omega_r-\omega_d$ und $\Delta_{\text{q,eff}}=\omega_{\text{q,eff}}-\omega_d$.\\
	
	Des Weiteres wird der unitäre Operator (Gl. \ref{equ:U_3}) auf der Gl. \ref{equ:tildeH_d} angewendet. Es folgt\\
	\begin{align}
		\tilde{\hat{H}}^{(3)}=\Delta_r \hat{a}^{\dag}\hat{a}+\left[\frac{\Delta_{\text{q,eff}}}{2}+\chi\left(\hat{a}^{\dag}\hat{a}+\frac{1}{2}\right)\right]\tilde{\hat{\sigma}}_z.
	\end{align}\\
	Wenn wir uns im mit $\omega_{d}$ rotierendem Bezugssystem befinden und $\omega_r=\omega_d$ wählen, hat der Hamilton die gleiche Formel wie Gl. $\ref{equ:H_3}$. Daraus schließen wir, dass die Dynamik in beiden Fällen, also ob eine externe resonanten Ansteuerung vorkommt, vergleichbar ist. Im Folgenden betrachten wir der Einfachheit halber die Situation ohne externes treibendes Feld.\\
	
	\section{Fidelity in verschiedenen Basen}
	\label{fidelity}
	
	Zum Auslesen des gekoppelten Flussqubit-QFP-Systems benutzten wir die gleiche Methode wie in \cite{Pommerening2020}. Der Superoperator der Messung (siehe Gl.\ref{equ:E_x} und Gl.\ref{equ:g_x_form} in Kapitel \ref{messmethode}) ist dann\\
	\begin{align}
		\begin{split}
			\mathcal{E}_{\pm}(\rho(0))&=\tr_{\text{res}}E_{\pm}U(t_m)\rho(0)\otimes\ket{\alpha} \bra{\alpha}\hat{U}^\dagger(t_m)\\
			&=\sum_{n,m=0}^{\infty}\ket{m}\bra{m}E_x\ket{n}\bra{n} U\left(t_m\right)\ket{n}\bra{n}\rho \ket{\alpha}\braket{\alpha|m}\bra{m} \hat{U}^{\dagger}\left(t_m\right)\\
			&=\sum_{n,m=0}^{\infty}g_{x}\left(m,n\right)\bra{n} U\left(t_m\right)\ket{n}\rho(0) \bra{m} \hat{U}^{\dagger}\left(t_m\right)\ket{m},
		\end{split}
		\label{equ:messoper}
	\end{align}\\
	mit \footnote{$\alpha$ ist reell und positiv; $\beta=\abs{\beta}\mathrm{e}^{i\theta_{\text{eff}}}\equiv r \mathrm{e}^{i\theta_{\text{eff}}}$, $\overline{\beta}=r \mathrm{e}^{-i\theta_{\text{eff}}}$ und für $x=+$ ist $\theta$ von $-\pi$ bis $0$ sowie für $x=-$ ist $\theta$ von $0$ bis $\pi$.}\\
	\begin{align*}
		\begin{split}
			g_{\pm}\left(m,n\right)&=\bra{m}E_x\ket{n}\braket{n|\alpha}\braket{\alpha|m}\\
			&=\frac{1}{\pi}\int_{\Omega_{x}}\mathrm{d}^2\beta \braket{m|\beta}\braket{\beta|n}\braket{n|\alpha}\braket{\alpha|m}\\
			&=\frac{1}{\pi}\int_{\Omega_{x}}\mathrm{d}	^2\beta 	\mathrm{e}^{-\abs{\beta}^2}\frac{\beta^m\overline{\beta}^n}{\sqrt{m! n!}} \mathrm{e}^{-\alpha^2}\frac{\alpha^{n+m}}{\sqrt{m! n!}}\\
			&= \frac{\mathrm{e}^{-\alpha^2}}{\pi} \int_{0}^{\infty}r\mathrm{d}r \mathrm{e}^{-r^2}r^{m+n}\frac{\alpha^{n+m}}{m!n!}\int_{0}^{\pi}\mathrm{d}\theta \mathrm{e}^{-ix\left(m-n\right)\theta}\\
			&=\frac{\mathrm{e}^{-\alpha^2}}{\pi} \int_{0}^{\infty}r\mathrm{d}r \mathrm{e}^{-r^2}r^{m+n}\frac{\alpha^{n+m}}{m!n!}\left(\pi \delta_{mn}-\frac{ix}{m-n}\left[1-\mathrm{e}^{-ix\left(m-n\right)\pi}\right]\right)\\
			&=\mathrm{e}^{-\alpha^2} \left(\frac{\alpha^{2n}}{\left(n!\right)^2}\frac{\Gamma\left(n\right)}{2}-\frac{ix}{\pi}\frac{ \alpha^{n+m} \Gamma\left(\frac{m+n}{2}+1\right)}{ m! n! \left(m-n\right)} \text{odd}\left(m-n\right)\right)\\
			&=\mathrm{e}^{-\alpha^2} \left(\frac{\alpha^{2n}}{2n!}-\frac{ix}{\pi}\frac{ \alpha^{n+m} \Gamma\left(\frac{m+n}{2}+1\right)}{ m! n! \left(m-n\right)} \text{odd}\left(m-n\right)\right).
		\end{split}
	\end{align*}\\
	Wobei\\
	\[ x\in \{\pm\}\quad \text{und}\quad \text{odd}\left(n\right) = \left\{ \begin{array}{ll}
		1, & \mbox{$n$ ist eine ungerade ganze Zahl}\\
		0, & \mbox{sonst}.\end{array} \right. \]\\
	
\begin{figure}[ht]
		\centering
		\includegraphics[width=0.7\textwidth]{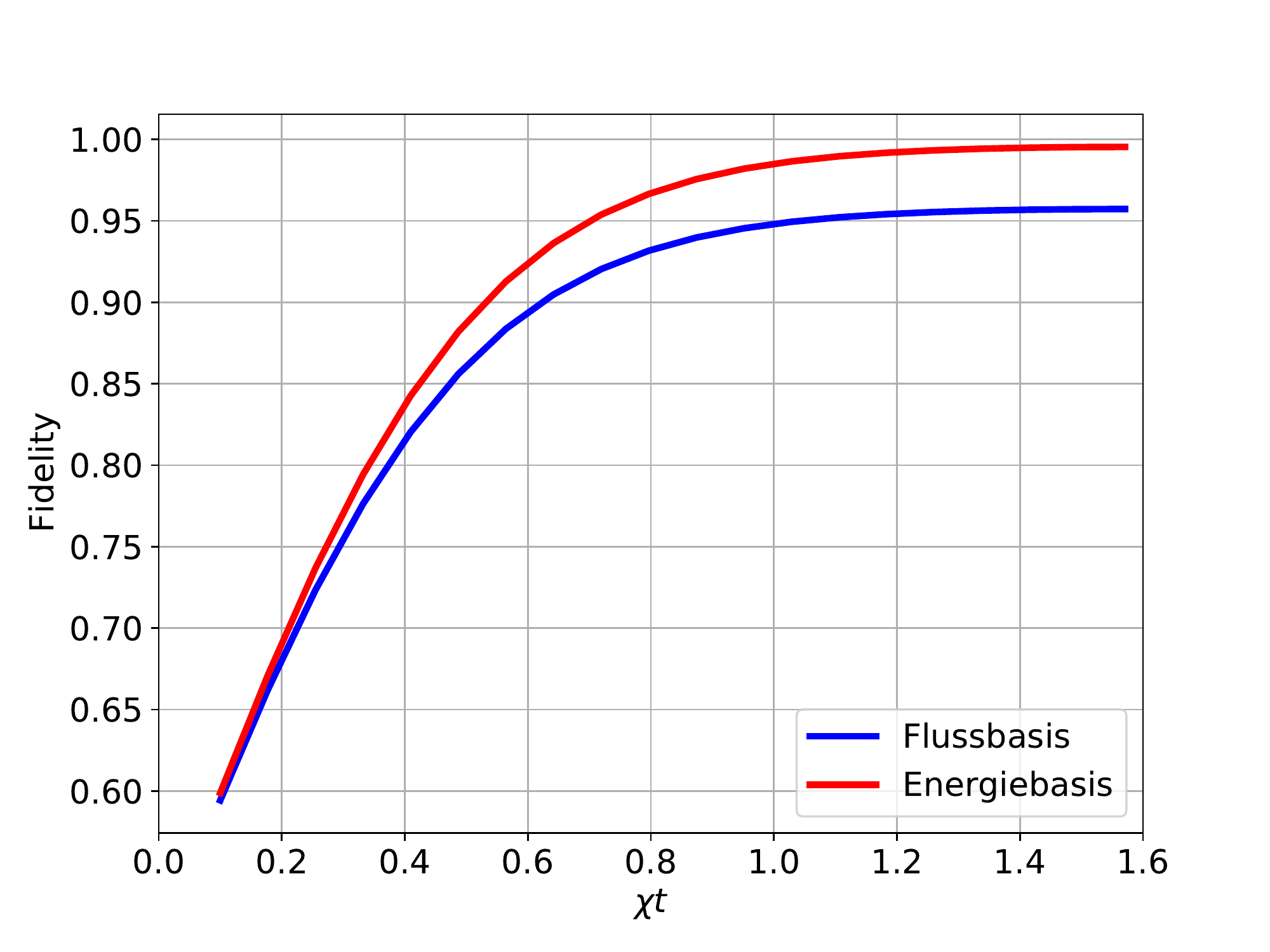}
		\caption{Fidelity in Abhängigkeit von $\chi t$ in der Flussbasis (blau) und Energiebasis (rot) mit $N=27$, $\Delta_{q}/\epsilon_{q}=1,~ \alpha=1,~ \delta/g=8,~ \eta =1.25$. In beiden Basen wird die Fidelity mit steigendem $\chi t$ höher und erreicht einen Sättigungswert.}
		\label{fig:1q_chit_fq}
	\end{figure}

\begin{figure}[ht]
		\centering
		\includegraphics[width=0.7\textwidth]{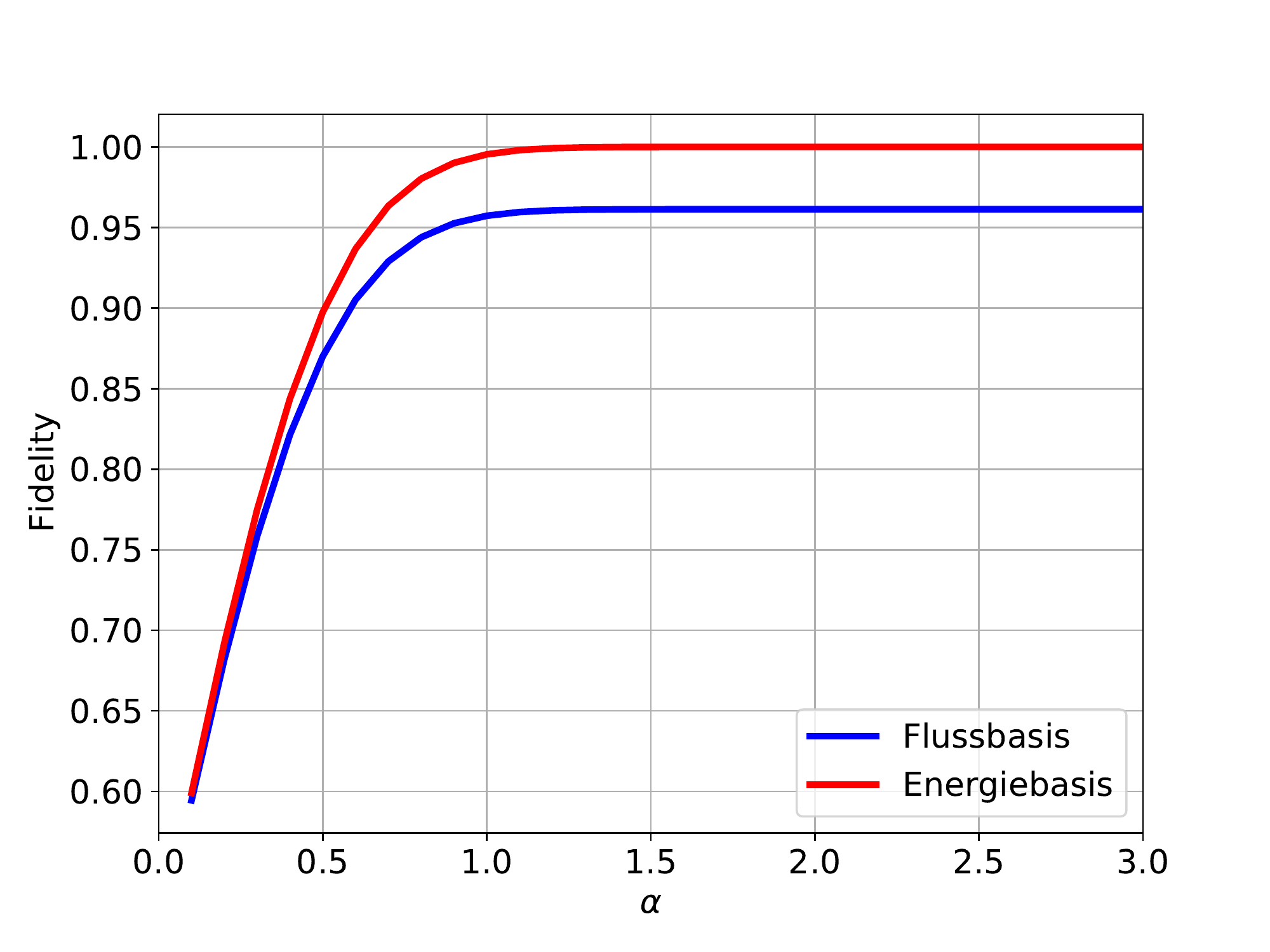}
		\caption{Fidelity in Abhängigkeit von $\alpha$ in der Fluss (blau)- und Energiebasis (rot) mit $N=27$, $\Delta_{q}/\epsilon_{q}=1,~ \delta/g=8,~ \eta =1.25$ und $t_m = \pi/2\chi $. Die Fidelity wird mit steigendem $\alpha$ in beiden Basen höher und erreicht einen Sättigungswert.}
		\label{fig:1q_alpha_fq}
	\end{figure}

	Der Unterschied zwischen Quantenzuständen wird durch die Fidelity beschrieben. Die Fidelity für zwei Dichtematrizen $\rho$ und $\sigma$ lässt sich definieren als \cite{Jozsa94}\\
	\begin{align*}
		\mathcal{F}\left(\rho, \sigma\right)=\left(\Tr\sqrt{\sqrt{\rho}\sigma\sqrt{\rho}}\right)^2.
	\end{align*}
	Im Fall, dass $\rho_0$ einen reinen Zustand und $\sigma$ einen gemischten Zustand darstellen, reduziert sich die Definition auf\\
	\begin{align*}
		\mathcal{F}\left(\rho_0, \sigma\right)&=\left(\Tr\left[\sqrt{\ket{\psi_0}\bra{\psi_0}\sigma\ket{\psi_0}\bra{\psi_0}}\right]\right)^2\\
		&=\bra{\psi_0}\sigma\ket{\psi_0}\left(\Tr\left[\sqrt{\ket{\psi_0}\bra{\psi_0}}\right]\right)^2\\
		&=\bra{\psi_0}\sigma\ket{\psi_0}.
	\end{align*}\\
	Wenn beide Zustände rein sind, reduziert sich die Definition auf den quadratischen Überlapp zwischen den Zuständen:\\
	\begin{align}
		\mathcal{F}\left(\rho_0, \sigma_0\right)=\abs{\braket{\psi_{\rho_0}|\psi_{\sigma_0}}}^2.
	\end{align}\\
	Sei $\ket{\psi_0}=\alpha\ket{0}+\beta\ket{1}$ der Anfangszustand des Qubits, dann ist die Fidelity\\
	\begin{align}
		\mathcal{F}=\bra{\psi_0}\mathcal{E}_{\pm}(\rho(0))\ket{\psi_0}.
		\label{equ:fi_2}
	\end{align}\\
	Die numerische Untersuchung der Fidelity wird in Abhängigkeit von $\chi t$ und $\alpha$ in Abb. \ref{fig:1q_chit_fq} und Abb. \ref{fig:1q_alpha_fq} respektive dargestellt.\\
	
	Mit steigendem $\chi t$ wird die Fidelity höher (siehe Abb. \ref{fig:1q_chit_fq}).
	Wenn die Wechselwirkungszeit zwischen QFP und Resonator zu kurz ist, ist die Fidelity niedrig. Zur Messzeit $t_m=\pi/2\chi$ \cite{Pommerening2020} erreicht die Fidelity in beiden Basen das Maximum. Energiebasis ist immer besser als Flussbasis.\\

	Das Maximum der Fidelity wird bei $\alpha\approx1$ erreicht (siehe Abb. \ref{fig:1q_alpha_fq}). Für kleines $\alpha$ ist die Fidelity in beiden Basen niedrig. Für gewählte Messzeit $t_m=\pi/2\chi$ wird die Energiebasis immer besser als die Flussbasis.\\
	
	\clearpage

	In diesem Teil haben wir die Modellierung der Messung eines Flussqubits erstellt und die Fidelity in der Fluss- und Energiebasis numerisch untersucht. Durch die adiabatische Durchführung vom QFP-Annealing wird der Zustand des Flussqubits im QFP gespeichert. Die numerische Simulation hat gezeigt, dass der Speicherprozess für Fidelitys $\gtrsim 0.75$ in der Energiebasis höher ist. Nach dem QFP-Annealing wird das verschränkte System durch einen effektiven Flussqubit-Hamiltonoperator beschrieben. Es wird dann durch einen Resonator ausgelesen. In der Theorie wurde gezeigt, dass die Fidelity bei Messungen in der Energiebasis höher ist, als in der Flussbasis. Das wird in den numerischen Ergebnissen bestätigt.
	
	\clearpage
	\thispagestyle{empty}
	
	\part{Modellierung der Messung von zwei Flussqubits mittels QFP}
	
	\clearpage
	\thispagestyle{plain}
	
	\vspace*{\fill}
	
	In diesem Teil werden zwei gekoppelte Flussqubits FQ1 und FQ2 betrachtet. FQ1 (respektive FQ2) ist an ein QFP1 (respektive QFP2) gekoppelt. Zum Auslesen von FQ2 führen wir QFP2-Annealing adiabatisch durch. Dann wird das QFP2 mittels Resonator gemessen.\\
	
	Zwei Modelle werden dabei berücksichtigt. Im ersten Modell führen wir bei dem mit FQ1 verbundenen QFP1 kein Annealing durch. Sobald das Signal von FQ2 sowie die Wechselwirkung zwischen FQ1 und FQ2 im QFP2 verschränkt, berücksichtigen wir nur den Ausleseprozess von QFP2.\\
	
	Im Unterschied zu dem ersten Modell wird QFP1-Annealing bei dem zweiten Modell durchgeführt. QFP1 und QFP2 wechselwirken dann miteinander. Die Fidelity beim Auslesen von FQ2  wird in beiden Modellen theoretisch analysiert und mittels QuTiP \cite{qutip1}\cite{qutip2} numerisch simuliert.\\
	
	\vspace{15em}
	\vspace*{\fill}
	
	\clearpage
	\thispagestyle{empty}

	\chapter{Messung von FQ2 ohne QFP1-Annealing}
	\label{cha:mq2}
	
	In diesem Modell hat QFP1 kein Annealing. Außerdem nehmen wir an, dass QFP2-Annealing adiabatisch durchgeführt wird und das FQ2-Signal schon im QFP2 verschränkt ist. Im Folgenden wird nur der Ausleseprozess von QFP2 untersucht. Die Fidelity wird in verschiedenen Basen theoretisch untersucht und numerisch simuliert.\\
	
	\begin{figure}[ht]
		\centering
		\includegraphics[width=0.7\textwidth]{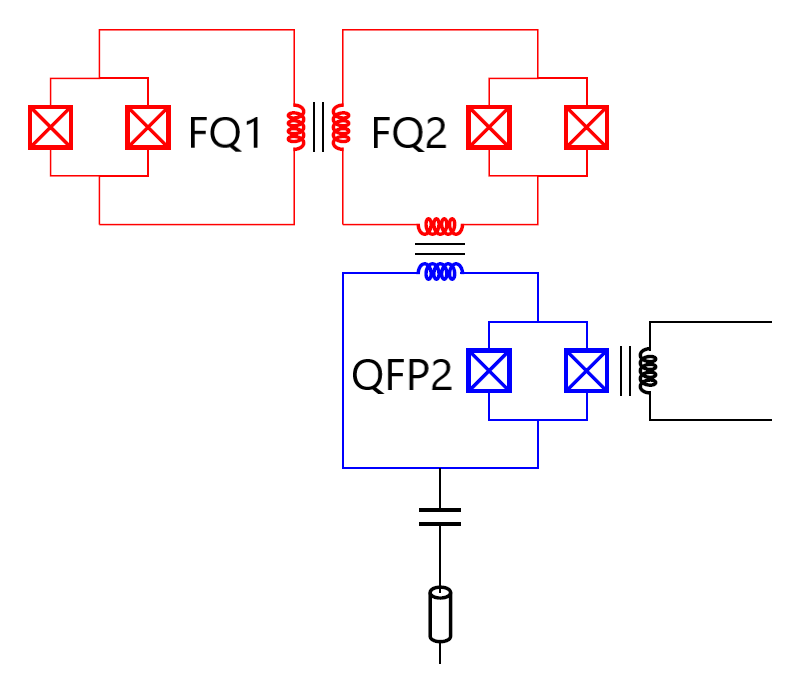}
		\caption{Modellierung der Messung von FQ2 ohne QFP1-Annealing. Zwei Flussqubits (rot) sind gekoppelt. FQ2 wird mittels QFP2 (blau) gemessen. Durch QFP2-Annealing wird das Signal von FQ2 im QFP2 gespeichert. Dann wird QFP2 durch einen Resonator ausgelesen.}
	\end{figure}
	
	\section{Modellierung}
	
	Sobald das FQ2-Signal und die Wechselwirkung zwischen FQ1 und FQ2 im QFP2 verschränkt sind, berücksichtigen wir nur das Auslesen von QFP2.\\
	
	Das verschränkte FQ2-QFP2-System lässt sich durch $\hat{H}_{\text{eff,2}}$ beschreiben. Der gemeinsame Hamiltonoperator lässt sich in der Flussbasis darstellen:\\
	\begin{align}
		&\hat{H}=\hat{H}_{\text{eff,2}}+J\hat{\sigma}_1^z\hat{\sigma}_2^z+g\hat{\sigma}_2^z\left(\hat{a}^{\dag}+\hat{a}\right)+\omega_r \hat{a}^{\dag}\hat{a},\\
		&\hat{H}_{\text{eff,2}}=-\frac{1}{2}\left(\epsilon_{\text{eff,2}} \hat{\sigma}_2^z+\Delta_{\text{eff,2}}\hat{\sigma}_2^x\right),
	\end{align}\\
	wobei $J$ der Kopplungsstärke zwischen beiden Flussqubits und $g$ der FQ2-Resonator-Kopplungsstärke entsprechen. Im Vergleich zum Hamiltonoperator für das FQ2 ohne QFP ist die effektive Qubit-Magnetenergie vergrößert: $\epsilon_{\text{eff,2}}>\epsilon_{2}$, während die effektive Tunnelrate verkleinert ist: $\Delta_{\text{eff,2}}=e^{-\eta} \Delta_{2}$. $\eta$ ist dabei vom QFP2 abhängig.\\
	
	In der FQ2-Energiebasis erhalten wir den Hamiltonoperator im mit $\omega_r$ rotierenden Bezugssystems:\\
	\begin{align}
		\begin{split}
			\tilde{\hat{H}}=&-\frac{\delta_{\text{eff,2}}}{2}\tilde{\hat{\sigma}}_2^z+J\hat{\sigma}_1^z\left[\cos(\theta_{\text{eff,2}}) \tilde{\hat{\sigma}}_2^z-\sin(\theta_{\text{eff,2}}) \tilde{\hat{\sigma}}_2^x\right]\\
			&+g\left[\cos(\theta_{\text{eff,2}}) \tilde{\hat{\sigma}}_2^z-\sin(\theta_{\text{eff,2}}) \tilde{\hat{\sigma}}_2^x\right]\left(\hat{a}^{\dag}+\hat{a}\right),
		\end{split}
	\end{align}\\
	wobei $\delta_{\text{eff,2}}=\omega_{\text{eff,2}}-\omega_r$ die effektive Verstimmung zwischen QFP2- und Resonatorfrequenz ist. $\omega_{\text{eff,2}}=\sqrt{\epsilon_{\text{eff,2}}^2+\Delta_{\text{eff,2}}^2}$ ist dabei die effektive FQ2-Frequenz. Die Mischungswinkel $\theta_{\text{eff,2}}$ ist durch $\tan(\theta_{\text{eff,2}})=\Delta_{\text{eff,2}}/\epsilon_{\text{eff,2}}$ definiert.\\
	
\begin{figure}[ht]
		\centering
		\includegraphics[width=0.7\textwidth]{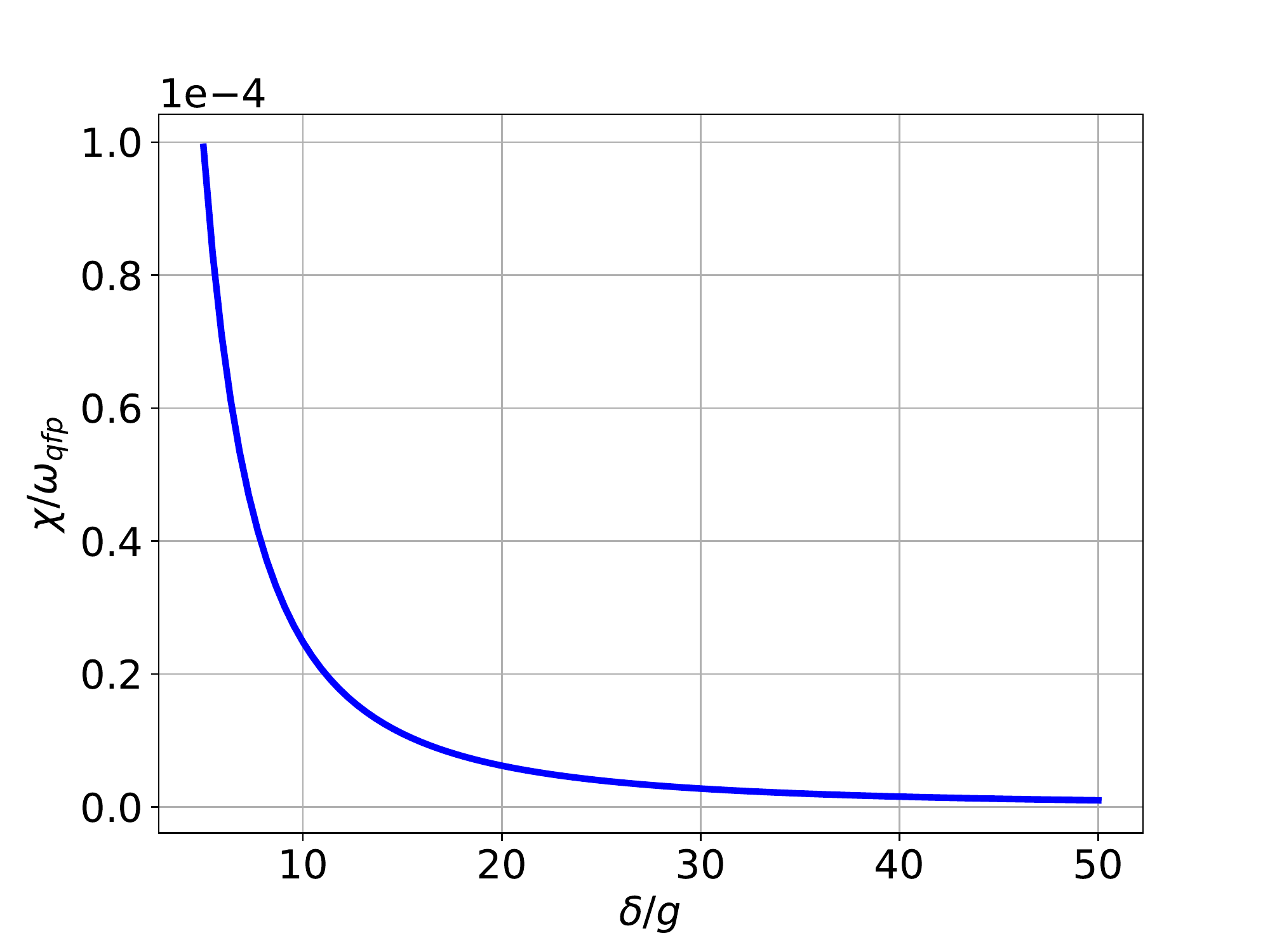}
		\caption{$\chi/\omega_{\text{qfp}}$ in Abhängigkeit von $\delta/g$ mit $\Delta_2/\epsilon_2=0.3$ und $\eta=1.25$.}
		\label{fig:2qo_chidelta}
	\end{figure}

	Durch Anwenden des Verschiebungsoperators $\hat{D}(\alpha)$ mit $\alpha=-g\cos(\theta_{\text{eff,2}})\tilde{\hat{\sigma}}_2^z/\omega_r$ folgt unter der Drehwellennäherung analog zur Gl. \ref{equ:Htilde1}\\
	\begin{align}
		\begin{split}
			\tilde{\hat{H}}^{(1)}=&-\frac{\delta_{\text{eff,2}}}{2}\tilde{\hat{\sigma}}_2^z+J\hat{\sigma}_1^z\left[\cos(\theta_{\text{eff,2}}) \tilde{\hat{\sigma}}_2^z-\sin(\theta_{\text{eff,2}}) \tilde{\hat{\sigma}}_2^x\right]\\
			&-g\sin(\theta_{\text{eff,2}}) \left(\hat{a}^{\dag}\tilde{\hat{\sigma}}_2^-+\hat{a}\tilde{\hat{\sigma}}_2^+\right).
		\end{split}
		\label{equ:Hq2}
	\end{align}\\
	Um den Term der Wechselwirkung zwischen FQ2 und Resonator zu eliminieren, wenden wir $\hat{U}=\exp\left(\lambda\left(\hat{a}\tilde{\hat{\sigma}}_2^+-\hat{a}^{\dag}\tilde{\hat{\sigma}}_2^-\right)\right)$ mit $\lambda=g\sin(\theta_{\text{eff,2}})/\delta_{\text{eff,2}}$ auf die Gl. \ref{equ:Hq2} an. So ergibt sich der Hamiltonoperator in der FQ2-Energiebasis\\
	\begin{align}
		\begin{split}
			\tilde{\hat{H}}^{(2)}\approx&-\frac{\hat{\delta}_{\text{eff2,n}}}{2}\tilde{\hat{\sigma}}_2^z+J\hat{\sigma}_1^z\left[\cos(\theta_{\text{eff,2}}) \tilde{\hat{\sigma}}_2^z-\sin(\theta_{\text{eff,2}}) \tilde{\hat{\sigma}}_2^x\right],
		\end{split}
		\label{equ:Ht2_q2eig}
	\end{align}\\
	wobei $\hat{\delta}_{\text{eff2,n}}=\delta_{\text{eff,2}}+\chi\left(2\hat{n}+1\right)$ mit $\hat{n} =\hat{a}^{\dag}\hat{a}$.  $\chi=g^2\sin[2](\theta_{\text{eff,2}})/ \delta_{\text{eff,2}}$ ist dabei die dispersive Kopplungsstärke. Sie wird in Abhängigkeit von $\delta/g$ mit $\delta=\omega_2 -\omega_r$ der Verstimmung zwischen FQ2- und Resonatorfrequenz in Abb. \ref{fig:2qo_chidelta} dargestellt. Mit steigendem $g$, also fallendem $\delta/g$ wird $\abs{\chi}$ größer.\\
	
	Der Term $\lambda J\cos(\theta_{\text{eff,2}})$ wird wegen $\{\lambda, J\}\ll \{\omega_{\text{eff,2}},~\omega_r\}$ vernachlässigt. Den aus $-J\hat{\sigma}_1^z\sin(\theta_{\text{eff,2}})U\tilde{\hat{\sigma}}_2^x\hat{U}^\dagger$ herausgekommenen Term $-\lambda J\hat{\sigma}_1^z\sin(\theta_{\text{eff,2}})\left(\hat{a}^{\dag}+\hat{a}\right)\tilde{\hat{\sigma}}_2^z$ kann man ignorieren, weil wir im Bezugssystem des verschobenen kohärenten Zustands sind \cite{Blais2007}.\\
	Um die Fidelity in verschiedenen Basen zu untersuchen, stellen wir auch den Hamiltonoperator (Gl. \ref{equ:Ht2_q2eig}) in der Flussbasis dar\\
	\begin{align}
		\begin{split}
			\hat{H}^{(\text{f})}\approx-\frac{\hat{\delta}_{\text{eff2,n}}}{2}\left[\cos(\theta_{\text{eff,2}}) \hat{\sigma}_2^z+\sin(\theta_{\text{eff,2}}) \hat{\sigma}_2^x\right]+J\hat{\sigma}_1^z\hat{\sigma}_2^z.
		\end{split}
		\label{equ:Hf}
	\end{align}\\

	Der Hamiltonoperator lässt sich zusätzlich in der FQ1-Energiebasis (respektive FQ1-FQ2-Energiebasis) umschreiben\\
	\begin{align}
		\begin{split}
			\tilde{\hat{H}}^{(3)}\approx&-\frac{\hat{\delta}_{\text{eff2,n}}}{2}\tilde{\hat{\sigma}}_2^z+J\left[\cos(\theta_1) \tilde{\hat{\sigma}}_1^z-\sin(\theta_1) \tilde{\hat{\sigma}}_1^x\right]\left[\cos(\theta_{\text{eff,2}}) \tilde{\hat{\sigma}}_2^z-\sin(\theta_{\text{eff,2}}) \tilde{\hat{\sigma}}_2^x\right],
		\end{split}
		\label{equ:ht3_q1q2}
	\end{align}\\
	wobei $\theta_1$ durch $\tan(\theta_1)=\epsilon_1/\Delta_{1}$ definiert ist.\\
	
	Sobald wir den Hamiltonoperator erhalten, können wir die Fidelity durch die gleiche Methode wie in \cite{Pommerening2020} analog zu Abschnitt \ref{fidelity} untersuchen. Der Superoperator der Messung lautet\\
	\begin{align}
		\mathcal{E}_{\pm}(\rho(0))&=\tr_{\text{res}}\tr_{\text{FQ1}} E_{\pm}U(t_m)\rho(0)\otimes\ket{\alpha} \bra{\alpha}\hat{U}^\dagger(t_m).
		\label{equ:sup_2q}
	\end{align}\\
	Im Vergleich zu vorher (Gl. \ref{equ:messoper}) wird bei der Gl. \ref{equ:sup_2q} zusätzlich FQ1 ausgespurt, denn wir interessieren uns nur für FQ2.\\
	
\begin{figure}[ht]
		\centering
		\includegraphics[width=0.7\textwidth]{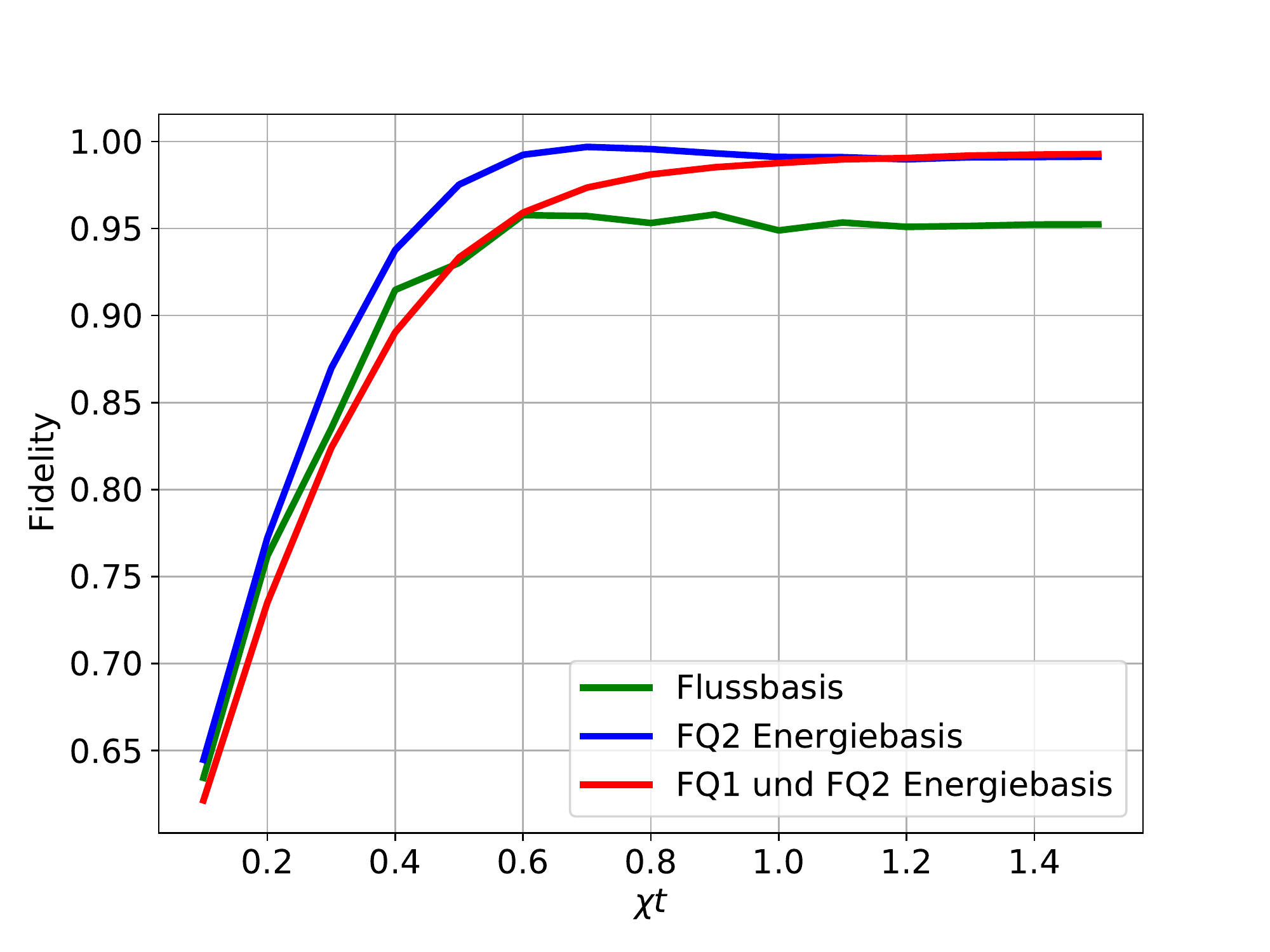}
		\caption{Fidelity in Abhängigkeit von $\chi t$ in der Flussbasis (grün), FQ2-Energiebasis (blau)  und FQ1-FQ2-Energiebasis (rot) mit $N=27$, $\eta=1.25$, $J=0.05(\omega_{2}-\omega_{1})$, $\Delta_{2}/\epsilon_2 = 1$, $\delta/g=8$ und $\alpha=1$. Die Fidelity wird mit steigender Messzeit höher.}
		\label{fig:2qo_chit_fq2q1q2}
	\end{figure}

\begin{figure}[ht]
		\centering
		\includegraphics[width=0.7\textwidth]{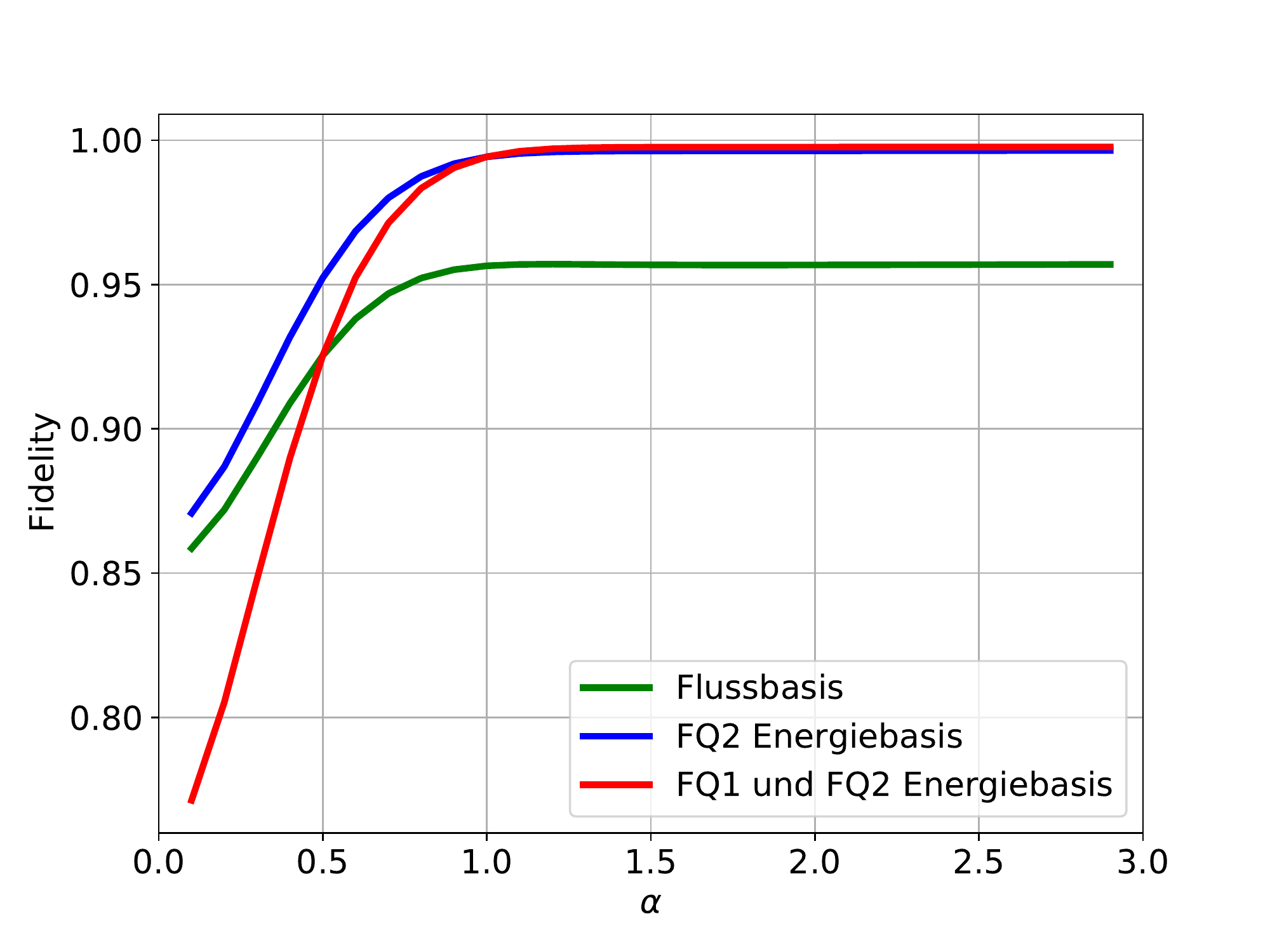}
		\caption{Fidelity in Abhängigkeit von $\alpha$ in der Flussbasis (grün), FQ2-Energiebasis (blau)  und FQ1-FQ2-Energiebasis (rot) mit $N=27$, $\eta=1.25$, $J=0.05(\omega_{2}-\omega_{1})$, $\Delta_{2}/\epsilon_2 = 1$, $\delta/g=8$ und $t_m=\pi/2\chi$. Die Fidelity wird mit steigendem $\alpha$ höher und erreicht einen Sättigungswert.}
		\label{fig:2qo_alpha_fq2q1q2}
	\end{figure}

\begin{figure}[ht]
		\centering
		\includegraphics[width=0.7\textwidth]{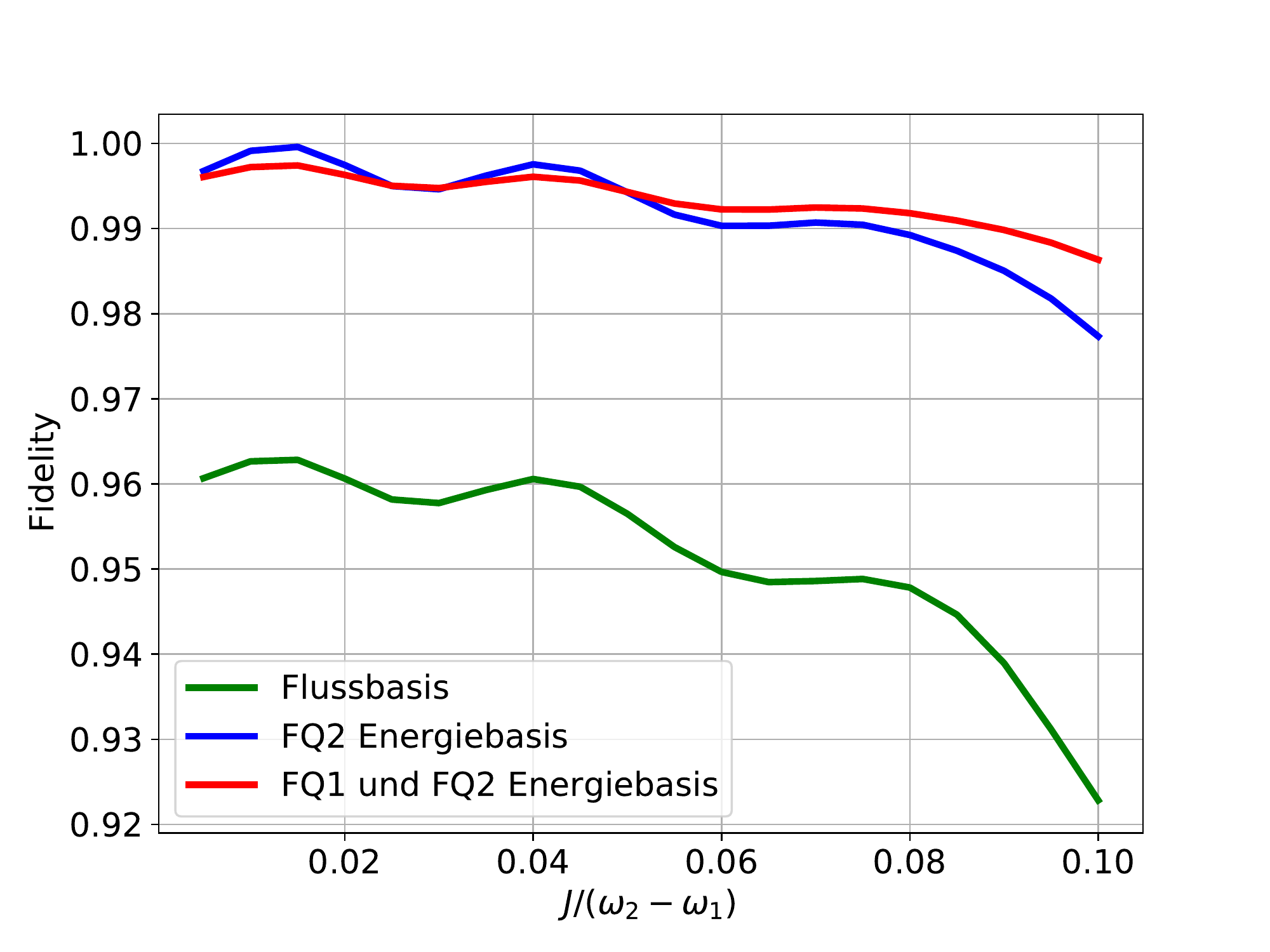}
		\caption{Fidelity in Abhängigkeit von $J/(\omega_{2}-\omega_{1})$ in der Fluss- (grün), FQ2-Energiebasis (blau)  und FQ1-FQ2-Energiebasis (rot) mit $N=27$, $\eta=1.25$, $t_m=\pi/2\chi$, $\Delta_{2}/\epsilon_2 = 1$, $\delta/g=8$ und $\alpha=1$. Die Fidelity oszilliert in Abhängigkeit von $J/(\omega_{2}-\omega_{1})$, wobei der Mittelwert der Oszillation mit steigendem $J/(\omega_{2}-\omega_{1})$ niedriger wird.}
		\label{fig:2qo_J_fq2q1q2}
	\end{figure}

	Die Fidelity in drei Basen wird in Abhängigkeit von $\chi t$, $\alpha$ und $J/(\omega_{2}-\omega_{1})$ in Abb. \ref{fig:2qo_chit_fq2q1q2}, Abb. \ref{fig:2qo_alpha_fq2q1q2} und Abb. \ref{fig:2qo_J_fq2q1q2} respektive dargestellt.\\
	
	Zur Messzeit $t_m=\pi/4\chi$ wird das Maximum der Fidelity in der FQ2-Energiebasis erreicht (siehe Abb. \ref{fig:2qo_chit_fq2q1q2}). Dann wird die Fidelity leicht niedriger. Im Gegensatz zur Abb. \ref{fig:1q_alpha_fq} tritt hier keine Sättigung auf. Eine mögliche Ursache dafür könnte die Wechselwirkung zwischen FQ1 und FQ2 sein. Die Qubits formen Dressed-Zustände und wir messen nicht qubit-spezifisch sondern am gekoppelten Zwei-Qubit-System, was nur das gleiche ist, wenn alles QND ist. So wird man bei einer langen Messzeit in einer FQ1-FQ2-Energiebasis effektiver messen. Im Vergleich dazu wird das Maximum der Fidelity in der FQ1-FQ2-Energiebasis langsamer erreicht. Flussbasis ist bei $\chi t\lesssim 0.5$ besser als die FQ1-FQ2-Energiebasis. Dann kommt das Basis-Crossover von den beiden. Basis-Crossover heißt hier, dass die Messbasis, welche die Messung besser annähert als die andere, wechselt.
	
	Bei $\alpha \approx 1$ wird das Maximum der Fidelity in allen Basen erreicht und bleibt dann konstant (siehe Abb. \ref{fig:2qo_alpha_fq2q1q2}). Das Basis-Crossover zwischen Flussbasis und FQ1-FQ2-Energiebasis kommt bei $\alpha\approx0.5$ vor, während das zwischen beiden Energiebasen nicht einfach zu erkennen ist. Es liegt wohl an der Wahl der Parameter $t_m=\pi/2\chi$ (siehe Abb. \ref{fig:2qo_chit_fq2q1q2}) und $J=0.05(\omega_{2}-\omega_{1})$ (siehe Abb. \ref{fig:2qo_J_fq2q1q2}).\\

	Zum Schluss wird die Fidelity in Abhängigkeit von $J/(\omega_2-\omega_1)$ in Abb. \ref{fig:2qo_J_fq2q1q2} dargestellt. Bei $t_m=\pi/2\chi$ und $\alpha=1$ ist die Fidelity sowohl in der FQ2-Energiebasis als auch in der FQ1-FQ2-Energiebasis immer höher als in der Flussbasis. Höhere Kopplung zwischen FQ1 und FQ2 wird zur gegenseitigen Beeinflussung während der Messung führen. Die Fidelity wird dann niedriger. Das Basis-Crossover zwischen beiden Energiebasen kommt bei $J\approx0.05(\omega_{2}-\omega_{1})$ vor. Dann wird die
	
	\clearpage
	
	Fidelity in der FQ1-FQ2-Energiebasis höher.\\
	
	In diesem Kapitel haben wir die Modellierung der Messung von FQ2 ohne QFP1-Annealing erstellt. Der Hamiltonoperator wird in der FQ2-Energiebasis (Gl. \ref{equ:Ht2_q2eig}), der Flussbasis (Gl. \ref{equ:Hf}) und der FQ1-FQ2-Energiebasis (Gl. \ref{equ:ht3_q1q2}) dargestellt. Beim Superoperator der Messung von FQ2 wird im Unterschied zu vorher noch zusätzlich das FQ1 ausgespurt.  Die Fidelity wird in verschiedenen Basen numerisch simuliert. Das Basis-Crossover kommt in verschiedenen Darstellungen vor.\\
	
	\section{Gegenüberstellung der verschiedenen Basen}
	
	Im Folgenden wird die Gültigkeit der Drehwellennäherung in verschiedenen Basen mit der in \cite{Pommerening2020} benutzten Methode (siehe Kapitel \ref{messmethode}) untersucht. Wir fangen zuerst mit dem Hamiltonoperator in der FQ2-Energiebasis $\tilde{\hat{H}}^{(2)}$ (Gl. $\ref{equ:Ht2_q2eig}$) an. Dann wird der Hamiltonoperator in der FQ1-FQ2-Energiebasis $\tilde{\hat{H}}^{(3)}$ (Gl. $\ref{equ:ht3_q1q2}$) diskutiert.\\
	
	\subsection{FQ2-Energiebasis}
	
	In der Bare-FQ2-Energiebasis haben wir den Hamiltonoperator in der Form von\\
	\begin{align}
		\tilde{\hat{H}}^{(2)}=-\frac{\hat{\delta}_{\text{eff2,n}}}{2}\tilde{\hat{\sigma}}_2^z +J_{zz}\hat{\sigma}_1^z\tilde{\hat{\sigma}}_2^z - J_{zx} \hat{\sigma}_1^z \tilde{\hat{\sigma}}_2^x,
		\label{equ:h2}
	\end{align}\\
	mit $J_{zx}= J\sin(\theta_{\text{eff,2}})$ und $J_{zz}= J\cos(\theta_{\text{eff,2}})$.
	Die entsprechende Matrixformel ist\\
	\begin{align}
		\begin{split}
			\tilde{\hat{H}}^{(2)}&=\begin{bmatrix}
					-\frac{\hat{\delta}_{\text{eff2,n}}}{2}+J_{zz} & -J_{zx} & 0 & 0\\
					-J_{zx} &\frac{\hat{\delta}_{\text{eff2,n}}}{2}-J_{zz} & 0 & \\
					0 & 0 & -\frac{\hat{\delta}_{\text{eff2,n}}}{2}-J_{zz} & J_{zx}\\
					0 & 0 & J_{zx} &\frac{\hat{\delta}_{\text{eff2,n}}}{2}+J_{zz}
				\end{bmatrix}.
		\end{split}
		\label{equ:Hohnedeltaqfpmf}
	\end{align}\\

	Um die Berechnung zur Diagonalisierung des Hamiltonoperators zu vereinfachen und mathematisch besser darzustellen, wird in der folgenden Berechnung $\hat{n}=\hat{a}^{\dag}\hat{a}$ in $\hat{\delta}_{\text{eff2,n}}=\delta_{\text{eff,2}}+\chi\left(2\hat{n}+1\right)$ durch dessen Erwartungswert  $n=\expval{\hat{a}^{\dag}\hat{a}}$ ersetzt, also $\hat{\delta}_{\text{eff2,n}} \rightarrow \delta_{\text{eff2,n}}$. Diese Linearisierung ist möglich, denn wir arbeiten nur mit den Basen der Flussqubits (ohne Resonator). So verhält sich der Term $\hat{a}^{\dag}\hat{a}$ in dieser Basis wie eine Konstante. Wenn man zurück zum Hamiltonoperator kommt, sollte man dann statt des Erwartungswertes den Operator benutzen, also $\delta_{\text{eff2,n}} \rightarrow \hat{\delta}_{\text{eff2,n}}$.\\
	
	Durch die Diagonalisierung erhalten wir die Eigenwerte\\
	\begin{align}
		\begin{split}
			E^{(2)}&=\big\{ -\omega_{n-},~
			\omega_{n-},~-\omega_{n+},
			~\omega_{n+} \big\},
		\end{split}
	\end{align}\\
	mit $\omega_{n\pm}=\sqrt{\left(\delta_{\text{eff2, n}}/2\pm J_{zz}\right)^2+J_{zx}^2}$ und die Eigenzustände\\
	\begin{align*}
		\begin{split}
			\ket{\overline{00}}&=\cos(\frac{\theta_{n-}}{2})\ket{\tilde 0\tilde 0}+\sin(\frac{\theta_{n-}}{2})\ket{\tilde 0\tilde 1},\ket{\overline{01}}=-\sin(\frac{\theta_{n-}}{2})\ket{\tilde 0\tilde 0}+\cos(\frac{\theta_{n-}}{2})\ket{\tilde 0\tilde 1},\\
			\ket{\overline{10}}&=\cos(\frac{\theta_{n+}}{2})\ket{\tilde1\tilde 0}+\sin(\frac{\theta_{n+}}{2})\ket{\tilde1\tilde 1}, \ket{\overline{11}}=-\sin(\frac{\theta_{n+}}{2})\ket{\tilde1\tilde 0}+\cos(\frac{\theta_{n+}}{2})\ket{\tilde1\tilde 1},
		\end{split}
	\end{align*}\\
	wobei $\theta_{n\pm}$ durch $\tan(\theta_{n\pm})=\mp J_{zx}/\left(\delta_{\text{eff2,n}}/2 \pm J_{zz}\right)$ definiert sind.
	Mit $\overline{\hat{\sigma}}_{2}^{z}=\ket{\overline{00}}\bra{\overline{00}}-\ket{\overline{01}}\bra{\overline{01}}+\ket{\overline{10}}\bra{\overline{10}}-\ket{\overline{11}}\bra{\overline{11}}$ usw. können wir die Beziehung des Operators in den Bare- und Dressed-Basen untersuchen. So bekommen wir\\
	\begin{align}
		\mbox{\normalsize$
			\begin{bmatrix}
				\hat{\sigma}_2^z+\hat{\sigma}_1^z\tilde{\hat{\sigma}}_2^z\\
				\hat{\sigma}_{2}^{x}+\hat{\sigma}_{1}^{z}\tilde{\hat{\sigma}}_2^x\\
				\hat{\sigma}_2^z-\hat{\sigma}_1^z\tilde{\hat{\sigma}}_2^z\\
				\hat{\sigma}_{2}^{x}-\hat{\sigma}_{1}^{z}\tilde{\hat{\sigma}}_2^x\\
			\end{bmatrix}=\begin{bmatrix}
				\cos(\theta_{0-}) & -\sin(\theta_{0-}) & 0 &0\\
				\sin(\theta_{0-}) & \cos(\theta_{0-}) & 0 & 0\\
				0 &0 & \cos(\theta_{0+}) & -\sin(\theta_{0+})\\
				0 & 0 & \sin(\theta_{0+}) & \cos(\theta_{0+})
			\end{bmatrix} \begin{bmatrix}
				\overline{\hat{\sigma}}_{2}^{z}+\overline{\hat{\sigma}}_{1}^{z}\overline{\hat{\sigma}}_{2}^{z}\\
				\overline{\hat{\sigma}}_{2}^{x}+\overline{\hat{\sigma}}_{1}^{z}\overline{\hat{\sigma}}_{2}^{x}\\
				\overline{\hat{\sigma}}_{2}^{z}-\overline{\hat{\sigma}}_{1}^{z}\overline{\hat{\sigma}}_{2}^{z}\\
				\overline{\hat{\sigma}}_{2}^{x}-\overline{\hat{\sigma}}_{1}^{z}\overline{\hat{\sigma}}_{2}^{x}\\
			\end{bmatrix}.
			$}
	\end{align}\\
	Dadurch kann man den Hamiltonoperator (\ref{equ:h2}) in der Dressed-Basis, also Dressed-FQ2-Energiebasis, darstellen:\\
	\begin{align}
		\begin{split}
			\overline{\hat{H}}^{(2)}=&-\frac{\omega_{n-}}{2}\left[\cos(\theta_{n-}-\theta_{0-})\left(\overline{\hat{\sigma}}_{2}^{z}+\overline{\hat{\sigma}}_{1}^{z}\overline{\hat{\sigma}}_{2}^{z}\right)+ \sin(\theta_{n-}-\theta_{0-}) \left(\overline{\hat{\sigma}}_{2}^{x}+\overline{\hat{\sigma}}_{1}^{z}\overline{\hat{\sigma}}_{2}^{x}\right)\right] \\
			&-\frac{\omega_{n+}}{2} \left[\cos(\theta_{n+}-\theta_{0+}) \left(\overline{\hat{\sigma}}_{2}^{z}-\overline{\hat{\sigma}}_{1}^{z}\overline{\hat{\sigma}}_{2}^{z}\right)+\sin(\theta_{n+}-\theta_{0+}) \left(\overline{\hat{\sigma}}_{2}^{x}-\overline{\hat{\sigma}}_{1}^{z}\overline{\hat{\sigma}}_{2}^{x}\right)\right].
			\label{equ:h2d}
		\end{split}
	\end{align}\\
	Mit Hilfe von\\
	\begin{align}
		\omega_{n\pm}\cos(\theta_{n\pm}-\theta_{0\pm})&\rightarrow\left(\hat{\delta}_{\text{eff2,n}}/2 \pm J_{zz}\right)\left(\delta_{\text{eff,2}}/2 \pm J_{zz}\right)+J_{zx}^2\\
		\omega_{n\pm}\sin(\theta_{n\pm}-\theta_{0\pm})&\rightarrow J_{zx}~\chi \left(\hat{a}^{\dag}\hat{a}+\frac{1}{2}\right)
	\end{align}\\
	kann man $\delta_{\text{eff2,n}}$ im Hamiltonoperator (Gl.\ref{equ:h2d}) zum $\hat{\delta}_{\text{eff2,n}}$ transformieren.
	Die entsprechende Matrixform in der Dressed-Basis $\{\ket{\overline{00}},\ket{\overline{01}}, \ket{\overline{10}}, \ket{\overline{11}}\}$ ist\\
	\begin{align*}
		\mbox{\small$\overline{\hat{H}}^{(2)}=\begin{bmatrix}
				-\frac{\omega_{n-}}{2}\cos(\theta_{n-}-\theta_{0-}) & -\frac{\omega_{n-}}{2}\sin(\theta_{n-}-\theta_{0-}) & 0 & 0\\
				-\frac{\omega_{n-}}{2}\sin(\theta_{n-}-\theta_{0-}) & \frac{\omega_{n-}}{2}\cos(\theta_{n-}-\theta_{0-}) & 0 & 0\\
				0 & 0 & -\frac{\omega_{n+}}{2}\cos(\theta_{n+}-\theta_{0+})  & -\frac{\omega_{n+}}{2}\sin(\theta_{n+}-\theta_{0+})\\
				0 & 0 & -\frac{\omega_{n+}}{2}\sin(\theta_{n+}-\theta_{0-}) & \frac{\omega_{n+}}{2}\cos(\theta_{n+}-\theta_{0+})\\
			\end{bmatrix}$}.
	\end{align*}\\
	Die Blockdiagonalmatrix $\overline{\hat{H}}^{(2)}$ lässt sich durch $\overline{\hat{H}}^{(2)}_{l}\oplus \overline{\hat{H}}^{(2)}_{r}$ untersuchen.
	Die Bedingung für die Gültigkeit der RWA ist\\
	\begin{align}
		\left\Vert\frac{\sin(\theta_{n\pm}-\theta_{0\pm})}{\cos(\theta_{n\pm}-\theta_{0\pm})}\right\Vert=\left\Vert\frac{J_{zx}~\chi \left(\hat{a}^{\dag}\hat{a}+\frac{1}{2}\right) }{\left(\hat{\delta}_{\text{eff2,n}}/2 \pm J_{zz}\right)\left(\delta_{\text{eff,2}}/2 \pm J_{zz}\right)+J_{zx}^2}\right\Vert \ll 1.
		\label{bed:mo2d}
	\end{align}\\
	Im Vergleich dazu erhalten wir von dem Hamiltonoperator in der Bare-FQ2-Basis (Gl. \ref{equ:h2}) sowie der entsprechenden Matrixform (Gl. \ref{equ:Hohnedeltaqfpmf}) die Bedingung als\\
	\begin{align}
		\left\Vert\frac{\sin(\theta_{n\pm})}{\cos(\theta_{n\pm})}\right\Vert = \left\Vert\frac {J_{zx}}{\left(\hat{\delta}_{\text{eff2,n}}/2 \pm J_{zz}\right)}\right\Vert \ll 1.
		\label{bed:mo2b}
	\end{align}\\
	Wenn die Beziehung\\
	\begin{align}
		\left\Vert\tan(\theta_{n\pm}-\theta_{0\pm})\right\Vert =\left\Vert\tan(\theta_{n\pm})\right\Vert,
	\end{align}\\
	entspricht\\
	\begin{align}
		\left\Vert \chi \left(\hat{a}^{\dag}\hat{a}+\frac{1}{2}\right) \right\Vert=\sqrt{\left(\delta_{\text{eff,2}}/2 \pm J_{zz}\right)^2+J_{zx}^2},
		\label{bed:fq2o_fq2}
	\end{align}\\
	erfüllt ist, macht es keinen Unterschied.\\
	
	\begin{figure}[ht]
		\centering
		\includegraphics[width=0.7\textwidth]{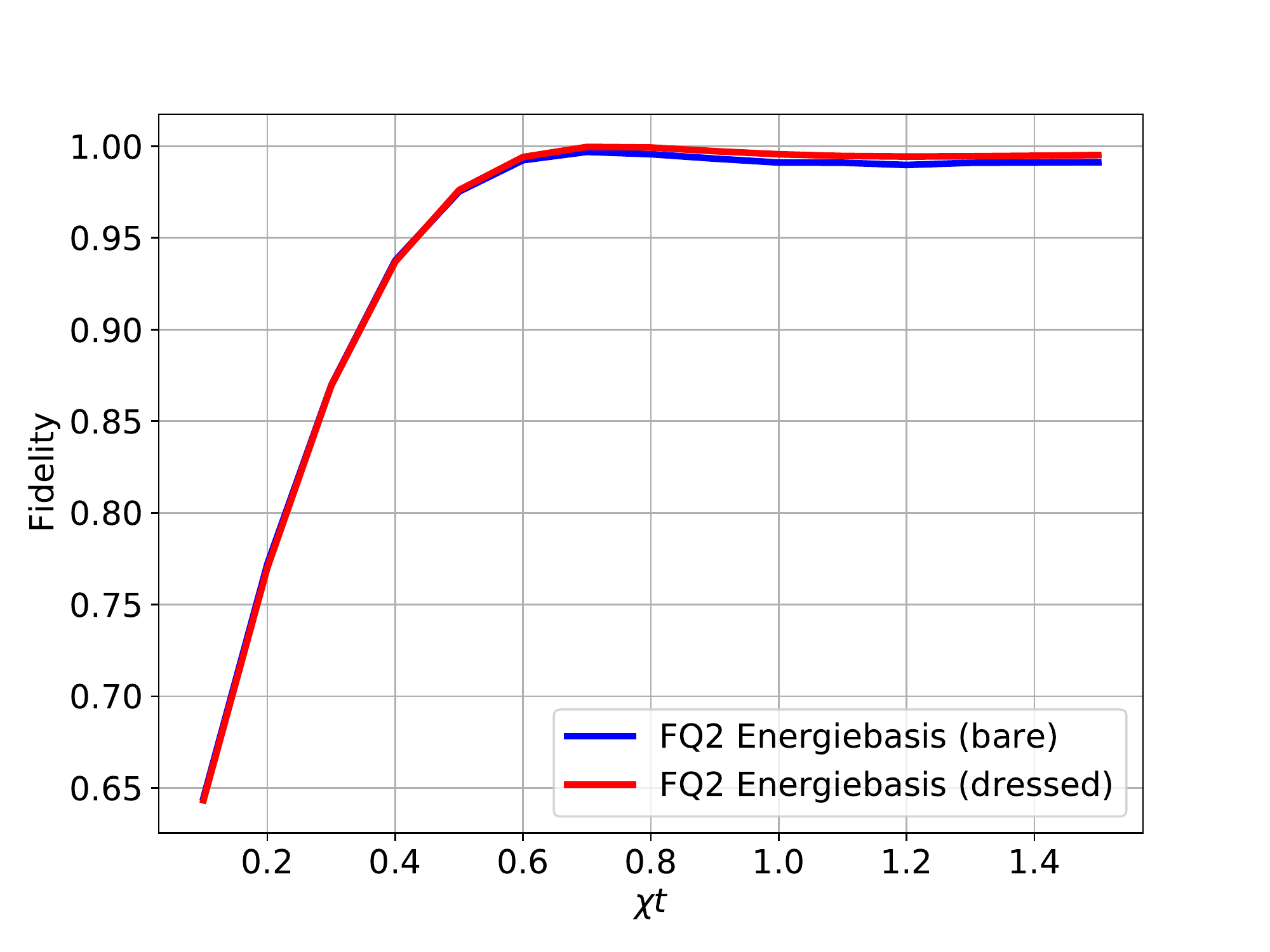}
		\caption{Fidelity in Abhängigkeit von $\chi t$ in der Bare- (blau) und Dressed- (rot) FQ2-Energiebasis mit $N=27$, $\eta=1.25$, $J=0.05(\omega_{2}-\omega_{1})$, $\Delta_{2}/\epsilon_2 = 1$, $\delta/g=8$ und $\alpha=1$. Die Fidelity wird mit steigendem $\chi t$ höher.}
		\label{fig:2qo_chit_q2}
	\end{figure}
	
\begin{figure}[ht]
		\centering
		\includegraphics[width=0.7\textwidth]{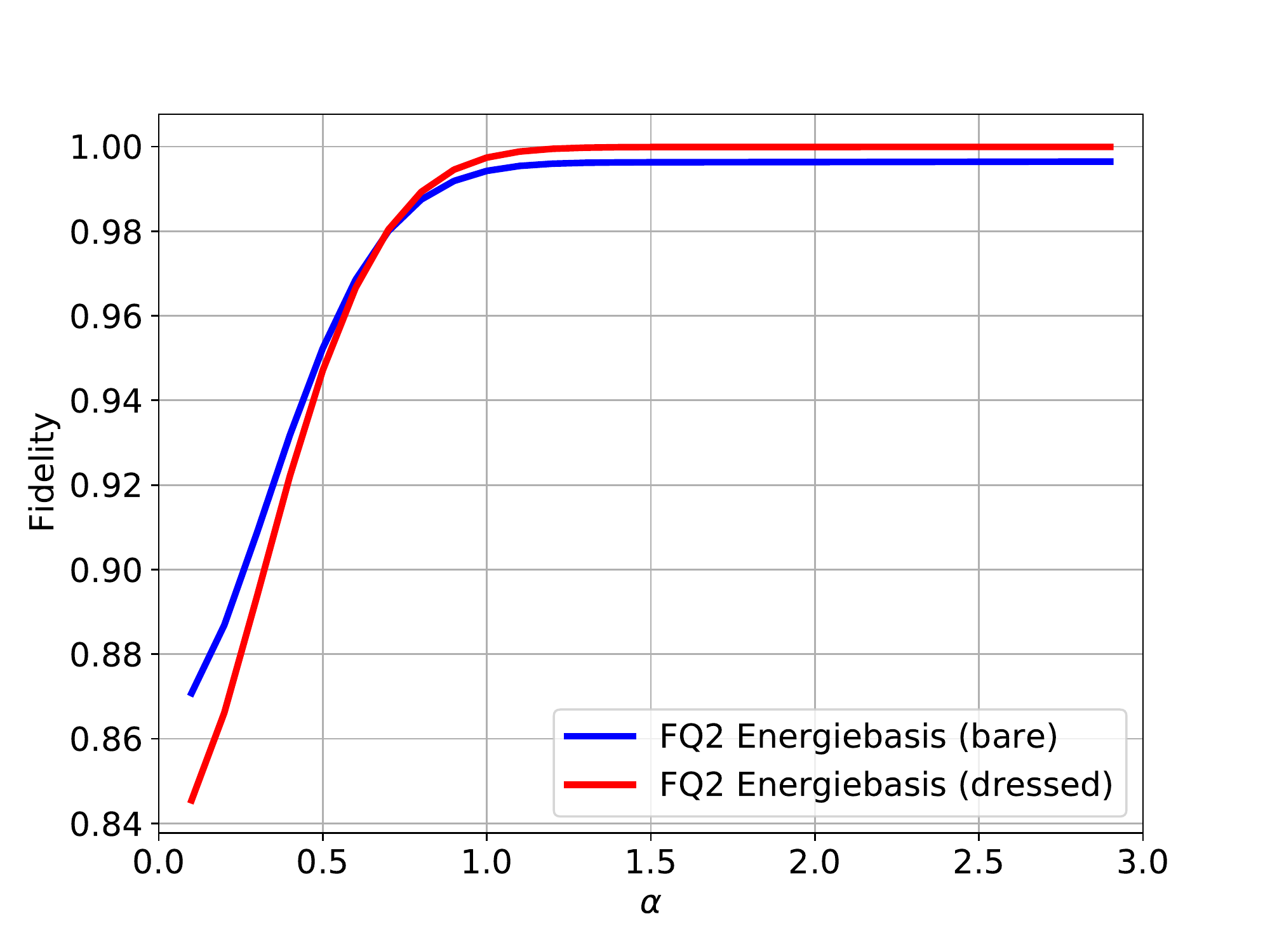}
		\caption{Fidelity in Abhängigkeit von $\alpha$ in der Bare- (blau) und Dressed- (rot) FQ2-Energiebasis mit $N=27$, $\eta=1.25$, $J=0.05(\omega_{2}-\omega_{1})$, $\Delta_{2}/\epsilon_2 = 1$, $\delta/g=8$ und $t_m=\pi/2\chi$. Die Fidelity wird mit steigendem $\alpha$ höher und erreicht einen Sättigungswert.}
		\label{fig:2qo_alpha_q2}
	\end{figure}

\begin{figure}[ht]
		\centering
		\includegraphics[width=0.7\textwidth]{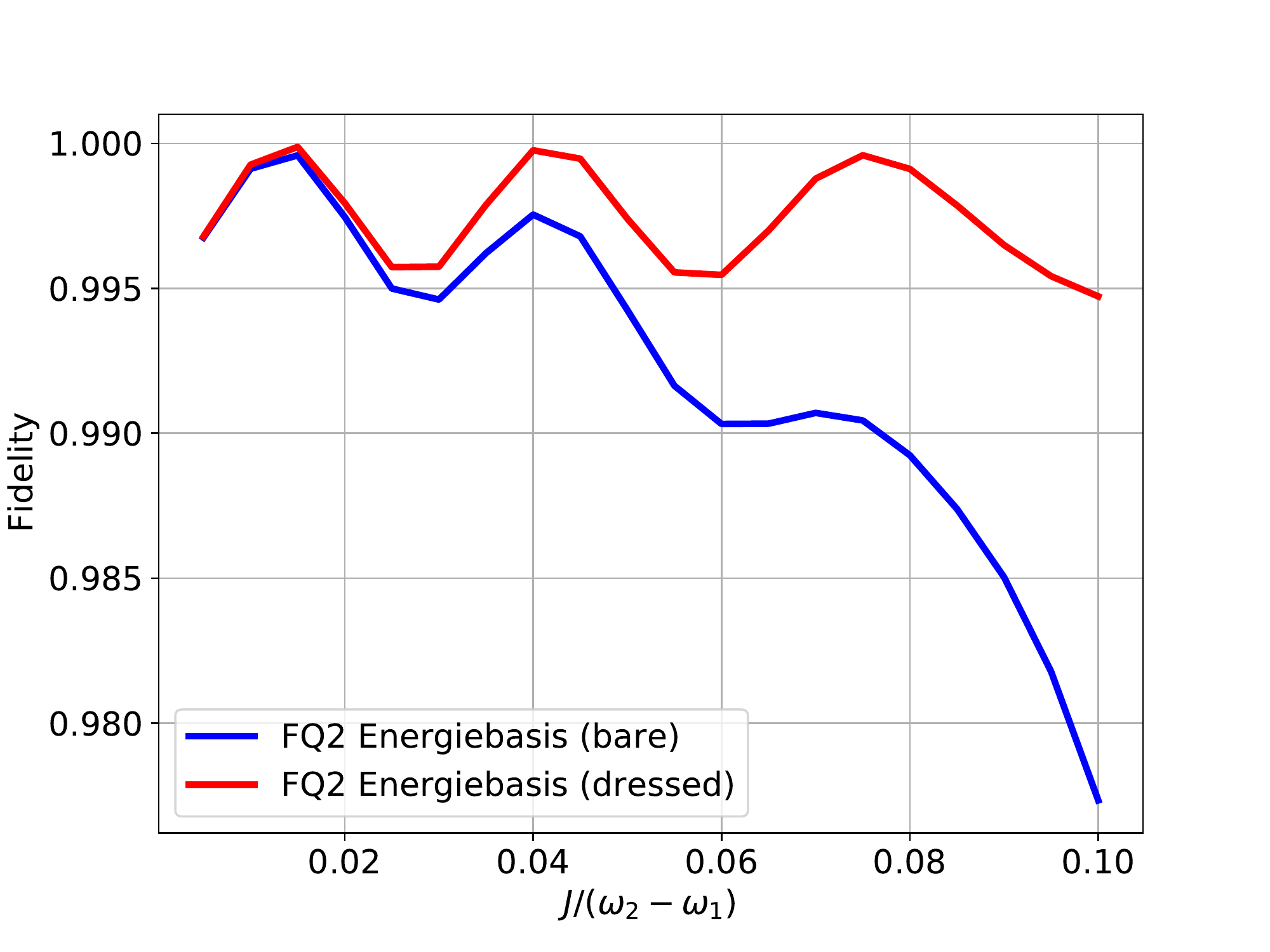}
		\caption{Fidelity in Abhängigkeit von $J/(\omega_{2}-\omega_{1})$ in der Bare- (blau) und Dressed- (rot) FQ2-Energiebasis mit $N=27$, $\eta=1.25$, $t_m=\pi/2\chi$, $\Delta_{2}/\epsilon_2 = 1$, $\delta/g=8$ und $\alpha=1$. Die Fidelity wird in der Bare-Basis mit steigendem $J$ niedriger.}
		\label{fig:2qo_J_q2}
	\end{figure}

	Die numerische Untersuchung der Fidelity in der Bare- und Dressed-Basis wird in Abhängigkeit von $\chi t$, $\alpha$ und $J/(\omega_{2}-\omega_{1})$ in Abb. \ref{fig:2qo_chit_q2}, Abb. \ref{fig:2qo_alpha_q2} und Abb. \ref{fig:2qo_J_q2} respektive dargestellt.\\
	
	Wenn die Messzeit zu kurz ist, wird die Fidelity in beiden Basen niedrig sein (siehe Abb. \ref{fig:2qo_chit_q2}). Zur Messzeit $t_m \approx \pi/4\chi $ erreicht die Fidelity in beiden Basen das Maximum. Danach wird sie in der Dressed-Basis höher als in der Bare-Basis. Aber in beiden Basen wird die Fidelity leicht gesenkt. Dies bedeutet, mit steigender Messzeit misst man in der FQ1-FQ2-Energiebasis effektiver (siehe Abb. \ref{fig:2qo_chit_fq2q1q2}). \\
	
	Bei kleinem $\alpha$ ist die Bare-Basis besser. Ein Basis-Crossover findet bei $\alpha\approx0.7$ statt. Sobald $\alpha>0.7$ wird die Fidelity in der Dressed-Basis höher. Das heißt, mit steigendem $\alpha$ und fester Messzeit (hier $\pi/2\chi$) misst man langsam in einer Dressed-Basis.\\
	
	
	\clearpage
	
	Wie die Kopplungsstärke zwischen beiden Flussqubits auf die Fidelity beeinflusst, wird in Abb. \ref{fig:2qo_J_q2} gezeigt. Für gewählte Parameter, $t_m=\pi/2\chi$ und $\alpha=1$, wird die Dressed-Basis gültiger als die Bare-Basis. Die Kopplungsstärke wirkt mehr auf die Bare-Basis ein.\\
	
	Aus der numerischen Simulation kann man schließen, dass die Bare-Basis für kleines $\alpha$ besser als die Dressed-Basis ist und die Messqualität in der Bare-Basis empfindlicher auf die Kopplung zwischen beiden Flussqubits reagiert. Im Vergleich dazu wird die Fidelity in der Dressed-Basis für großes $\alpha$ höher und oszilliert zwar in Abhängigkeit von $J$, aber der Mittelwert der Oszillation sinkt nur leicht. Die Oszillationen sind möglicherweise transiente Oszillationen. Der Grund dafür könnte sein, dass wir nichtadiabatisch den Arbeitspunkt ändern und genügend Energieniveaus haben. \\
	
	\subsection{FQ1-FQ2-Energiebasis}
	\label{sec:hinq1q2}
	
	Der Hamiltonoperator in der FQ1-FQ2-Energiebasis (Gl. \ref{equ:ht3_q1q2}) wird im Folgenden untersucht. Aufgrund der Komplexität des Hamiltonoperators (\ref{equ:ht3_q1q2}) betrachten wir analytisch zwei spezielle Fälle für den Wechselwirkungsterm von FQ1 und FQ2.\\
	\subsection*{1. Fluss-dominiertes Regime}
		\begin{align}
			\{\abs{\cos(\theta_1)}, \abs{\cos(\theta_{\text{eff,2}})}\} \gg  \{\abs{\sin(\theta_1)}, \abs{\sin(\theta_{\text{eff,2}})}\}
			\label{bed:c12gs}
		\end{align}\\
		Unter dieser Bedingung können wir den Wechselwirkungsterm von FQ1 und FQ2 näherungsweise durch 
		$J_{zz}\tilde{\hat{\sigma}}_1^z\tilde{\hat{\sigma}}_2^z$ mit $J_{zz}=J\cos(\theta_1)\cos(\theta_{\text{eff,2}})$ ersetzen. So ergibt sich\\
		\begin{align}
			\begin{split}
				\tilde{\hat{H}}^{(3)}_{zz}\approx-\frac{\hat{\delta}_{\text{eff2,n}}}{2}\tilde{\hat{\sigma}}_2^z+J_{zz} \tilde{\hat{\sigma}}_1^z\tilde{\hat{\sigma}}_2^z.
			\end{split}
			\label{equ:Ht2_zz_q1q2}
		\end{align}\\
	Dies nennen wir eine zz-Wechselwirkungs-Näherung (zz-WW-Näherung). Unter der Näherung ist $\tilde{\hat{H}}^{(3)}_{zz}$ eine Diagonalmatrix. So ist die FQ1-FQ2-Energiebasis schon ihre Dressed-Basis und die Bedingung für eine gute Drehwellennäherung in \cite{Pommerening2020} wird immer erfüllt. Daher wird die Fidelity unter der Bedingung Gl. \ref{bed:c12gs} dem Hamiltonoperator (Gl. \ref{equ:Ht2_zz_q1q2}) gemäß höher.\\
	
\begin{figure}[ht]
		\centering
		\includegraphics[width=0.7\textwidth]{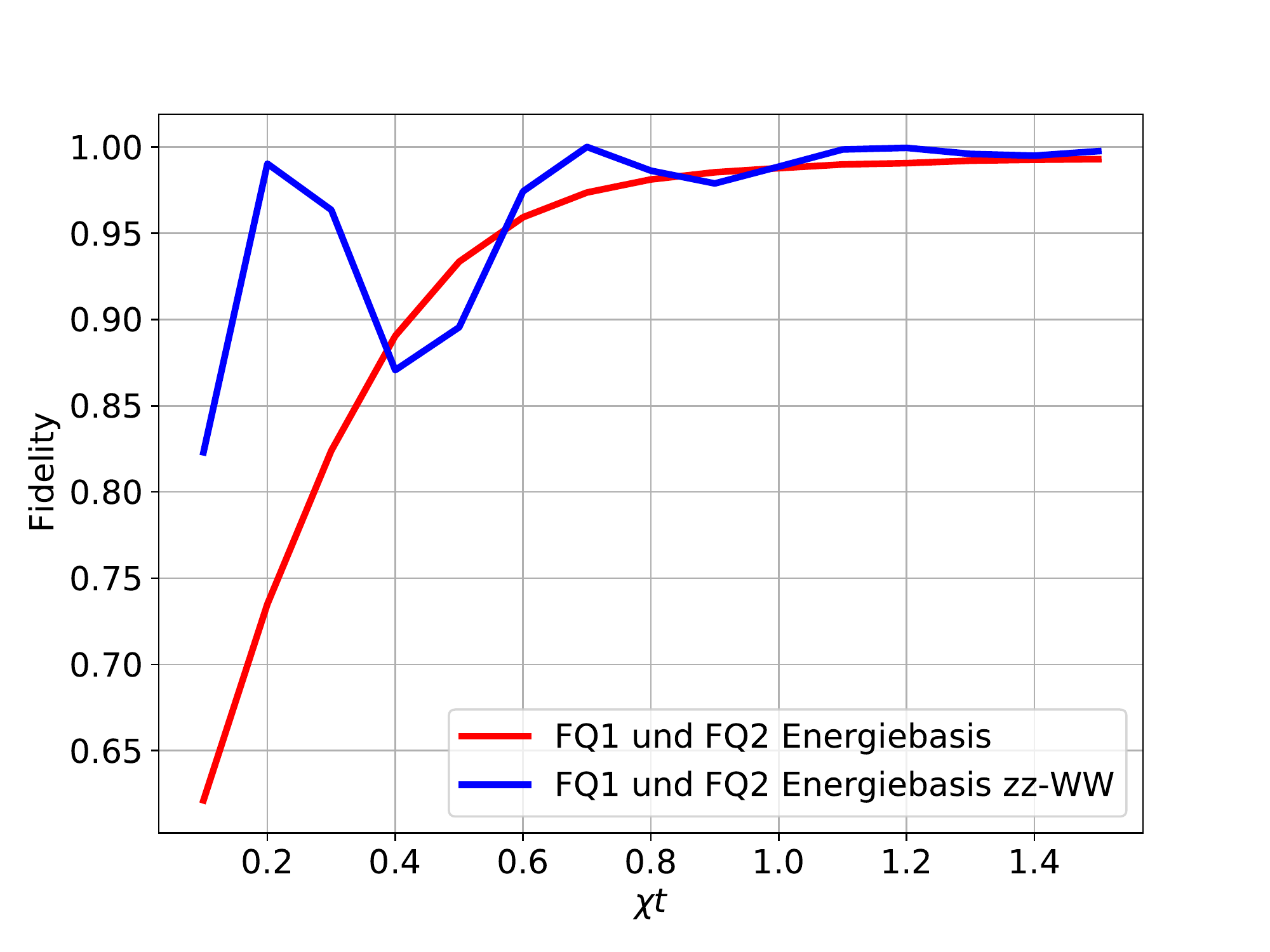}
		\caption{Fidelity in Abhängigkeit von $\chi t$ in der FQ1-FQ2-Energiebasis (rot) sowie unter der zz-WW-Näherung (blau) mit $N=27$, $\eta=1.25$, $J=0.05(\omega_{2}-\omega_{1})$, $\Delta_{2}/\epsilon_2 = 1$, $\delta/g=8$ und $\alpha=1$. Die Fidelity unter der zz-WW-Näherung wird für kurze Messdauer wesentlich höher als in der FQ1-FQ2-Energiebasis, während sich die Werte für längere Messzeit angleichen.}
		\label{fig:2qo_chit_fq1q2zz}
	\end{figure}

\begin{figure}[ht]
		\centering
		\includegraphics[width=0.7\textwidth]{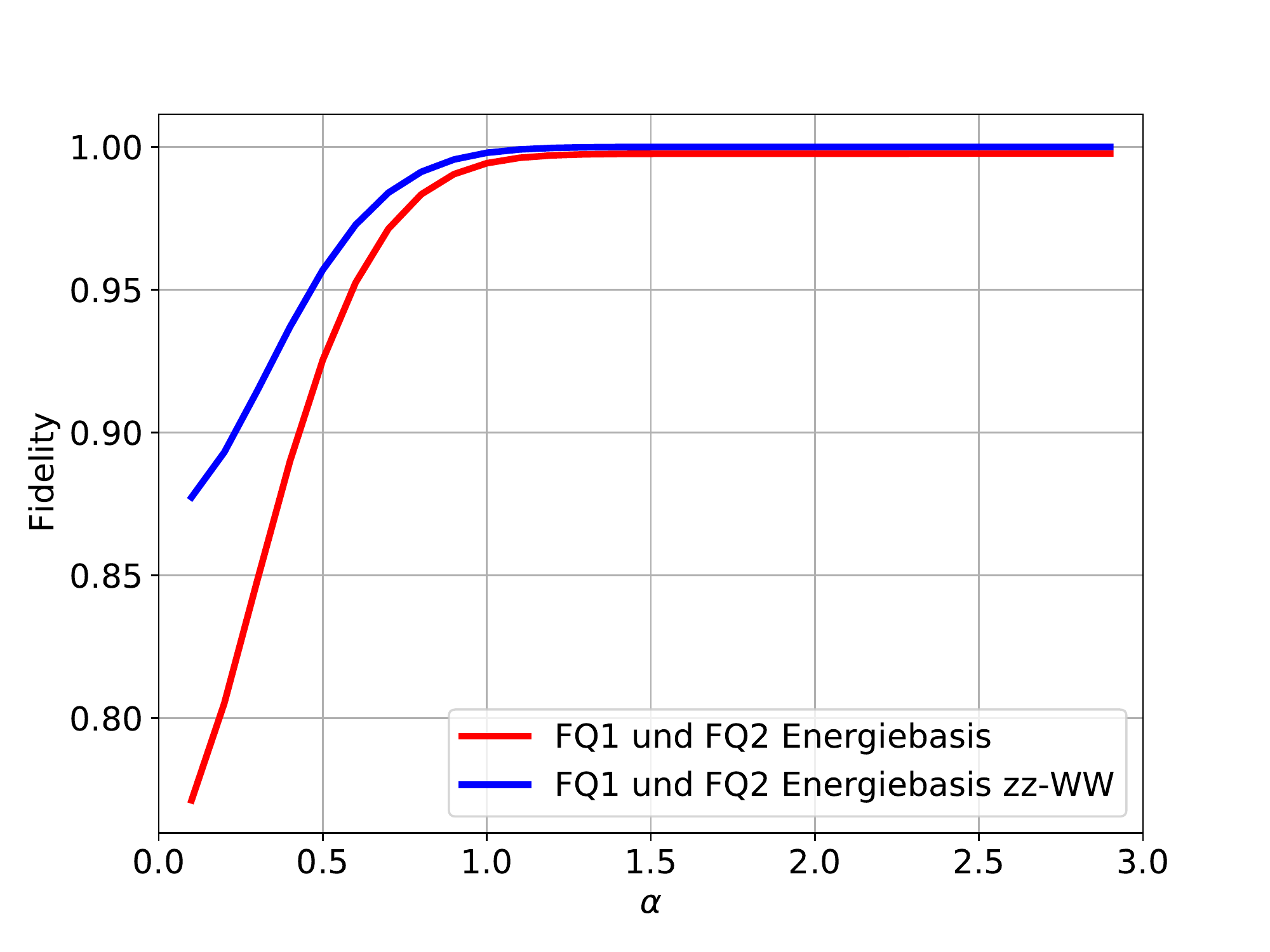}
		\caption{Fidelity in Abhängigkeit von $\alpha$ in der FQ1-FQ2-Energiebasis (rot) sowie unter der zz-WW-Näherung (blau) mit $N=27$, $\eta=1.25$, $J=0.05(\omega_{2}-\omega_{1})$, $\Delta_{2}/\epsilon_2 = 1$, $\delta/g=8$ und $t_m=\pi/2\chi$. Die Fidelity wird mit steigendem $\alpha$ höher und erreicht einen Sättigungswert.}
		\label{fig:2qo_alpha_fq1q2zz}
	\end{figure}

\begin{figure}[ht]
		\centering
		\includegraphics[width=0.7\textwidth]{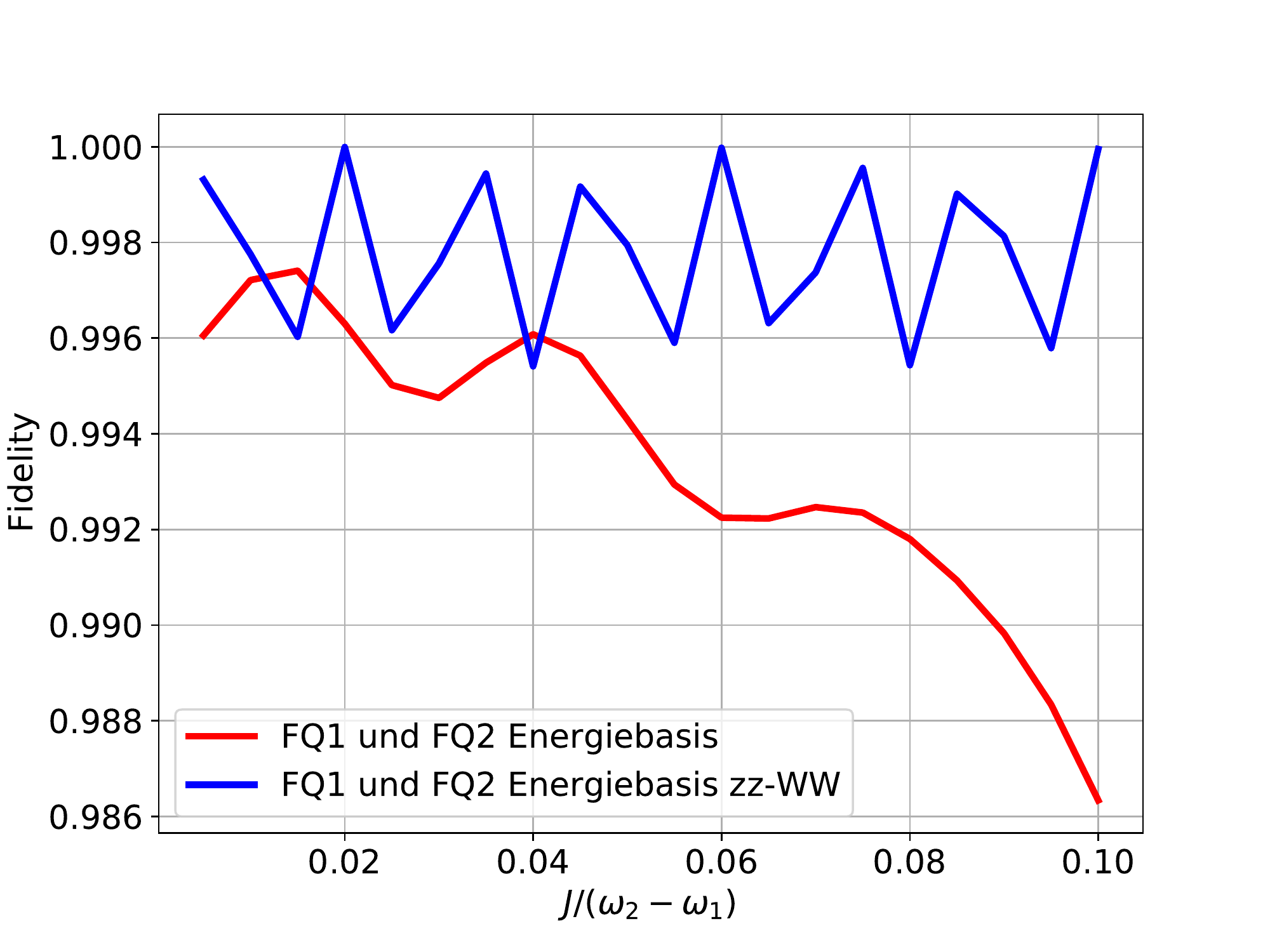}
		\caption{Fidelity in Abhängigkeit von $J/(\omega_{2}-\omega_{1})$ in der FQ1-FQ2-Energiebasis (rot) sowie unter der zz-WW-Näherung (blau) mit $N=27$, $\eta=1.25$, $t_m=\pi/2\chi$, $\Delta_{2}/\epsilon_2 = 1$, $\delta/g=8$ und $\alpha=1$. Die Fidelity in der FQ1-FQ2-Energiebasis wird mit steigendem $J/(\omega_{2}-\omega_{1})$ niedriger. Unter der zz-WW-Näherung oszilliert die Fidelity mit ungefähr konstantem Mittelwert.}
		\label{fig:2qo_J_fq1q2zz}
	\end{figure}

	Die numerische Untersuchung der Fidelity wird in Abhängigkeit von $\chi t$, $\alpha$ und $J/(\omega_{2}-\omega_{1})$ in Abb. \ref{fig:2qo_chit_fq1q2zz}, Abb. \ref{fig:2qo_alpha_fq1q2zz} und Abb. \ref{fig:2qo_J_fq1q2zz} respektive dargestellt.\\
	
	In der FQ1-FQ2-Energiebasis steigt die Fidelity in Abhängigkeit von $\chi t$ monoton und erreicht einen Sättigungswert. Unter der zz-WW-Näherung oszilliert die Fidelity gedämpft um die in der FQ1-FQ2-Energiebasis (siehe Abb. \ref{fig:2qo_chit_fq1q2zz}), während die in der FQ1-FQ2-Energiebasis monoton steigt und eine Sättigung erreicht. Zur Messzeit $t_m \approx\pi/2\chi$ wird die Fidelity in beiden Basen am höchsten. Die Oszillation führt dazu, dass für kurze Messdauer eine hohe Fidelity erreicht werden kann.\\
	
	Die Fidelity unter der zz-Näherung ist für jedes $\alpha$ höher. Für kleines $\alpha$ ist der
	
	\clearpage
	
	Unterschied größer. Bei $\alpha>1$ ist die Fidelity in beiden Basen kaum verschieden (siehe Abb. \ref{fig:2qo_alpha_fq1q2zz}).\\
	
	Unter der zz-WW-Näherung oszilliert die Fidelity in Abhängigkeit von $J/(\omega_{2}-\omega_{1})$ um den Wert $0.998$ (siehe Abb. \ref{fig:2qo_J_fq1q2zz}). Im Vergleich dazu sinkt die Fidelity in der FQ1-FQ2-Energiebasis mit steigender Kopplungsstärke $J$, wobei in der FQ1-FQ2-Energiebasis die Fidelity auch für niedrige $J$ geringer ist. Hier gibt es wieder transiente Oszillationen (vgl. Abb. \ref{fig:2qo_J_q2}). Es liegt daran, dass wir nichtadiabatisch den Arbeitspunkt ändern und genügend Energieniveaus haben.\\

	Insgesamt kann man sagen, dass die Fidelity unter der zz-Näherung höher ist. So könnte man, die zz-Näherung bei Design eines Flussqubits berücksichtigen, um eine hohe Fidelity zu erreichen.\\

\begin{figure}[ht]
			\centering
			\includegraphics[width=0.7\textwidth]{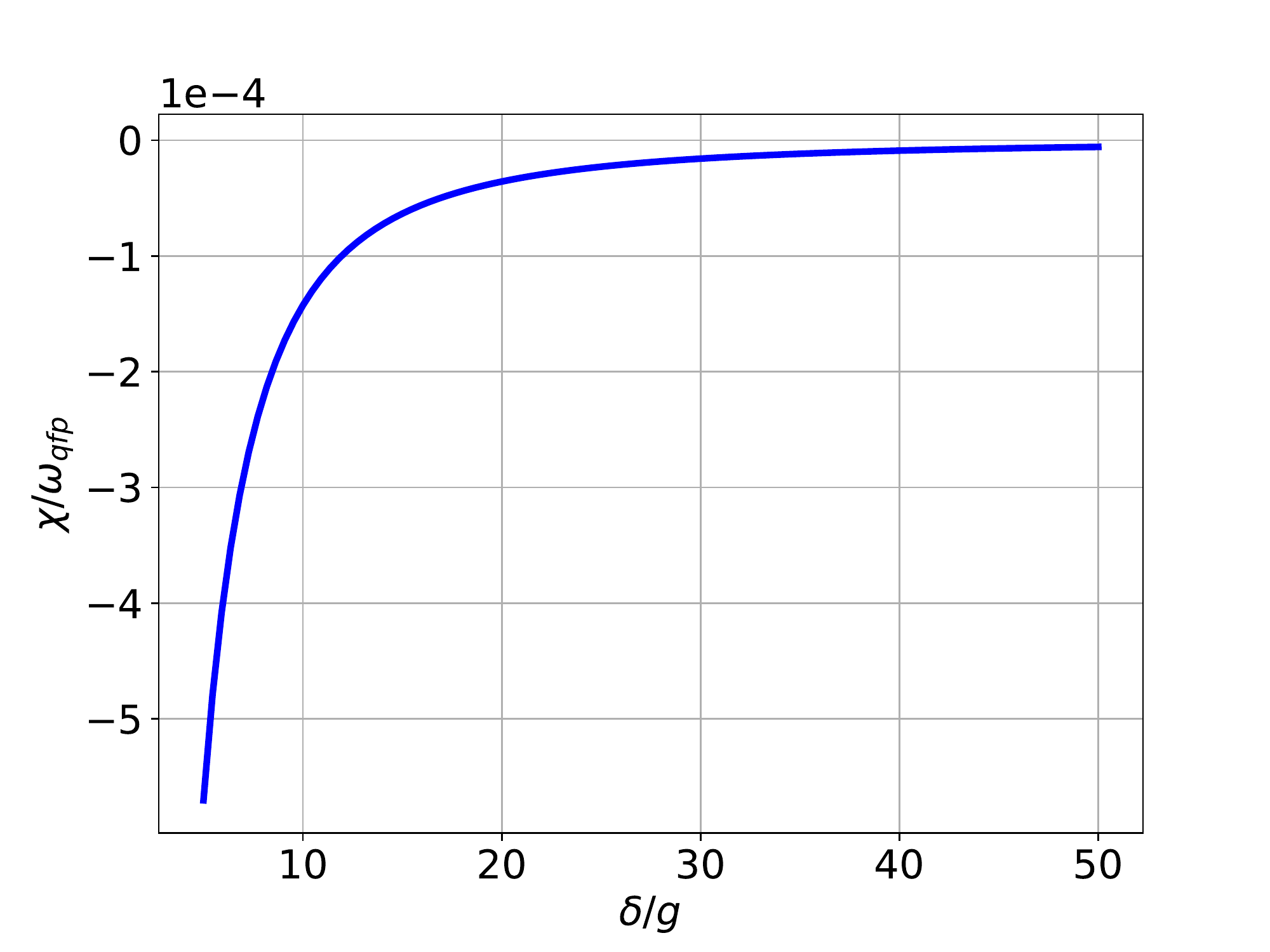}
			\caption{$\chi/\omega_{\text{qfp}}$ in Abhängigkeit von $\delta/g$ mit $\Delta_2/\epsilon_{2}=10$ und $\eta=1.25$.}
			\label{fig:2qo_chideltaxx}
		\end{figure}

	\subsection*{2. Tunnel-dominiertes Regime}	

		\begin{align}
			\{\abs{\cos(\theta_1)}, \abs{\cos(\theta_{\text{eff,2}})}\} \ll \{\abs{\sin(\theta_1)}, \abs{\sin(\theta_{\text{eff,2}})}\}
			\label{bed:c12ks}
		\end{align}\\
		In dieser Situation erhalten wir den Wechselwirkungsterm als
		$J_{xx}\tilde{\hat{\sigma}}_1^x\tilde{\hat{\sigma}}_2^x$ (xx-WW-Näherung) mit $J_{xx}=J\sin(\theta_1)\sin(\theta_{\text{eff,2}})$. Der Hamiltonoperator lässt sich in der FQ1-FQ2-Energiebasis darstellen:\\
		\begin{align}
			\begin{split}
				\tilde{\hat{H}}^{(3)}_{xx}\approx-\frac{\hat{\delta}_{\text{eff2,n}}}{2}\tilde{\hat{\sigma}}_2^z+J_{xx} \tilde{\hat{\sigma}}_1^x\tilde{\hat{\sigma}}_2^x.
			\end{split}
			\label{equ:Hq4_xx}
		\end{align}\\
		Um $\tilde{\hat{H}}^{(3)}_{xx}$ zu diagonalisieren, wenden wir den Transformationsoperator\\
		\begin{align*}
			U_r =\begin{pmatrix}
				\cos(\frac{\theta_{n-}}{2})& 0 & 0 & -\sin(\frac{\theta_{n-}}{2})\\
				0& \cos(\frac{\theta_{ n+}}{2}) & -\sin(\frac{\theta_{ n+}}{2})&0\\	
				0 & \sin(\frac{\theta_{ n+}}{2})& \cos(\frac{\theta_{ n+}}{2})&0\\
				\sin(\frac{\theta_{n-}}{2})&0&0&\cos(\frac{\theta_{n-}}{2})
			\end{pmatrix},
		\end{align*}\\
		auf die Gl. \ref{equ:Hq4_xx} an.
		Die Mischungswinkel $\theta_{n\pm}$ sind durch $\tan(\theta_{n\pm })=\pm2J_{xx}/\delta_{\text{eff2,n}} $ definiert.
		So bekommen wir die Eigenwerte\\
		\begin{align}
			E^{(3)}_{xx}= U_r \tilde{\hat{H}}^{(4)}_{xx} U_r^\dagger = \text{diag}(-\omega_n,~ -\omega_n,~ \omega_n,~ \omega_n),
		\end{align}\\
		mit $\omega_n = \sqrt{J_{xx}^2+\left(\delta_{\text{eff2,n}}/2\right)^2}$. $\chi/\omega_{\text{qfp}}$ wird unter der Bedingung $\ref{bed:c12ks}$ in Abhängigkeit von $\delta/g$ in Abb. \ref{fig:2qo_chideltaxx} geplottet.\\
		
		Die Beziehung zwischen Bare-und Dressed-Basis lässt sich darstellen:\\
		\begin{align*}
			\begin{split}
				\ket{\overline{00}}&=\cos(\frac{\theta_{0+}}{2})\ket{\tilde0\tilde 0}+\sin(\frac{\theta_{0+}}{2})\ket{\tilde1\tilde 1}, \ket{\overline{01}}=\cos(\frac{\theta_{0-}}{2})\ket{\tilde0\tilde 1}+\sin(\frac{\theta_{0-}}{2})\ket{\tilde1\tilde 0},\\
				\ket{\overline{10}}&=-\sin(\frac{\theta_{0-}}{2})\ket{\tilde0\tilde 1}+\cos(\frac{\theta_{0-}}{2})\ket{\tilde1\tilde 0}, \ket{\overline{11}}=-\sin(\frac{\theta_{0+}}{2})\ket{\tilde0\tilde 0}+\cos(\frac{\theta_{0+}}{2})\ket{\tilde1\tilde 1}.
			\end{split}
		\end{align*}\\
		Die Bare-Basis bedeutet hier, dass beide Flussqubits in ihrer eigenen Energiebasis sind. Die Dressed-Basis ist dann die Eigenbasis der beiden Flussqubit-Energiebasen (ohne Resonator).\\
		
		Mit $\overline{\hat{\sigma}}_{1}^{z}=\ket{\overline{00}}\bra{\overline{00}}+\ket{\overline{01}}\bra{\overline{01}}-\ket{\overline{10}}\bra{\overline{10}}-\ket{\overline{11}}\bra{\overline{11}}$ usw. können wir dann die Beziehung des Operators in der Bare- und Dressed-Basis untersuchen. So bekommen wir\\
		\begin{align}
			\mbox{\small$
				\begin{bmatrix}
					\overline{\hat{\sigma}}_{1}^{z}+\overline{\hat{\sigma}}_{2}^{z}\\
					\overline{\hat{\sigma}}_{1}^{x}\overline{\hat{\sigma}}_{2}^{x}-\overline{\hat{\sigma}}_{1}^{y}\overline{\hat{\sigma}}_{2}^{y}\\
					\overline{\hat{\sigma}}_{1}^{z}-\overline{\hat{\sigma}}_{2}^{z}\\
					\overline{\hat{\sigma}}_{1}^{x}\overline{\hat{\sigma}}_{2}^{x}+\overline{\hat{\sigma}}_{1}^{y}\overline{\hat{\sigma}}_{2}^{y}\\
				\end{bmatrix}=\begin{bmatrix}
					\phantom{0}\cos(\theta_{0-}) & \sin(\theta_{0-}) & 0 &0\\
					-\sin(\theta_{0-}) & \cos(\theta_{0-}) & 0 & 0\\
					0 &0 & \phantom{0}\cos(\theta_{0+}) & \sin(\theta_{0+})\\
					0 & 0 & -\sin(\theta_{0+}) & \cos(\theta_{0+})
				\end{bmatrix} \begin{bmatrix}
					\tilde{\hat{\sigma}}_{1}^{z}+\tilde{\hat{\sigma}}_{2}^{z}\\
					\tilde{\hat{\sigma}}_{1}^{x}\tilde{\hat{\sigma}}_{2}^{x}-\tilde{\hat{\sigma}}_{1}^{y}\tilde{\hat{\sigma}}_{2}^{y}\\
					\tilde{\hat{\sigma}}_{1}^{z}-\tilde{\hat{\sigma}}_{2}^{z}\\
					\tilde{\hat{\sigma}}_{1}^{x}\tilde{\hat{\sigma}}_{1}^{x}+\tilde{\hat{\sigma}}_{1}^{y}\tilde{\hat{\sigma}}_{2}^{y}\\
				\end{bmatrix},
				$}
			\label{u_tran}
		\end{align}\\
		Dadurch kann man den Hamiltonoperator (\ref{equ:Hq4_xx}) in der Dressed-Basis darstellen:\\
		\begin{align}
			\begin{split}
				\overline{H}_{xx}^{(3)}=\frac{\omega_n}{2} &\left[\cos(\theta_{n-}-\theta_{0-}) \left(\overline{\hat{\sigma}}_{1}^{z}+\overline{\hat{\sigma}}_{2}^{z}\right)+ \sin(\theta_{n-}-\theta_{0-}) \left(\overline{\hat{\sigma}}_{1}^{x}\overline{\hat{\sigma}}_{2}^{x}-\overline{\hat{\sigma}}_{1}^{y}\overline{\hat{\sigma}}_{2}^{y}\right)\right.\\
				&\left.+\cos(\theta_{n+}-\theta_{0+})\left(\overline{\hat{\sigma}}_{1}^{z}-\overline{\hat{\sigma}}_{2}^{z}\right)+ \sin(\theta_{n+}-\theta_{0+}) \left(\overline{\hat{\sigma}}_{1}^{x}\overline{\hat{\sigma}}_{2}^{x}+\overline{\hat{\sigma}}_{1}^{y}\overline{\hat{\sigma}}_{2}^{y}\right)\right].
			\end{split}
			\label{equ:Ht2_xx_q1q2}
		\end{align}\\
		Mit Hilfe von\\
		\begin{align}
			\omega_{n}\cos(\theta_{n\pm}-\theta_{0\pm})&\rightarrow \hat{\delta}_{\text{eff2,n}} \delta_{\text{eff,2}}/4 + J_{xx}^2\\
			\omega_{n}\sin(\theta_{n\pm}-\theta_{0\pm})&\rightarrow J_{xx}~\chi \left(\hat{a}^{\dag}\hat{a}+\frac{1}{2}\right)
		\end{align}\\
		kann man $\delta_{\text{eff2,n}}$ im Hamiltonoperator (Gl.\ref{equ:h2d}) zum $\hat{\delta}_{\text{eff2,n}}$ transformieren.
		Die Bedingung der Gültigkeit der RWA in der Dressed-Basis ist\\
		\begin{align}
			\left\Vert\frac{\sin(\theta_{n\pm}-\theta_{0\pm})}{\cos(\theta_{n\pm}-\theta_{0\pm})}\right\Vert = \left\Vert \frac{J_{xx}~\chi \left(\hat{a}^{\dag}\hat{a}+\frac{1}{2}\right)}{\hat{\delta}_{\text{eff2,n}} \delta_{\text{eff,2}}/4 + J_{xx}^2}\right\Vert  \ll 1.
		\end{align}\\
		Im Vergleich dazu haben wir die Bedingung der Gültigkeit der RWA in der Bare-Basis\\
		\begin{align}
			\left\Vert\frac{\sin(\theta_{n\pm})}{ \cos(\theta_{n\pm})}\right\Vert = \left\Vert \frac{J_{xx}}{\hat{\delta}_{\text{eff2,n}}/2} \right\Vert \ll 1.
		\end{align}\\
		Wenn die Bedingung\\
		\begin{align}
			\left\Vert\tan(\theta_{n\pm}-\theta_{0\pm})\right\Vert =\left\Vert\tan(\theta_{n\pm})\right\Vert,
		\end{align}\\
		entspricht\\
		\begin{align}
			\left\Vert \chi \left(\hat{a}^{\dag}\hat{a}+\frac{1}{2}\right) \right\Vert=\sqrt{\left(\delta_{\text{eff,2}}/2\right)^2+J_{xx}^2},
			\label{bed:q1q2xx}
		\end{align}\\
		erfüllt ist, macht es keinen Unterschied.\\
		
		So haben wir die Bedingung (Gl. \ref{bed:q1q2xx}) zum Basis-Crossover zwischen Bare-und Dressed-Basis in der FQ1-FQ2-Energiebasis für den Fall $J_{xx} \tilde{\hat{\sigma}}_1^x\tilde{\hat{\sigma}}_2^x$ hergeleitet.\\
		
\begin{figure}[ht]
			\centering
			\includegraphics[width=0.7\textwidth]{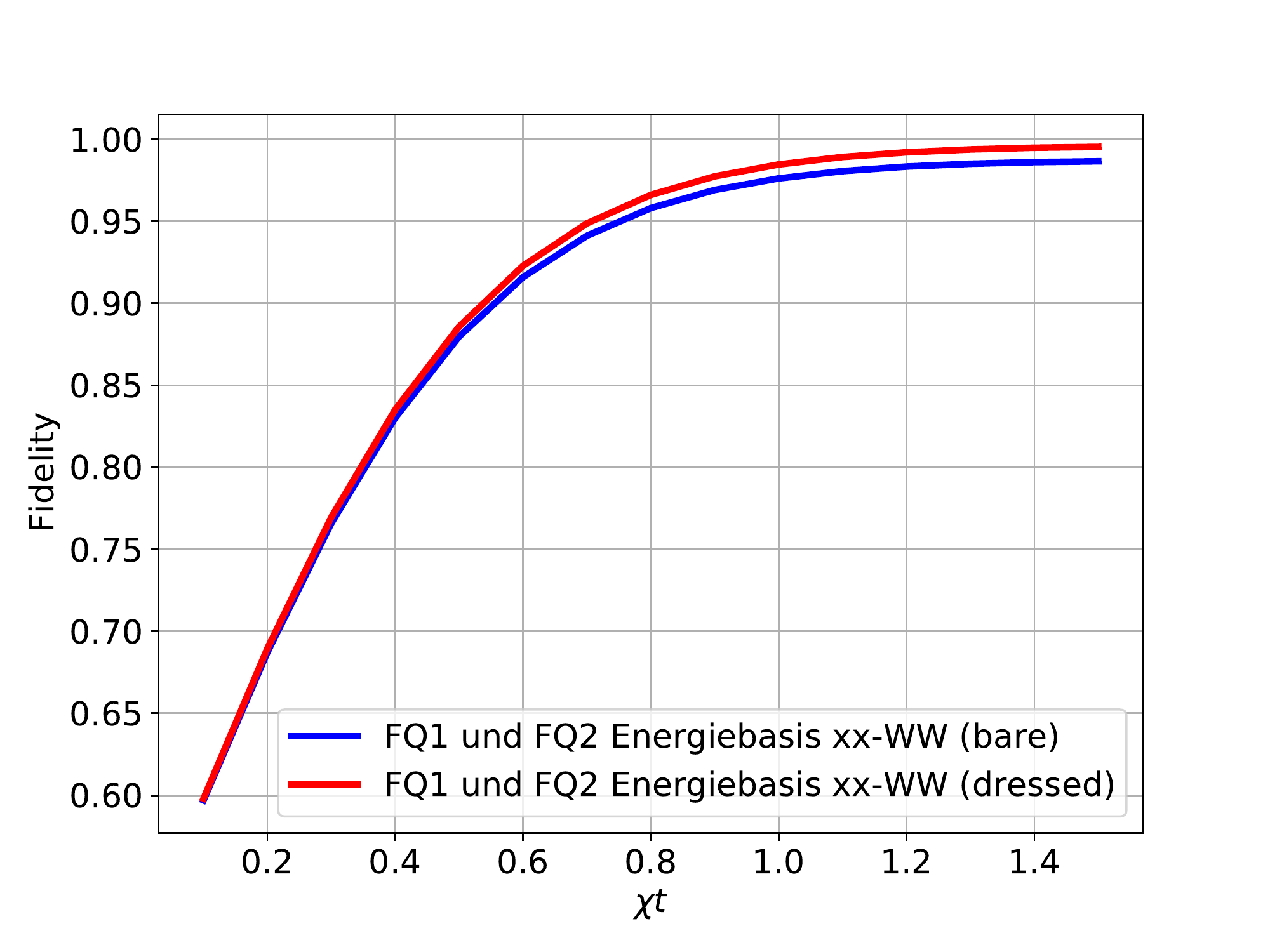}
			\caption{Fidelity in Abhängigkeit von $\chi t$ in der Bare- (blau) und Dressed- (rot) FQ1-FQ2-Energiebasis unter der xx-WW-Näherung mit $N=27$, $\eta=1.25$, $J=0.05(\omega_{2}-\omega_{1})$, $\Delta_{2}/\epsilon_2 = 10$, $\delta/g=8$ und $\alpha=1$. Mit steigendem $\chi t$ wird die Fidelity in beiden Basen höher  und erreicht einen Sättigungswert.}
			\label{fig:2qo_chit_q1q2xx}
		\end{figure}

\begin{figure}[ht]
			\centering
			\includegraphics[width=0.7\textwidth]{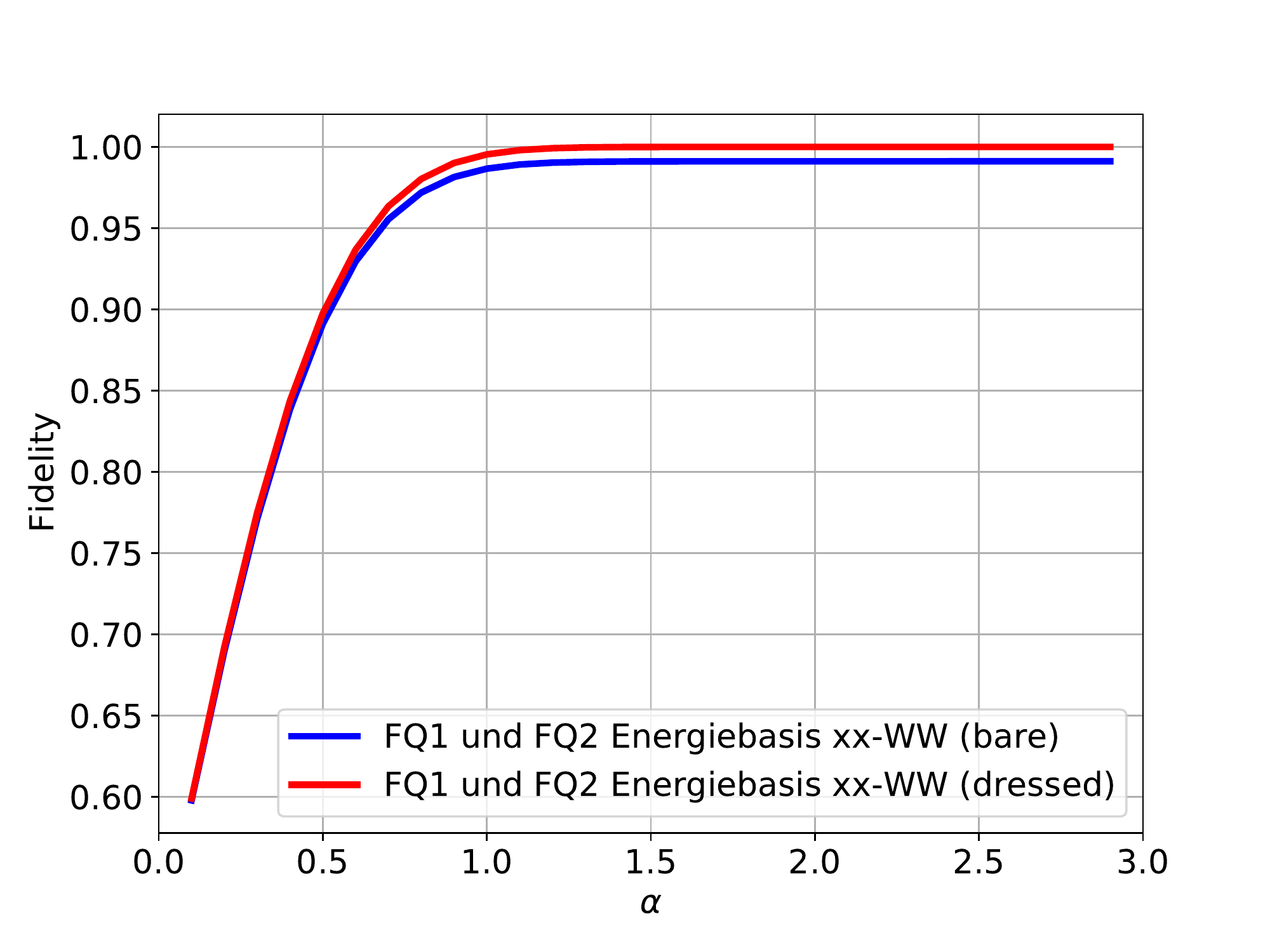}
			\caption{Fidelity in Abhängigkeit von $\alpha$ in der Bare- (blau) und Dressed- (rot) FQ1-FQ2-Energiebasis unter der xx-WW-Näherung mit $N=27$, $\eta=1.25$, $J=0.05(\omega_{2}-\omega_{1})$, $\Delta_{2}/\epsilon_2 = 10$, $\delta/g=8$ und $t_m=\pi/2\chi$. Die Fidelity wird mit steigendem $\alpha$ in beiden Basen höher und erreicht einen Sättigungswert.}
			\label{fig:2qo_alpha_fq1q2xx}
		\end{figure}

\begin{figure}[ht]
			\centering
			\includegraphics[width=0.7\textwidth]{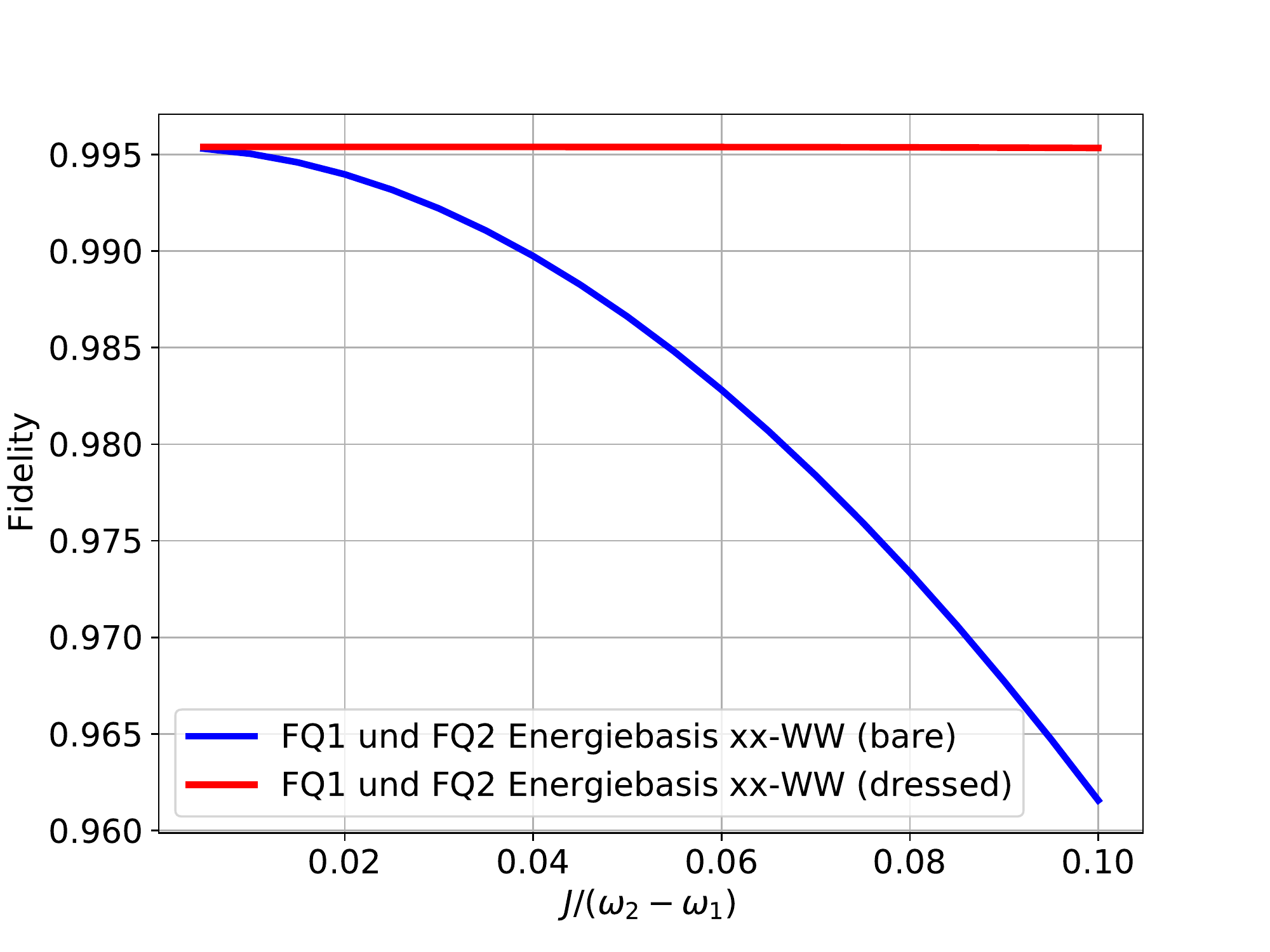}
			\caption{Fidelity in Abhängigkeit von $J/(\omega_{2}-\omega_{1})$ in der Bare- (blau) und Dressed- (rot) FQ1-FQ2-Energiebasis unter der xx-WW-Näherung mit $N=27$, $\eta=1.25$, $t_m=\pi/2\chi$, $\Delta_{2}/\epsilon_2 = 10$, $\delta/g=8$ und $\alpha=1$. Mit steigendem $J/(\omega_{2}-\omega_{1})$ wird die Fidelity in der Bare-Basis niedriger.}
			\label{fig:2qo_J_q1q2xx}
		\end{figure}

		Die numerische Untersuchung der Fidelity unter der xx-WW-Näherung wird in der Bare- und Dressed-Basis in Abhängigkeit von $\chi t$, $\alpha$ und $J/(\omega_{2}-\omega_{1})$ in Abb. \ref{fig:2qo_chit_q1q2xx}, Abb. \ref{fig:2qo_alpha_fq1q2xx} und Abb. \ref{fig:2qo_J_q1q2xx} entsprechend dargestellt.\\
		
		Zur Messzeit $t_m \approx \pi/2\chi$ wird das Maximum der Fidelity in beiden Basen (siehe Abb. \ref{fig:2qo_chit_q1q2xx}) erreicht. Dann bleibt die Fidelity konstant. Bei $\chi t>0.5$ wird die Fidelity in der Dressed-Basis langsam höher als in der bare.
	
		Für $\alpha\lesssim0.5$ ist die Fidelity in beiden Basen ungefähr gleich (siehe Abb. \ref{fig:2qo_alpha_fq1q2xx}). Mit steigendem $\alpha$ wird die Fidelity in der Dressed-Basis langsam höher als die in der bare. Auch hier ist der Sättigungswert für die Dressed-Basis höher als für die Bare-Basis.\\
	 
	 Je größer die Kopplungsstärke ist, desto niedriger wird die Fidelity in der Bare-Basis (siehe Abb. \ref{fig:2qo_J_q1q2xx}). Im Vergleich dazu bleibt die Fidelity in der Dressed-Basis konstant auf hohen Niveau.\\
	 
	 Durch die numerische Simulation lässt sich sagen, dass für gewählte Parameter die Fidelity in der Dressed-Basis immer höher als in der Bare-Basis ist. In der Bare-Basis sinkt die Fidelity, während die Fidelity in der Dressed-Basis robust gegenüber Änderung der Kopplungsstärke.\\

	Wir haben die Bedingung für eine gute Drehwellennäherung in der FQ1-FQ2-Energiebasis als zwei Fälle untersucht. Unter der zz-WW-Näherung ist der Hamiltonoperator in einer diagonalisierten Form. So ist die Bedingung direkt erfüllt. Bei der xx-WW-Näherung kommt die Bedingung für eine gute RWA vor. In beiden Fällen kann man eine hohe  Fidelity durch Optimierung der Parameter von Messzeit, $\alpha$ oder $J$ erreichen.\\
	
	\section{Vergleich und Analyse}
	
	Bei der Modellierung der Messung eines Flussqubits mittels QFP wird in der Energiebasis eine höhere Fidelity befunden, als in der Flussbasis. Der Grund liegt darin, dass die Bedingung für eine gute Drehwellennäherung in der Energiebasis immer erfüllt ist, während sie in der Flussbasis nur bei sehr kleiner Tunnelfähigkeit erfüllt ist.\\

	Im Vergleich dazu werden verschiedene Basen bei der Modellierung der Messung von FQ2 ohne QFP1-Annealing benutzt: FQ2-Energiebasis (Gl. \ref{equ:Ht2_q2eig}), Flussbasis (Gl. \ref{equ:Hf}) und FQ1-FQ2-Energiebasis (Gl. \ref{equ:ht3_q1q2}).\\

	In der FQ2-Energiebasis wird die Bedingung für eine gute Drehwellennäherung in den Bare- (Gl. \ref{bed:mo2b}) und Dressed- (Gl. \ref{bed:mo2d}) Basen untersucht. Ein Basis-Crossover trat bei $\alpha\approx0.7$ auf. Mit großem $\alpha$ wird die Fidelity in der Dressed-Basis höher.\\
	
	Im Unterschied dazu werden zwei Fälle in der FQ1-FQ2-Energiebasis betrachtet. Für eine zz-WW-Näherung ist der Hamiltonoperator in einer diagonalisierten Form. So ist die Bedingung direkt erfüllt. Bei der xx-WW-Näherung kommt die Bedingung zum Wechsel zwischen Bare- und Dressed-Basis (Gl. \ref{bed:q1q2xx}) vor.\\
	
\begin{figure}[ht]
		\centering
		\includegraphics[width=0.7\textwidth]{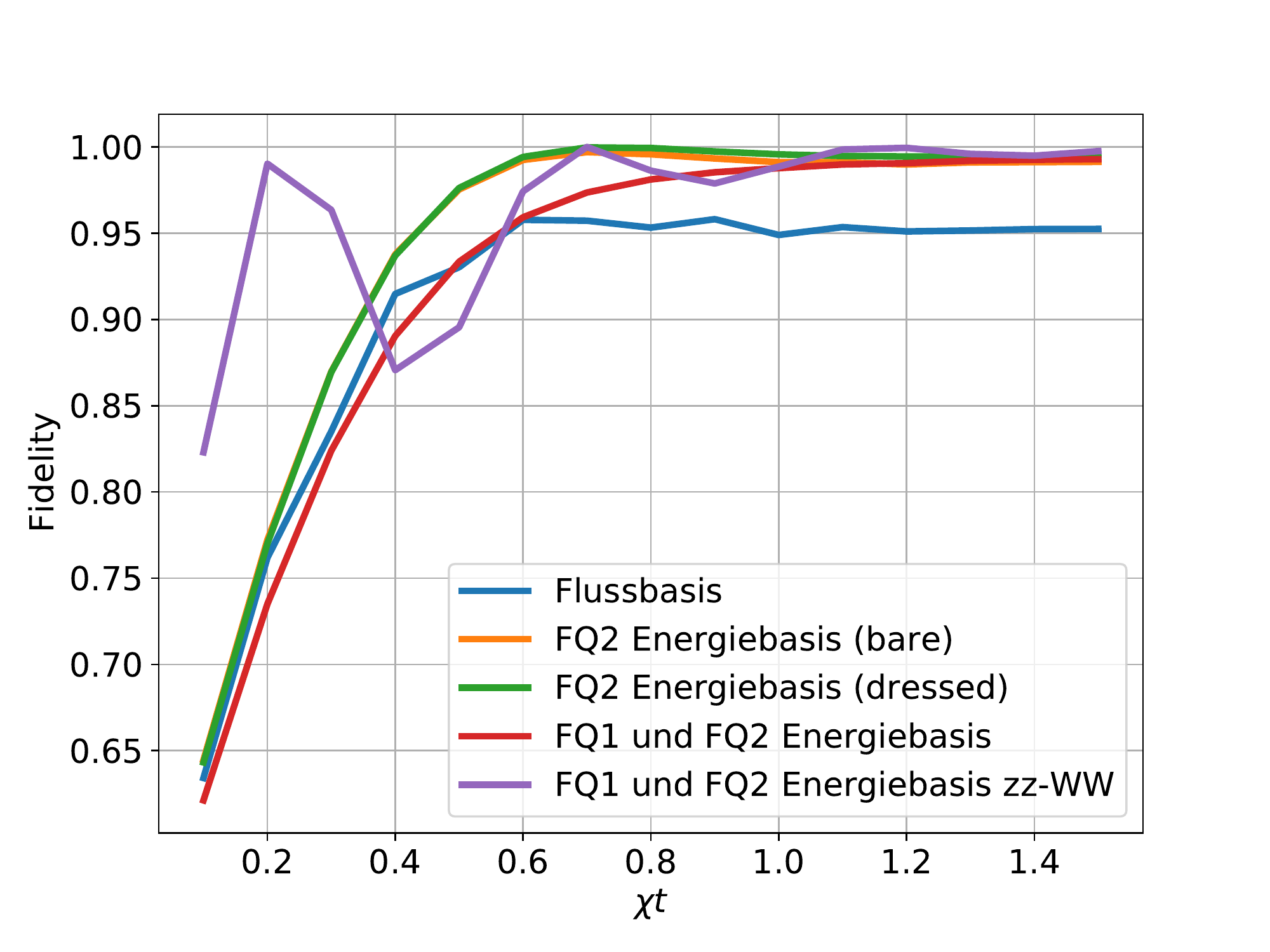}
		\caption{Fidelity in Abhängigkeit von $\chi t$ in verschiedenen Basen mit $N=27$, $\eta=1.25$, $J=0.05(\omega_{2}-\omega_{1})$, $\Delta_{2}/\epsilon_2 = 1$, $\delta/g=8$ und $\alpha=1$. Mit steigendem $\chi t$ wird die Fidelity höher und stabiler.}
		\label{fig:2qo_chit4}
	\end{figure}

	Zur Messzeit $t_m=\pi/2\chi$ wird das Maximum der Fidelity in allen Basen erreicht (siehe Abb. \ref{fig:2qo_chit4}). Flussbasis wird bei einer kurzen Messzeit gültiger als FQ1-FQ2-Energiebasis. Sobald $\chi t>1$, wird die Fidelity in allen Energiebasen höher als in der Flussbasis.\\
	\begin{figure}[ht]
		\centering
		\includegraphics[width=0.7\textwidth]{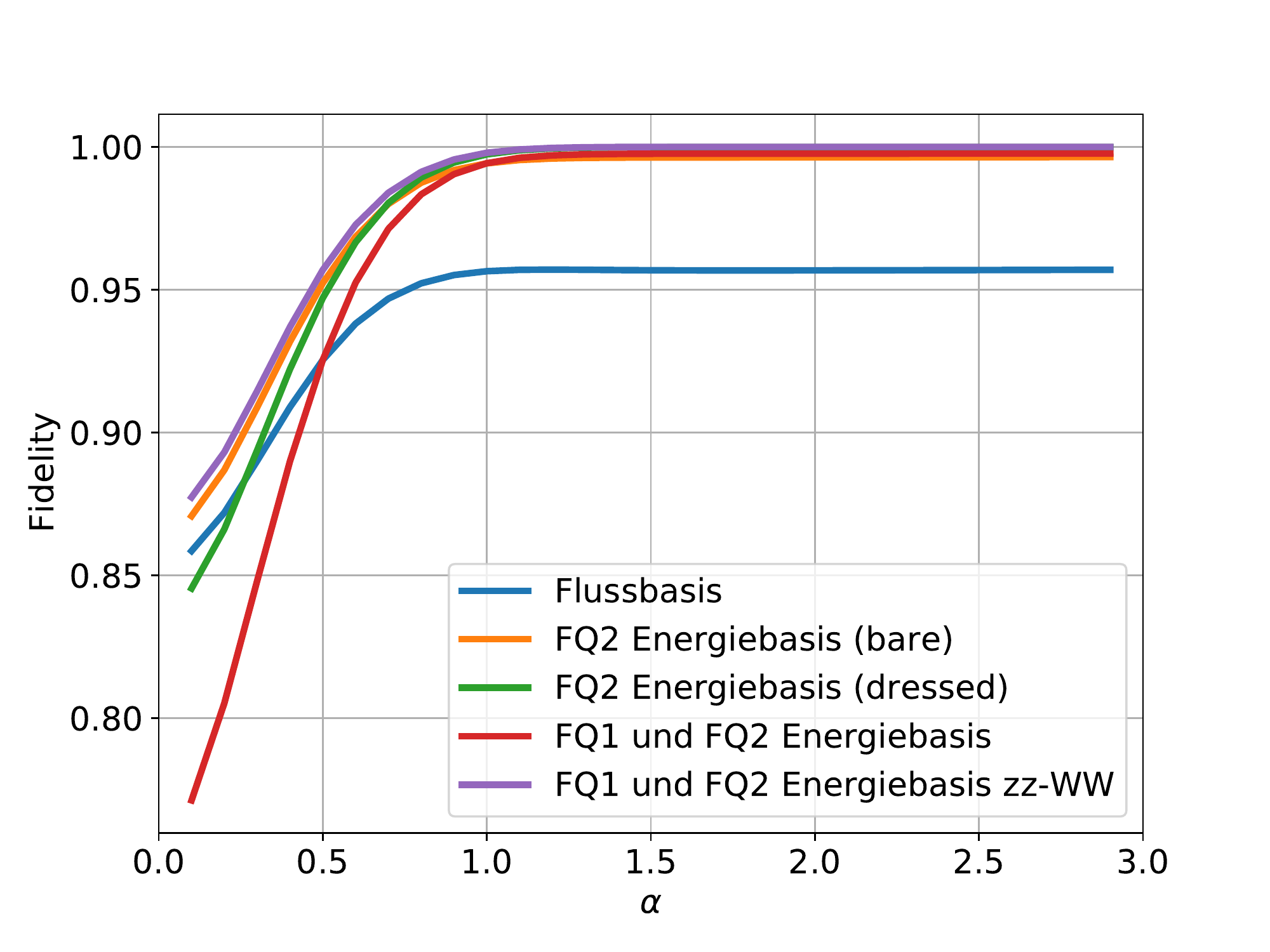}
		\caption{Fidelity in Abhängigkeit von $\alpha$ in verschiedenen Basen mit $N=27$, $\eta=1.25$, $J=0.05(\omega_{2}-\omega_{1})$, $\Delta_{2}/\epsilon_2 = 1$, $\delta/g=8$ und $t_m=\pi/2\chi$. Die Fidelity wird mit steigendem $\alpha$ in allen Basen höher und erreicht Sättigung.}
		\label{fig:2qo_alpha4}
	\end{figure}
	\begin{figure}[ht]
		\centering
		\includegraphics[width=0.7\textwidth]{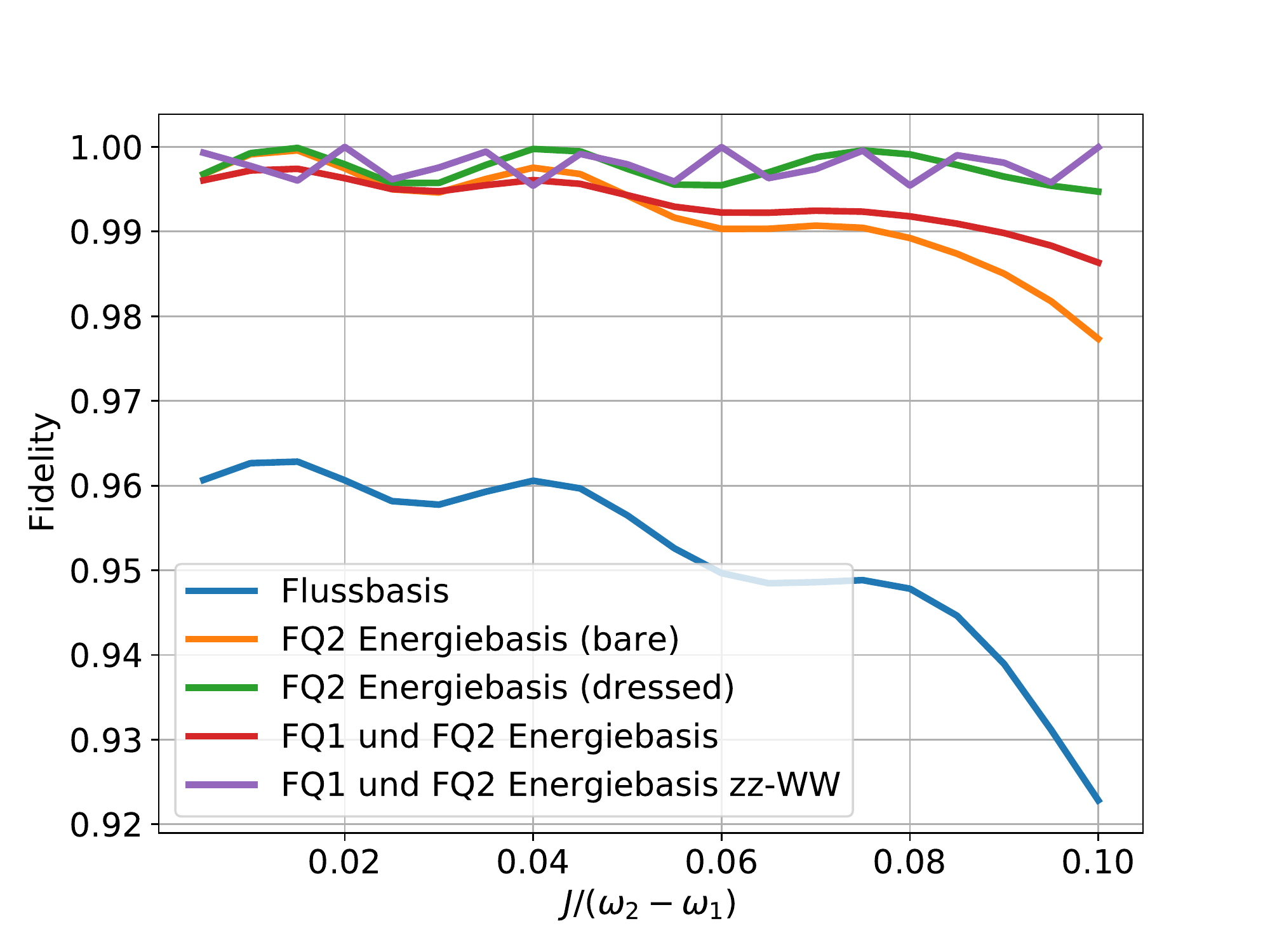}
		\caption{Fidelity in Abhängigkeit von $J/(\omega_{2}-\omega_{1})$ in verschiedenen Basen mit $N=27$, $\eta=1.25$, $t_m=\pi/2\chi$, $\Delta_{2}/\epsilon_2 = 1$, $\delta/g=8$ und $\alpha=1$. Mit steigendem $J/(\omega_{2}-\omega_{1})$ wird die Fidelity in der Flussbasis deutlich abgenommen als in den verschiedenen Energiebasen.}
		\label{fig:2qo_J4}
	\end{figure}

	Außerdem wird ein Maximum der Fidelity in allen Basen bei $\alpha\approx1$ erreicht (siehe	Abb. \ref{fig:2qo_alpha4}) und dann bleibt stabil. Für kleines $\alpha$ wird die Fidelity in der Flussbasis höher als in der Dressed-FQ2-Energiebasis und der FQ1-FQ2-Energiebasis. Sobald $\alpha > 0.5$, wird alle Energiebasen effektiver als die Flussbasis.\\
	
	\clearpage
	
	Die Flussbasis wird im Vergleich zu anderen Energiebasen mehr von der Kopplungsstärke zwischen beiden Flussqubits beeinflusst (siehe Abb. \ref{fig:2qo_J4}). Für die gewählte Parameter, $t_m=\pi/2\chi$ und $\alpha=1$, ist die Flussbasis immer schlechter als die anderen Energiebasen. Da die Fidelity für die Flussbasis in Abhängigkeit von $t$ und $\alpha$ insgesamt deutlich niedrigere Werte erreicht, ist es nicht sinnvoll, in der Flussbasis zu messen.\\

	Zusammenfassend kann man sagen, dass für die Messung eines Flussqubits die Energiebasis immer eine gute Wahl ist und das Basis-Crossover je nach der verschiedenen Bedingungen bei der Messung von FQ2 ohne QFP1-Annealing vorkommt. So ist es uns möglich, eine hohe  Fidelity durch die Optimierung der Parameter bei verschiedenen Situationen zu erreichen. Außerdem wird das Maximum der Fidelity zur Messzeit $\pi/4\chi$ erreicht. Im Vergleich dazu braucht es bei der Messung eines Flussqubit eine relative lange Messzeit.
	
	\clearpage
	\thispagestyle{empty}
	
	\chapter{Messung von FQ2 mit QFP1-Annealing}
	
\begin{figure}[ht]
		\centering
		\includegraphics[width=0.7\textwidth]{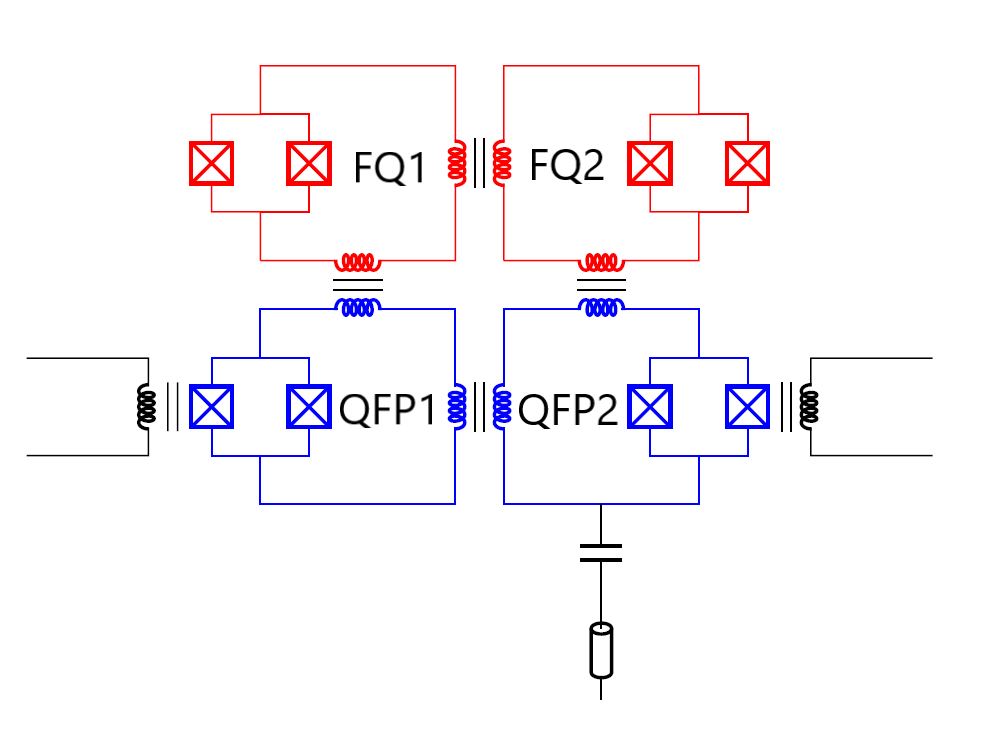}
		\caption{Modellierung der Messung von FQ2 mit QFP1-Annealing. Das Flussqubit-Signal wird durch Annealing im QFP gespeichert. Im Unterschied zu vorher werden die beiden QFPs miteinander wechselwirken. Dann wird QFP2 durch einen Resonator ausgelesen.}
		\label{fig:modell2	m}
	\end{figure}

	In diesem Modell kommt QFP1-Annealing vor (siehe Abb.\ref{fig:modell2	m}). Das heißt, die Signale von FQ1 und FQ2 werden durch QFP1- und QFP2-Annealing entsprechend im QFP1 und QFP2 verschränkt. QFP1 und QFP2 wechselwirken miteinander. Die Messung von QFP2 wird jetzt durch QFP1 beeinflusst. Der Resonator wird zum Auslesen von QFP2 benutzt. Analog zu vorher wird die Fidelity auch in verschiedenen Basen theoretisch analysiert und numerisch simuliert.\\
	
	\section{Modellierung}
	
	Das zu untersuchende Modell lässt sich in der Flussbasis beschreiben:
	
	\begin{align}
		&\hat{H}=\hat{H}_{\text{eff,1}}+\hat{H}_{\text{eff,2}}+J~\hat{\sigma}_1^z\hat{\sigma}_2^z
		+g_2\hat{\sigma}_2^z\left(\hat{a}^{\dag}+\hat{a}\right)+\omega_r \hat{a}^{\dag}\hat{a},\\
		&\hat{H}_{\text{eff,1}}=-\frac{1}{2}\left(\epsilon_{\text{eff,1}} \hat{\sigma}_1^z+\Delta_{\text{eff,1}}\hat{\sigma}_1^x\right),\\
		&\hat{H}_{\text{eff,2}}=-\frac{1}{2}\left(\epsilon_{\text{eff,2}} \hat{\sigma}_2^z+\Delta_{\text{eff,2}}\hat{\sigma}_2^x\right),
	\end{align}\\
	wobei $\omega_r$ der Resonatorfrequenz, $J$ der Kopplungsstärke zwischen beiden Flussqubits und $g_2$ der FQ2-Resonator-Kopplungsstärke entsprechen.\\
	
	In der FQ1-FQ2-Energiebasis erhalten wir\\
	\begin{align}
		\begin{split}
			\tilde{\hat{H}}=&-\frac{\omega_{\text{eff,1}}}{2}\tilde{\hat{\sigma}}_1^z-\frac{\omega_{\text{eff,2}}}{2}\tilde{\hat{\sigma}}_2^z
			+g_2\left[\cos(\theta_{\text{eff,2}}) \tilde{\hat{\sigma}}_2^z-\sin(\theta_{\text{eff,2}}) \tilde{\hat{\sigma}}_2^x\right]\left(\hat{a}^{\dag}+\hat{a}\right)+\omega_r \hat{a}^{\dag}\hat{a}\\
			&+J\left[\cos(\theta_{\text{eff,1}}) \tilde{\hat{\sigma}}_1^z-\sin(\theta_{\text{eff,1}}) \tilde{\hat{\sigma}}_1^x\right]\left[\cos(\theta_{\text{eff,2}}) \tilde{\hat{\sigma}}_2^z-\sin(\theta_{\text{eff,2}}) \tilde{\hat{\sigma}}_2^x\right],
		\end{split}
	\label{equ:hfq1fq2}
	\end{align}\\
	mit $\omega_{\text{eff,i}}=\sqrt{\epsilon_{\text{eff,i}}^2+\Delta_{\text{eff,i}}^2}$ und $\tan(\theta_{\text{eff,i}})=\Delta_{\text{eff,i}}/\epsilon_{\text{eff,i}}$, $i\in \{1,~2\}$.\\
	
	Durch Anwenden des Verschiebungsoperators $\hat{D}(\alpha)$ mit\\ \[\alpha=-g_2\cos(\theta_{\text{eff,2}}) \tilde{\hat{\sigma}}_2^z
	/\omega_r,\]\\
	ergibt sich der Hamiltonoperator unter der Drehwellennäherung\\
	\begin{align}
		\begin{split}
			\tilde{\hat{H}}^{(1)}=&-\frac{\omega_{\text{eff,1}}}{2}\tilde{\hat{\sigma}}_1^z-\frac{\omega_{\text{eff,2}}}{2}\tilde{\hat{\sigma}}_2^z
			-g_2\sin(\theta_{\text{eff,2}}) \left(\hat{a}^{\dag}\tilde{\hat{\sigma}}_2^-+\hat{a}\tilde{\hat{\sigma}}_2^+\right)+\omega_r \hat{a}^{\dag}\hat{a}\\
			&+J\left[\cos(\theta_{\text{eff,1}}) \tilde{\hat{\sigma}}_1^z-\sin(\theta_{\text{eff,1}}) \tilde{\hat{\sigma}}_1^x\right]\left[\cos(\theta_{\text{eff,2}}) \tilde{\hat{\sigma}}_2^z-\sin(\theta_{\text{eff,2}}) \tilde{\hat{\sigma}}_2^x\right].
		\end{split}
		\label{equ:H_3_q1q2}
	\end{align}\\
	Um den Wechselwirkungsterm zwischen Qubit und Resonator zu eliminieren, verwenden wir wieder den häufig benutzten unitären Operator \cite{Blais2004}\cite{Blais2007}\\
	\begin{align}
		\hat{U}=\exp\left[
		\lambda_2\left(\hat{a}\tilde{\hat{\sigma}}_2^+-\hat{a}^{\dag}\tilde{\hat{\sigma}}_2^-\right)\right],
	\end{align}\\
	mit $\lambda_2=g_2\sin(\theta_{\text{eff,2}})/\Delta_{\text{eff,2}}$ auf die Gl. \ref{equ:H_3_q1q2}. Es folgt\\
	\begin{align}
		\begin{split}
			\tilde{\hat{H}}^{(2)}=&-\frac{\omega_{\text{eff,1}}}{2}\tilde{\hat{\sigma}}_1^z
			-\left(\frac{\omega_{\text{eff,2}}+\chi}{2}+\chi\hat{a}^\dagger\hat{a}\right)\tilde{\hat{\sigma}}_2^z+\omega_r\hat{a}^\dagger\hat{a}\\
			&+J\left[\cos(\theta_{\text{eff,1}}) \tilde{\hat{\sigma}}_1^z-\sin(\theta_{\text{eff,1}}) \tilde{\hat{\sigma}}_1^x\right]\left[\cos(\theta_{\text{eff,2}}) \tilde{\hat{\sigma}}_2^z-\sin(\theta_{\text{eff,2}}) \tilde{\hat{\sigma}}_2^x\right].
		\end{split}
	\end{align}\\
	Wenn wir uns sowohl für beide Flussqubits als auch für den Feldoperator zu einem mit der Frequenz $\omega_r$ rotierenden Bezugssystem bewegen, ergibt sich der Hamiltonoperator in der FQ1-FQ2-Energiebasis \\
	\begin{align}
		\begin{split}
			\tilde{\hat{H}}^{(3)}=&-\frac{\delta_{\text{eff,1}}}{2}\tilde{\hat{\sigma}}_1^z-\frac{\hat{\delta}_{\text{eff2,n}}}{2}\tilde{\hat{\sigma}}_2^z
			\\
			&+J\left[\cos(\theta_{\text{eff,1}}) \tilde{\hat{\sigma}}_1^z-\sin(\theta_{\text{eff,1}}) \tilde{\hat{\sigma}}_1^x\right] \left[\cos(\theta_{\text{eff,2}}) \tilde{\hat{\sigma}}_2^z-\sin(\theta_{\text{eff,2}}) \tilde{\hat{\sigma}}_2^x\right],
		\end{split}
		\label{equ:ht3_q1q1_b}
	\end{align}\\
	wobei $\hat{\delta}_{\text{eff2,n}}=\delta_{\text{eff,2}}+\chi\left(2 \hat{n} +1\right)$, $\hat{n} = \hat{a}^{\dag}\hat{a}$ und $\chi=g_{2}^2\sin[2](\theta_{\text{eff,2}})/\delta_{\text{eff,2}}$.\\
	
	Zurück zur Flussbasis bekommen wir\\
	\begin{align}
		\begin{split}
			\hat{H}^{(f)}=&-\frac{\delta_{\text{eff,1}}}{2}\left[\cos(\theta_{\text{eff,1}}) \hat{\sigma}_1^z+\sin(\theta_{\text{eff,1}}) \hat{\sigma}_1^x\right]\\
			&-\frac{\hat{\delta}_{\text{eff2,n}}}{2}\left[\cos(\theta_{\text{eff,2}}) \hat{\sigma}_2^z+\sin(\theta_{\text{eff,2}}) \hat{\sigma}_2^x\right]
			+J~\hat{\sigma}_1^z\hat{\sigma}_2^z.
		\end{split}
		\label{equ:h2_q1q2_fb}
	\end{align}\\
	Hier haben wir wieder den gleichen Superoperator der Messung (Gl. \ref{equ:sup_2q}) mit dem Zeitentwicklungsoperator $\hat{U}(t)=e^{-i\hat{H}t}$ und der Fidelity (Gl. \ref{equ:fi_2}). \\
	
	So haben wir die Modellierung der Messung von FQ2 mit QFP1-Annealing erstellt. Der Hamiltonoperator wird in der FQ1-FQ2-Energiebasis und der Flussbasis dargestellt.\\
	
	\section{Gegenüberstellung der verschiedenen Basen}
	
	Wir betrachten hier auch analytisch zwei spezielle Situationen des Wechselwirkungsterms zwischen FQ1 und FQ2.\\
	\subsection*{1. Fluss-dominiertes Regime}
		\begin{align}
			\{\abs{\cos(\theta_{\text{eff,1}})}, \abs{\cos(\theta_{\text{eff,2}})}\} \gg \{\abs{\sin(\theta_{\text{eff,1}})}, \abs{\sin(\theta_{\text{eff,2}})}\}
			\label{bed:cc12ggss}
		\end{align}\\
		Unter dieser Bedingung können wir den Wechselwirkungsterm von FQ1 und FQ2 näherungsweise durch 
		$J_{zz}\tilde{\hat{\sigma}}_1^z\tilde{\hat{\sigma}}_2^z$ (zz-WW-Näherung) mit $J_{zz}=J\cos(\theta_{\text{eff,1}})\cos(\theta_{\text{eff,2}})$ ersetzen:\\
		\begin{align}
			\begin{split}
				\tilde{\hat{H}}^{(3)}_{zz}\approx-\frac{\delta_{\text{eff,1}}}{2}\tilde{\hat{\sigma}}_1^z-\frac{\hat{\delta}_{\text{eff2,n}}}{2}\tilde{\hat{\sigma}}_2^z
				+J_{zz} \tilde{\hat{\sigma}}_1^z\tilde{\hat{\sigma}}_2^z.
			\end{split}
			\label{ht33_zz}
		\end{align}\\
	\begin{figure}[ht]
		\centering
		\includegraphics[width=0.7\textwidth]{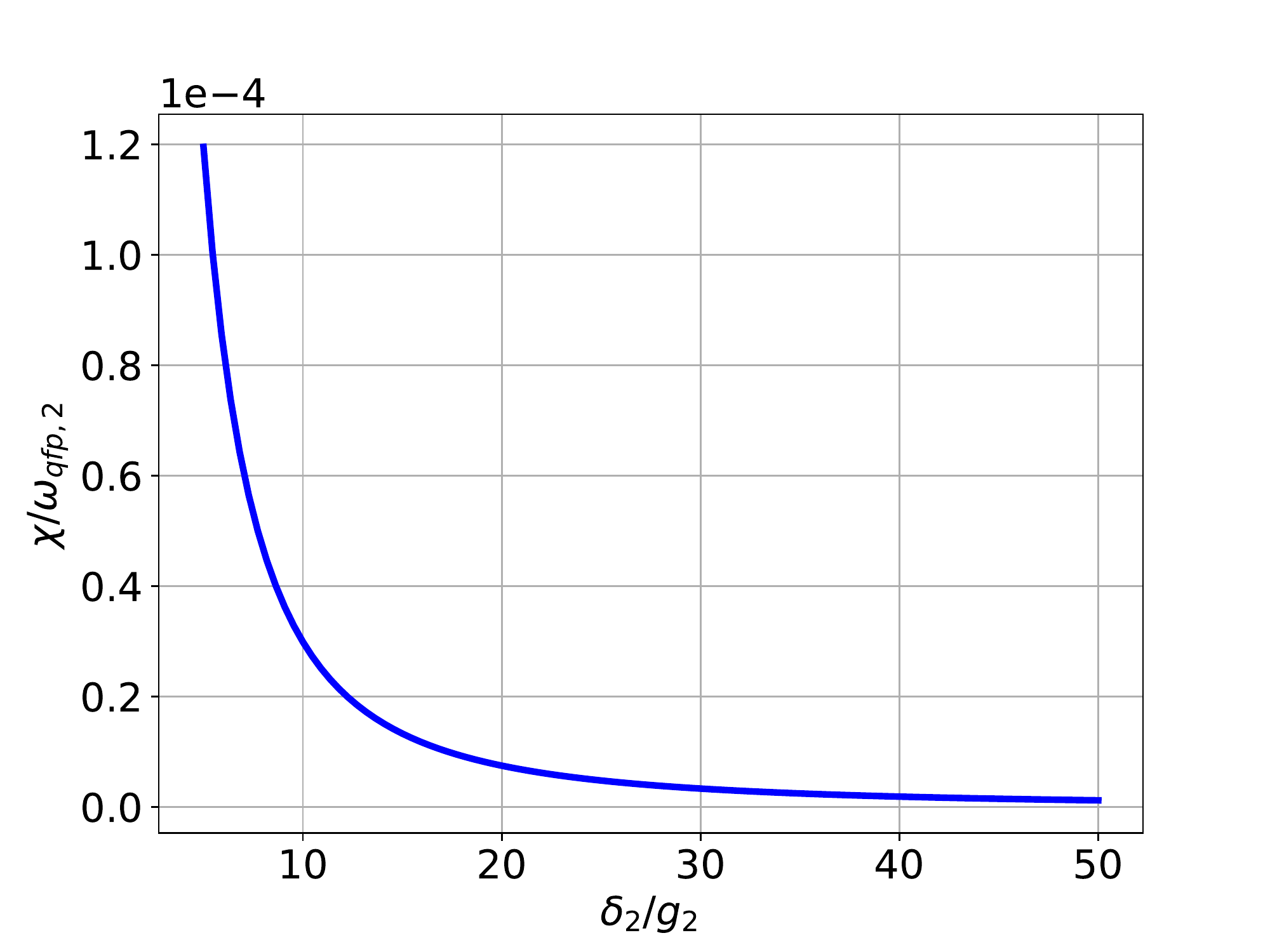}
		\caption{$\chi/\omega_{\text{qfp,2}}$ in Abhängigkeit von $\delta_2/g_2$ mit $\Delta_{2}/\epsilon_{2}=0.3$ und $\eta=1.25$.}
		\label{fig:2qm_chidelta}
	\end{figure}
	In diesem Fall ist der Hamiltonoperator schon diagonalisiert. Analog zu Abschnitt \ref{sec:hinq1q2} wird die Fidelity dem Hamiltonoperator (Gl. \ref{ht33_zz}) gemäß höher. $\chi$ wird in Abhängigkeit von $\delta_2/g_2$ mit $\delta_2=\omega_{2}-\omega_{r}$ der Verstimmung zwischen FQ2- und Resonatorfrequenz in Abb. \ref{fig:2qm_chidelta} dargestellt.\\

\begin{figure}[h]
		\centering
		\includegraphics[width=0.7\textwidth]{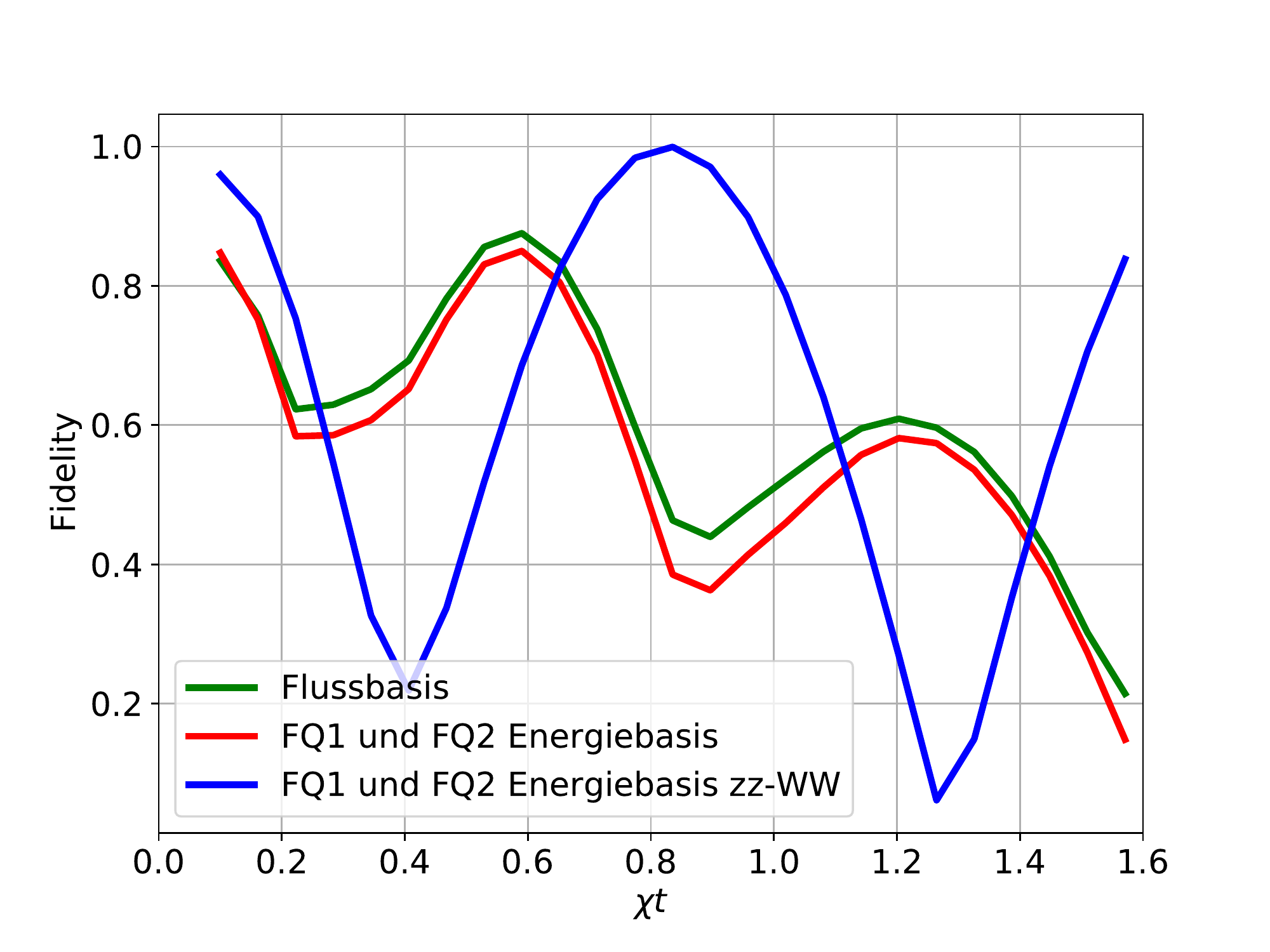}
		\caption{Fidelity in Abhängigkeit von $\chi t$ in der Flussbasis (grün), FQ1-FQ2-Energiebasis (rot) sowie unter der zz-WW-Näherung (blau) mit $N=21$, $\alpha=2$, $\eta=1.25$, $J=0.05(\omega_{2}-\omega_{1})$, $\Delta_{2}/\epsilon_2 = 0.5$ und $\delta_{2}/g_{2}=8$. Die Fidelity oszilliert mit steigendem $\chi t$ gedämpft mit Mittelwert.}
		\label{fig:2qm_chit}
	\end{figure}

\begin{figure}[h]
		\centering
		\includegraphics[width=0.7\textwidth]{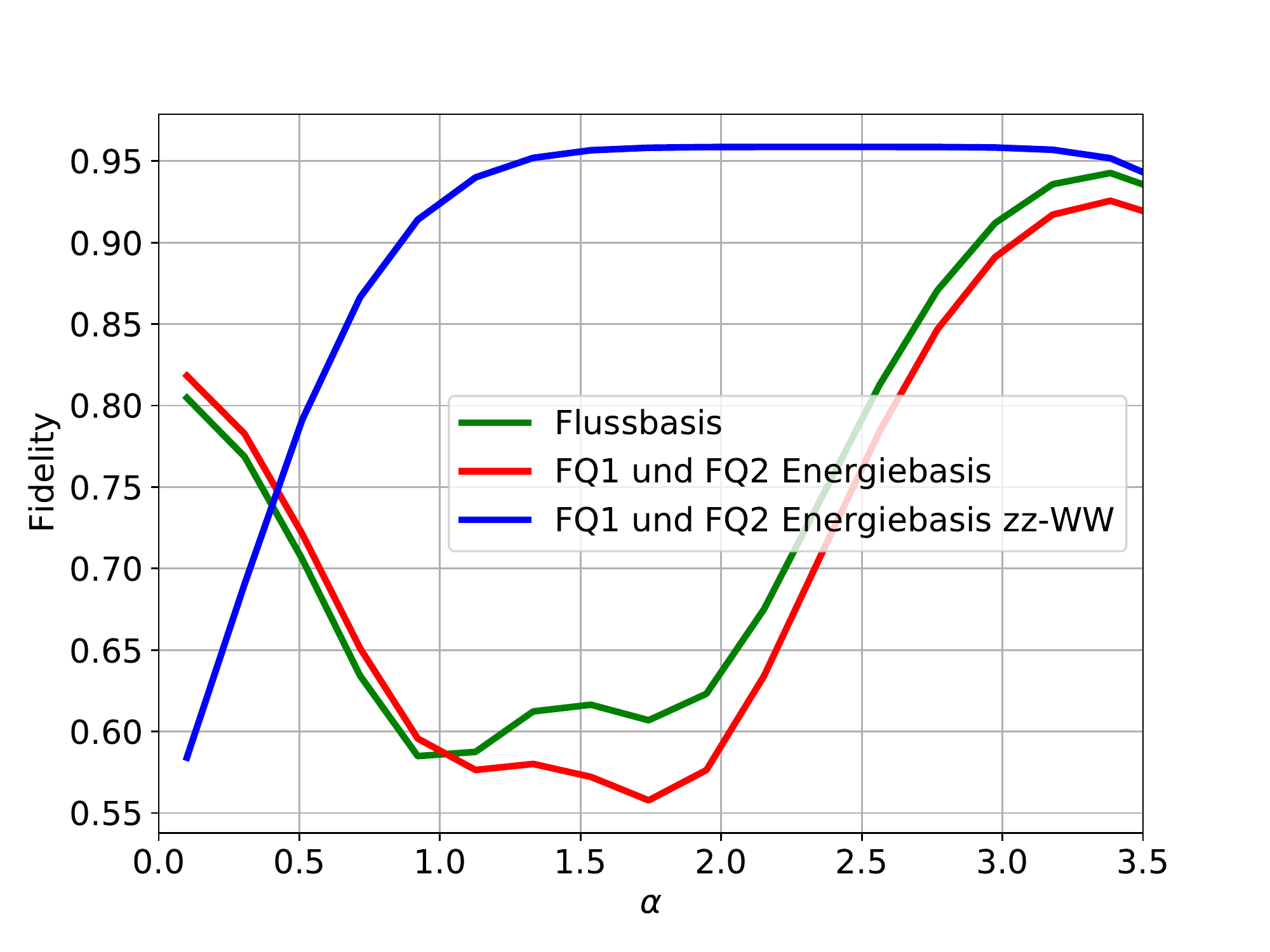}
		\caption{Fidelity in Abhängigkeit von $\alpha$ in der Flussbasis (grün), FQ1-FQ2-Energiebasis (rot) sowie unter der zz-WW-Näherung (blau) mit $N=21$, $\eta=1.25$, $J=0.05(\omega_{2}-\omega_{1})$, $\Delta_{2}/\epsilon_2 = 0.5$, $\delta_{2}/g_{2}=8$ und $t_m=\pi/4\chi$. Unter der zz-WW-Näherung wird die Fidelity mit steigendem $\alpha$ höher und erreicht einen Sättigungswert.}
		\label{fig:2qm_alpha}
	\end{figure}

\begin{figure}[h]
		\centering
		\includegraphics[width=0.7\textwidth]{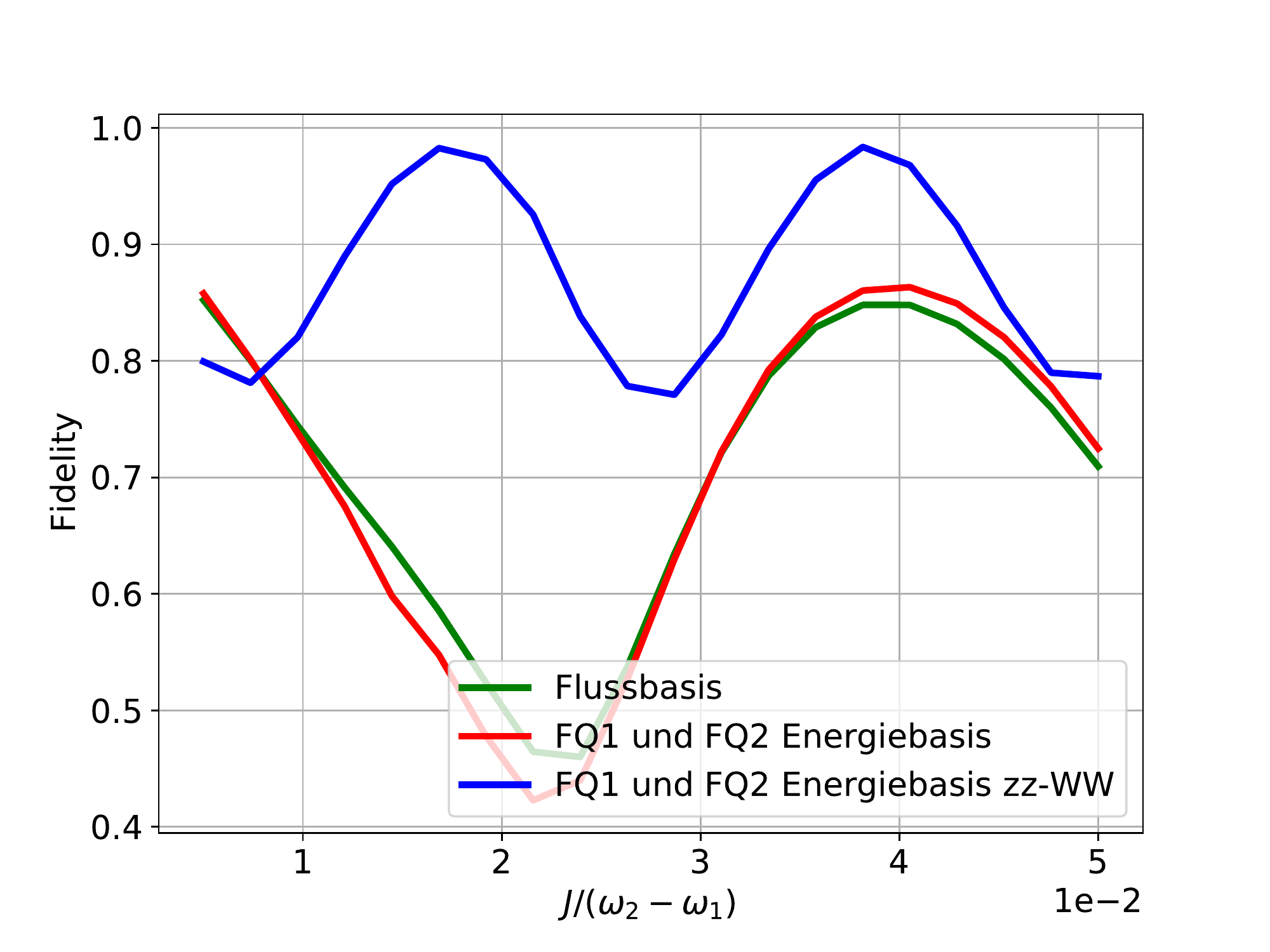}
		\caption{Fidelity in Abhängigkeit von $J/(\omega_{2}-\omega_{1})$ in der Flussbasis (grün), FQ1-FQ2-Energiebasis (rot) sowie unter der zz-WW-Näherung (blau) mit $N=21$, $\eta=1.25$, $\alpha=0.5$, $\Delta_{2}/\epsilon_2 = 0.5$, $\delta_{2}/g_{2}=8$ und $t_m=\pi/4\chi$. Die Fidelity verändert sich periodisch. Unter der zz-WW-Näherung ist sie höher.}
		\label{fig:2qm_J}
	\end{figure}

	Die numerische Untersuchung der Fidelity wird in Abhängigkeit von $\chi t$, $\alpha$ und $J/(\omega_{2}-\omega_{1})$ mit $\omega_{1}$ ($\omega_{2}$) der FQ1- (respektive FQ2-) Frequenz in Abb. \ref{fig:2qm_chit}, Abb. \ref{fig:2qm_alpha} und Abb. \ref{fig:2qm_J} respektive dargestellt.\\
	
	Beim $\chi t \approx 0.1$ ist die Fidelity in allen Basen höher als $0.8$ und unter der zz-WW-Näherung sogar $0.95$ (siehe Abb. \ref{fig:2qm_chit}). Es ermöglicht uns, eine hohe  Fidelity bei kurzer Messzeit zu erreichen. In diesem Modell spielt die Wechselwirkung zwischen QFP1 und QFP2 bei der Messung eine zentrale Rolle, sodass der Term mit $J$ in der diagonalisierter Form im Hamiltonoperator (Gl. \ref{equ:h2_q1q2_fb}) mehr von Bedeutung und die Bedingung für gute RWA besser erfüllt ist. Daher ist die Fidelity in der Flussbasis im Unterschied zu vorher (siehe Abb. \ref{fig:2qo_chit_fq2q1q2}) höher als in der FQ1-FQ2-Energiebasis. \\

	Bei der Untersuchung der Fidelity in Abhängigkeit von $\alpha$ wird eine Messzeit $t_m=\pi/4\chi$ gewählt. So wird die Fidelity unter der zz-WW-Näherung mit steigendem $\alpha$ höher und erreicht einen Sättigungswert (siehe Abb. \ref{fig:2qm_alpha}). Im Unterschied dazu ist die Fidelity in der Flussbasis und der FQ1-FQ2-Energiebasis für $\alpha\lesssim0.4$ erst
	
	\clearpage
	
	größer als die unter der zz-WW-Näherung. Sie fällt erst ab und erreicht ihr Maximum bei $\alpha\approx3.3$, ist aber niedriger als die unter der zz-WW-Näherung. Für großes $\alpha$ wird die Fidelity wieder etwas niedriger. Das liegt wahrscheinlich daran, dass die hier benutzten Näherungen nur für $\alpha^2\ll N$ gelten. Bei $\alpha\approx1$ kommt das Basis-Crossover zwischen Flussbasis und FQ1-FQ2-Energiebasis.\\
	
	Für kleine Kopplungsstärke ($J/(\omega_{2}-\omega_{1})\lesssim0.01$) aber auch $J/(\omega_{2}-\omega_{1})\approx0.05$ wird die Fidelity in allen Basen vergleichbar (siehe Abb. \ref{fig:2qm_J}). Unter der zz-WW-Näherung ist die Fidelity insgesamt höher, weil wir dafür eine bessere Messzeit ($t_m=\pi/4\chi$) gewählt haben (siehe Abb. \ref{fig:2qm_chit}). Im Unterschied zum Modell in Kapitel \ref{cha:mq2} ist hier die Kopplung zwischen beiden QFPs entscheidend, sodass die Dynamik des System unterschiedlich ist. \\
	
\subsection*{2. Tunnel-dominiertes Regime}
\begin{figure}[ht]
		\centering
		\includegraphics[width=0.7\textwidth]{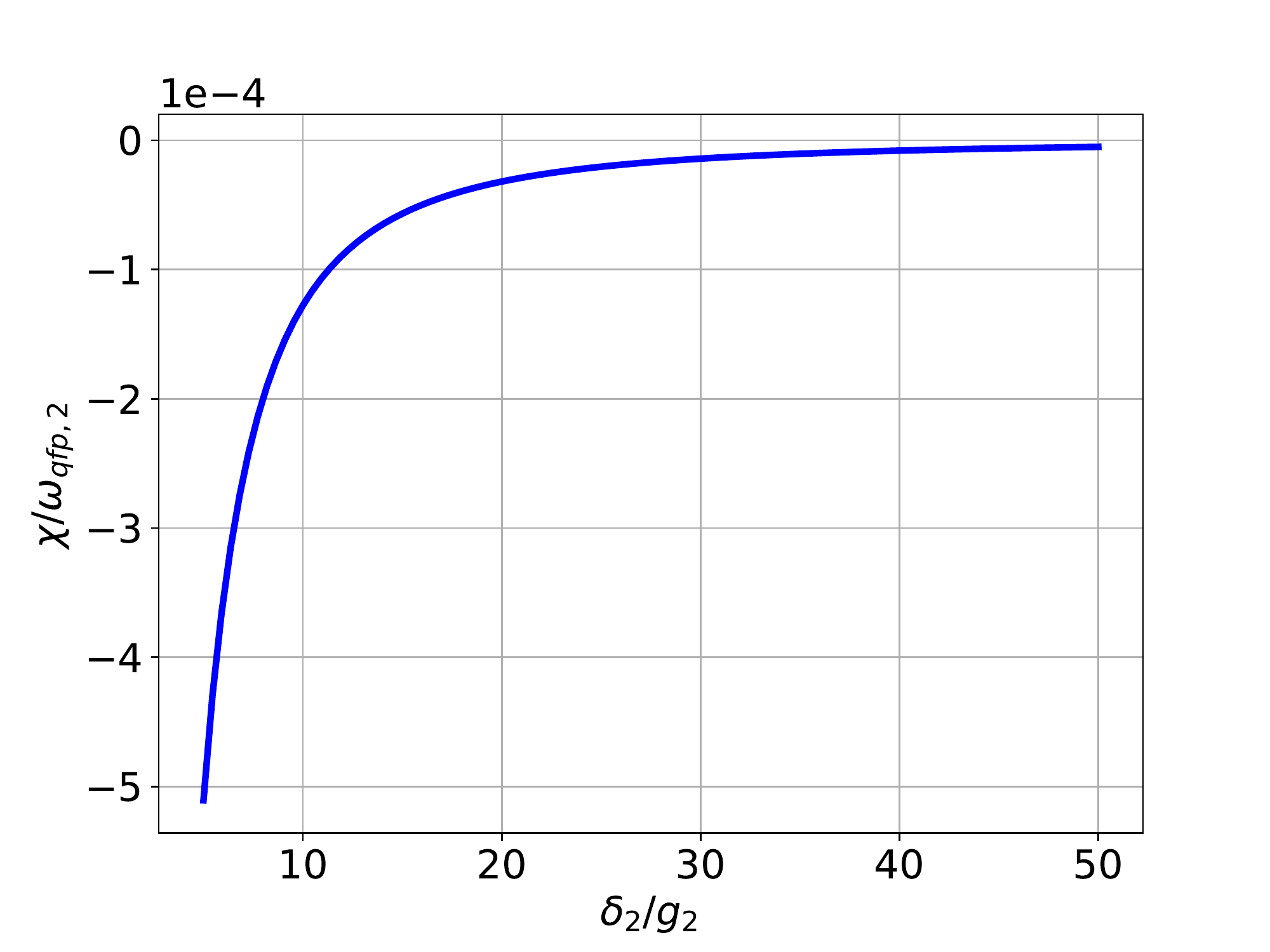}
		\caption{$\chi/\omega_{\text{qfp,2}}$ in Abhängigkeit von $\delta_2/g_2$ mit $\Delta_{2}/\epsilon_{2}=10$ und $\eta=1$.}
		\label{fig:2qm_chideltaxx}
	\end{figure}

		\begin{align}
			\{\abs{\cos(\theta_{\text{eff,1}})}, \abs{\cos(\theta_{\text{eff,2}})}\} \ll \{\abs{\sin(\theta_{\text{eff,1}})}, \abs{\sin(\theta_{\text{eff,2}})}\}
			\label{bed:cc12kss}
		\end{align}\\
		In dieser Situation haben wir den Wechselwirkungsterm von FQ1 und FQ2 in der Form von $J_{xx}\tilde{\hat{\sigma}}_1^x\tilde{\hat{\sigma}}_2^x$ (xx-WW-Näherung) mit $J_{xx}=J \sin(\theta_\text{eff,1})\sin(\theta_{\text{eff,2}})$. Der entsprechende Hamiltonoperator ist\\
		\begin{align}
			\begin{split}
				\tilde{\hat{H}}^{(3)}_{xx}=-\frac{\delta_{\text{eff,1}}}{2}\tilde{\hat{\sigma}}_1^z-\frac{\hat{\delta}_{\text{eff2,n}}}{2}\tilde{\hat{\sigma}}_2^z
				+J_{xx} \tilde{\hat{\sigma}}_1^x \tilde{\hat{\sigma}}_2^x.
			\end{split}
			\label{equ:ht3_xx_b}
		\end{align}\\
	$\chi$ wird in Abhängigkeit von $\delta_2/g_2$ mit $\Delta_2/\epsilon_2=10$ in Abb. \ref{fig:2qm_chideltaxx} dargestellt.

		Die Matrixform ist\\
		\begin{align}
			\tilde{\hat{H}}^{(3)}_{xx}=\begin{bmatrix}
				-\frac{\delta_{\text{eff,1}}+\hat{\delta}_{\text{eff2,n}}}{2} & 0 & 0 & J_{xx}\\
				0 & -\frac{\delta_{\text{eff,1}}-\hat{\delta}_{\text{eff2,n}}}{2} & J_{xx} & 0\\
				0 & J_{xx}& \frac{\delta_{\text{eff,1}}-\hat{\delta}_{\text{eff2,n}}}{2} & 0\\
				J_{xx} & 0 & 0 & \frac{\delta_{\text{eff,1}}+\hat{\delta}_{\text{eff2,n}}}{2}
			\end{bmatrix}.
		\end{align}\\
		Hier kann man eine gleiche Strategie wie die zweite Situation in Absatz \ref{sec:hinq1q2} verwenden. Analog dazu benutzen wir den Transformationsoperator zur Diagonalisierung\\
		\begin{align*}
			\hat{U}_r =\begin{pmatrix}
				\cos(\frac{\theta_{n+}}{2})& 0 & 0 & -\sin(\frac{\theta_{n+}}{2})\\
				0& \cos(\frac{\theta_{ n-}}{2}) & -\sin(\frac{\theta_{ n-}}{2})&0\\	
				0 & \sin(\frac{\theta_{ n-}}{2})& \cos(\frac{\theta_{ n-}}{2})&0\\
				\sin(\frac{\theta_{n+}}{2})&0&0&\cos(\frac{\theta_{n+}}{2})
			\end{pmatrix}.
		\end{align*}\\
		Die Mischungswinkel $\theta_{n\pm}$ sind unterschiedlich als vorher durch $\tan(\theta_{n\pm })=\\
		2J_{xx}/\left(\delta_{\text{eff2,n}}\pm \delta_{\text{eff,1}}\right)$ definiert.
		So bekommen wir die Eigenwerte\\
		\begin{align}
			E^{(3)}_{xx}= U_r \tilde{\hat{H}}^{(3)}_{xx} U_r^\dagger = \text{diag}(-\omega_{n+},~ -\omega_{n-},~ \omega_{n-},~ \omega_{n+}),
		\end{align}\\
		mit $\omega_{n\pm} = \sqrt{J_{xx}^2+\left(\delta_{\text{eff2,n}}\pm\delta_{\text{eff,1}}\right)^2/4}$.\\
		
		Analog zu der Gl. \ref{equ:Ht2_xx_q1q2} kann man den Hamiltonoperator (\ref{equ:ht3_xx_b}) in der Dressed-Basis darstellen:\\
		\begin{align*}
			\begin{split}
				\overline{H}_{xx}^{(3)}=&-\frac{\omega_{n+}}{2} \cos(\theta_{n+}+\theta_{0+}) \left(\overline{\hat{\sigma}}_{1}^{z}+\overline{\hat{\sigma}}_{2}^{z}\right)+\frac{\omega_{n+}}{2} \sin(\theta_{n+}+\theta_{0+}) \left(\overline{\hat{\sigma}}_{1}^{x}\overline{\hat{\sigma}}_{2}^{x}-\overline{\hat{\sigma}}_{1}^{y}\overline{\hat{\sigma}}_{2}^{y}\right)\\
				&+\frac{\omega_{n-}}{2}\cos(\theta_{n-}-\theta_{0-})\left(\overline{\hat{\sigma}}_{1}^{z}-\overline{\hat{\sigma}}_{2}^{z}\right)+\frac{\omega_{n-}}{2} \sin(\theta_{n-}-\theta_{0-}) \left(\overline{\hat{\sigma}}_{1}^{x}\overline{\hat{\sigma}}_{2}^{x}+\overline{\hat{\sigma}}_{1}^{y}\overline{\hat{\sigma}}_{2}^{y}\right).
			\end{split}
		\end{align*}\\
		Mit Hilfe von\\
		\begin{align}
			\omega_{n\pm}\cos(\theta_{n\pm}\pm\theta_{0\pm})&\rightarrow \left(\delta_{\text{eff,2}}\pm \delta_{\text{eff,1}}\right)\left(\hat{\delta}_{\text{eff2,n}} \pm\delta_{\text{eff,1}}\right)/4 +J_{xx}^2\\
			\omega_{n\pm}\sin(\theta_{n\pm}\pm\theta_{0\pm})&\rightarrow J_{xx}\left[\left(\delta_{\text{eff,2}}\pm \delta_{\text{eff,1}}\right)\pm\left(\hat{\delta}_{\text{eff2,n}}\pm \delta_{\text{eff,1}}\right)\right]/2
		\end{align}\\
		kann man $\delta_{\text{eff2,n}}$ im Hamiltonoperator (Gl.\ref{equ:h2d}) zum $\hat{\delta}_{\text{eff2,n}}$ transformieren.
		Die Bedingung der Gültigkeit der RWA ist\\
		\begin{align}
			\left\Vert\frac{\sin(\theta_{n\pm}\pm\theta_{0\pm})}{\cos(\theta_{n\pm}\pm\theta_{0\pm})}  \right\Vert = \left\Vert \frac{J_{xx}\left[\left(\delta_{\text{eff,2}}\pm \delta_{\text{eff,1}}\right)\pm\left(\hat{\delta}_{\text{eff2,n}}\pm \delta_{\text{eff,1}}\right)\right]/2}{\left(\delta_{\text{eff,2}}\pm \delta_{\text{eff,1}}\right)\left(\hat{\delta}_{\text{eff2,n}} \pm\delta_{\text{eff,1}}\right)/4 +J_{xx}^2} \right\Vert \ll 1.
		\end{align}\\
		Zusammen mit der Bedingung für eine gute RWA in der Bare-Basis\\
		\begin{align}
			\left\Vert\frac{\sin(\theta_{n\pm})}{ \cos(\theta_{n\pm})}\right\Vert = \left\Vert \frac{J_{xx}}{\left(\hat{\delta}_{\text{eff2,n}}\pm \delta_{\text{eff,1}}\right)/2} \right\Vert \ll 1,
		\end{align}\\
		ergibt sich die Gleichung für den Crossover-Punkt\\
		\begin{align}
			\left\Vert\tan(\theta_{n\pm}\pm\theta_{0\pm})\right\Vert=\left\Vert\tan(\theta_{n\pm})\right\Vert,
		\end{align}\\
	was\\
	\begin{align}
			\left\Vert\chi\left(\hat{a}^\dagger\hat{a}+\frac{1}{2}\right)+\left(\delta_{\text{eff,2}}+ \delta_{\text{eff,1}}\right)/2\right\Vert = J_{xx}
			\label{bed:2qmxx1}
	\end{align}\\
	und\\
	\begin{align}
		\left\Vert\chi\left(\hat{a}^\dagger\hat{a}+\frac{1}{2}\right)\right\Vert = \sqrt{\left(\delta_{\text{eff,2}}- \delta_{\text{eff,1}}\right)^2/4 + J_{xx}^2}
		\label{bed:2qmxx2}
	\end{align}
	entspricht.\\
		
	So haben wir die Bedingungen (Gl. \ref{bed:2qmxx1} und \ref{bed:2qmxx2}) zum Basis-Crossover zwischen Bare- und Dressed-Basis für den Fall $J_{xx}\tilde{\hat{\sigma}}_1^x\tilde{\hat{\sigma}}_2^x$ in der FQ1-FQ2-Energiebasis für die FQ2 Messung mit QFP1-Annealing theoretisch hergeleitet.\\
	
\begin{figure}[ht]
		\centering
		\includegraphics[width=0.7\textwidth]{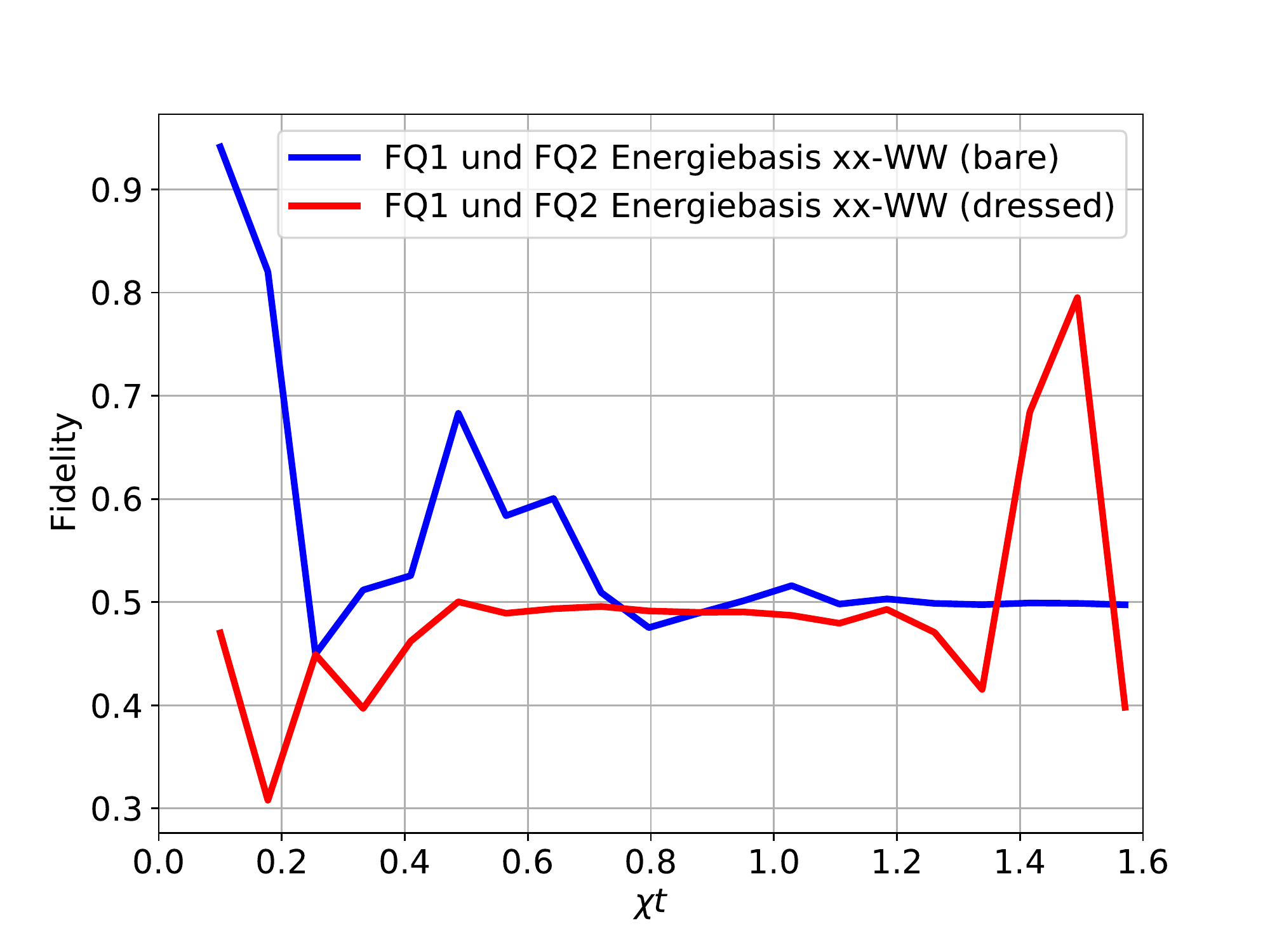}
		\caption{Fidelity in Abhängigkeit von $\chi t$ in der Bare- (blau) und Dressed- (rot) FQ1-FQ2-Energiebasis unter der xx-WW-Näherung mit $N=21$, $\eta=1$, $\alpha=2$, $J=0.05(\omega_{2}-\omega_{1})$, $\Delta_{2}/\epsilon_2 = 8$ und $\delta_{2}/g_{2}=8$. Je nach $\chi t$ unterscheidet sich die Fidelity in der Bare- und Dressed-Basis.}
		\label{fig:2qm_chitxx}
	\end{figure}

\begin{figure}[ht]
		\centering
		\includegraphics[width=0.7\textwidth]{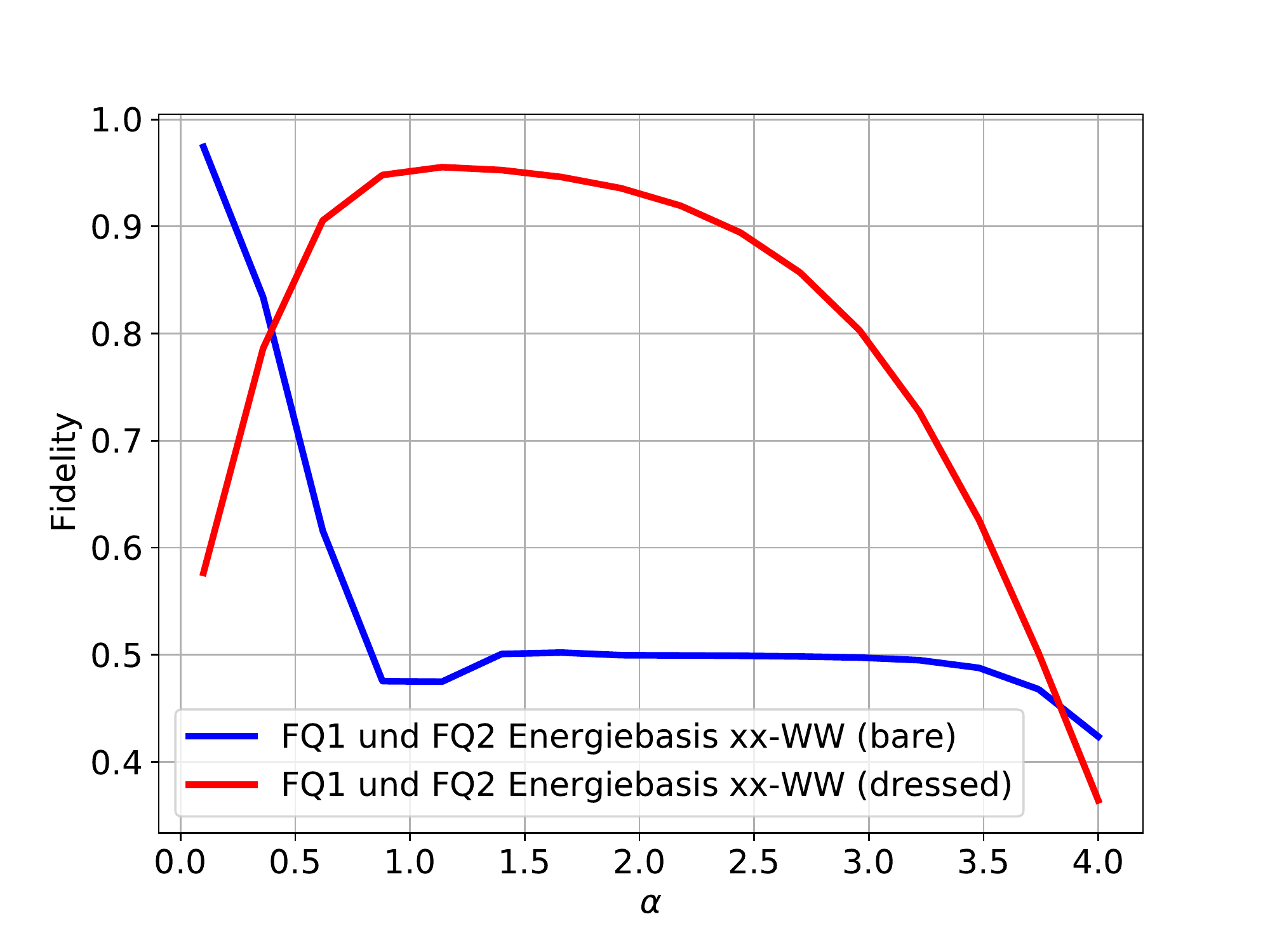}
		\caption{Fidelity in Abhängigkeit von $\alpha$ in der Bare- (blau) und Dressed- (rot) FQ1-FQ2-Energiebasis unter der xx-WW-Näherung mit $N=21$, $\Delta_{2}/\epsilon_{2}=10$, $\eta=1$, $J=0.05(\omega_{2}-\omega_{1})$ und $t_m=\pi/2\chi$. Mit steigendem $\alpha$ wird die Fidelity in der Dressed-Basis zuerst höher und dann niedriger.}
		\label{fig:2qm_alphaxx}
	\end{figure}

\begin{figure}[ht]
 		\centering
 		\includegraphics[width=0.7\textwidth]{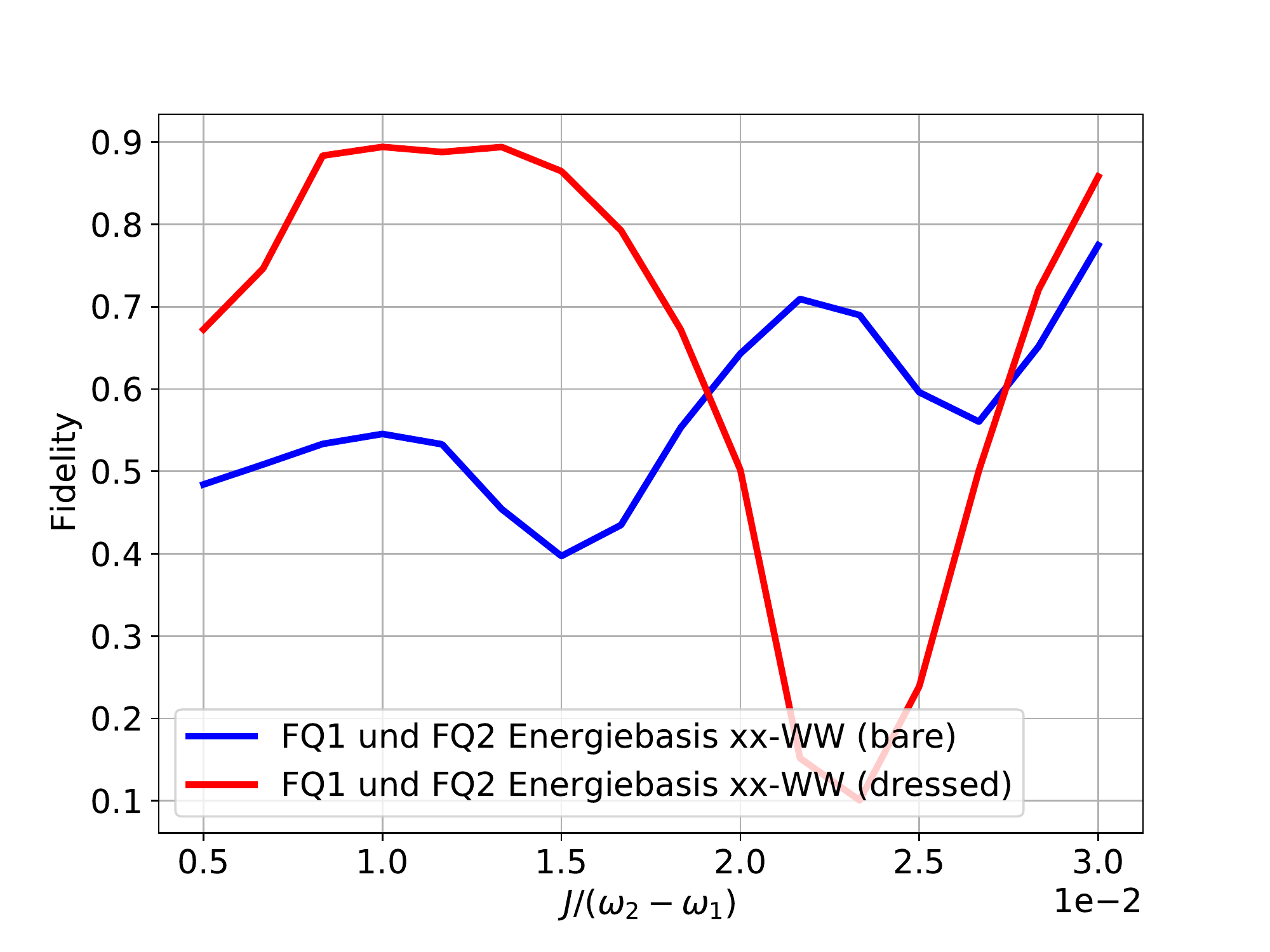}
 		\caption{Fidelity in Abhängigkeit von $J/(\omega_{2}-\omega_{1})$ in der Bare- (blau) und Dressed- (rot) FQ1-FQ2-Energiebasis unter der xx-WW-Näherung mit $N=21$, $\alpha=0.5$, $\Delta_{2}/\epsilon_{2}=0.8$, $\eta=1$, $\delta_{2}/g_{2}=8$ und $t_m=\pi/2\chi$. Die Amplitude der Schwingung von Fidelity ist in der Dressed-Basis größer als in der Bare-Basis.}
 		\label{fig:2qm_Jxx}
 	\end{figure}

	Die numerische Untersuchung der Fidelity wird in Abhängigkeit von $\chi t$, $\alpha$ und $J/(\omega_2-\omega_{1})$ in Abb. \ref{fig:2qm_chitxx}, Abb. \ref{fig:2qm_alphaxx} und Abb. \ref{fig:2qm_Jxx} respektive dargestellt.\\
	
	Bei kurzer Messzeit ($t \approx 0.1/\chi $) ist die Fidelity in der Bare-Basis sehr hoch, während die bei langer Messzeit ($ t \approx \pi/2\chi$) in der Dressed-Basis hoch ist (siehe Abb. \ref{fig:2qm_chitxx}). Dieses Ergebnis ist mit \cite{Pommerening2020} vergleichbar. So ist es möglich, eine hohe Fidelity in der Bare-Basis bei einer kurzen Messzeit zu erreichen.\\

	Für einen kleinen Wert von $\alpha$ ist die Fidelity in der Bare-Basis höher (siehe Abb. \ref{fig:2qm_alphaxx}). In der Dressed-Basis wird die Fidelity mit steigendem $\alpha$ höher und erreicht das Maximum. Wenn $\alpha$ zu groß ist, wird die Bedingung $\alpha^{2} \ll N$ nicht mehr erfüllt. Dann ist die Fidelity wieder niedriger. Im Vergleich dazu wird die Fidelity in der Bare-Basis mehr davon beeinflusst.\\
 	
 	Je nach Wert von $J/(\omega_{2}-\omega_{1})$ oszilliert die Fidelity in der Bare- und Dressed-Basis (siehe Abb. \ref{fig:2qm_Jxx}). Die Fidelity in der Dressed-Basis reagiert empfindlicher auf die Wechselwirkung zwischen beiden Flussqubits. Das Basis-Crossover zwischen Bare- und Dressed-Basis kann man gut erkennen.\\

 	Unter der xx-WW-Näherung kommt das Basis-Crossover deutlich vor. Es liegt wohl an der Wechselwirkung zwischen beiden QFPs. Bei kurzer Messzeit ist die Messung in der Bare-Basis vorteilhaft, während die bei einer langen Messzeit in der Dressed-Basis besser ist. Zur Messzeit $t_m=\pi/2\chi$ kann man auch eine hohe  Fidelity je nach kleinem oder großem $\alpha$ in der Bare- oder Dressed-Basis erreichen. Im Gegenteil zu vorher wird die Dressed-Basis mehr von der Wechselweisung zwischen beiden Flussqubits beeinflusst.\\

	Zwei Fälle werden in der Q1 und Q2 Energiebasis betrachtet. Bei der zz-WW-Näherung ist die Bedingung für eine gute Drehwellennäherung bereits erfüllt. Im anderen Fall, also xx-WW-Näherung, kommt zwei Bedingungen (Gl. \ref{bed:2qmxx1} und Gl. \ref{bed:2qmxx2}) für das Basis-Crossover von den Bare- und Dressed-Basis vor. Die Wechselwirkung zwischen beiden QFPs ermöglicht uns in beiden Fällen, eine hohe  Fidelity bei einer schnellen Messung zu erreichen.\\
	
	\section{Vergleich und Analyse}
	
	Für die Messung eines Flussqubits ist die Fidelity in der Energiebasis immer höher als in der Flussbasis. Im Vergleich dazu ist die Modellierung der Messung von FQ2 mit QFP1-Annealing wegen der Wechselwirkung zwischen beiden QFPs bei einer sehr kurzen Messzeit vorteilhaft.\\
	
	Bei der Modellierung der Messung von FQ2 ohne QFP1-Annealing gibt es verschiedene Messbasen: FQ2-Energiebasis (Gl. \ref{equ:Ht2_q2eig}), Flussbasis (Gl. \ref{equ:Hf}) und FQ1-FQ2-Energiebasis (Gl. \ref{equ:ht3_q1q2}). In der FQ2-Energiebasis kommt Bare- und Dressed-Basis vor. Das Basis-Crossover dazwischen trat bei der Darstellung der Fidelity bei $\alpha \approx 0.7$ auf.\\
	
	Im Vergleich dazu sind die Messbasen in der Modellierung der Messung von FQ2 mit QFP1-Annealing: FQ1-FQ2-Energiebasis (Gl. \ref{equ:ht3_q1q1_b}) und Flussbasis (Gl. \ref{equ:h2_q1q2_fb}). Unter der zz-WW-Näherung ist die Fidelity in der Flussbasis im Gegenteil zur Messung ohne QFP1-Annealing höher als in der FQ1-FQ2-Energiebasis. Dies lässt sich auch durch die Wechselwirkung von beiden QFPs erklären.\\
	
	Für die Untersuchung der Bedingung für eine gute Drehwellennäherung in der FQ1-FQ2-Energiebasis wurden in beiden Modellierungen zwei Fälle betrachtet. Unter der zz-WW-Näherung ist der Hamiltonoperator in einer diagonalisierten Form und die Bedingung schon direkt erfüllt. Bei der xx-WW-Näherung bekommt man die Bedingungen zum Basis-Crossover zwischen Bare- und Dressed-Basis. Ohne die Wechselwirkung zwischen QFPs ist die Fidelity zur Messzeit $t_m=\pi/2\chi$ in allen Energiebasen höher als in der Flussbasis. Dies ist vergleichbar mit der Messung eines einzelnen Flussqubits.
	Bei Anwesenheit der Wechselwirkung zwischen QFPs wird eine hohe  Fidelity bei einer schnellen Messung erreicht.

	\clearpage
	\thispagestyle{empty}
	
	\part{Zusammenfassung und Ausblick}
	
	\chapter*{Zusammenfassung und Ausblick}

	Ziel der vorliegenden Arbeit war es, das gekoppelte Flussqubit-QFP-System zu analysieren und die Fidelity der Messung in verschiedenen Basen zu vergleichen. Bedingungen für eine höhere Fidelity wurden dabei hergeleitet.
	Die Fidelity wurde in Abhängigkeit von der Messzeit, der Amplitude des kohärenten Zustands des Resonators sowie der Kopplungsstärke zwischen beiden Flussqubits in verschiedenen Basen numerisch simuliert.\\
	
	Ein QFP wird zwischen Flussqubit und Resonator eingefügt, um das Stromsignal zu verstärken und das Flussqubit vom Resonator zu isolieren. QFP-Annealing wird adiabatisch durchgeführt, um das Signal des daran gekoppelten Flussqubits erfolgreich zu speichern. Das gekoppelte Flussqubit-QFP-System kann nach dem QFP-Annealing durch einen effektiven Flussqubit-Hamiltonoperator beschrieben werden.\\
	
	Es wurde durch numerische Simulation gezeigt, dass nach dem QFP-Annealing das Flussqubit-Signal im QFP sowohl in der Flussbasis als auch in der Energiebasis effizient verschränkt werden kann. Das heißt, die Verschränkung hängt nicht von der lokalen Basis ab. Eine sehr hohe Fidelity kann in der Energiebasis mit kürzerer Messzeit erreicht werden, als in der Flussbasis.
	Der Überlapp zwischen Bare- und Dressed-Zustand des gekoppelten Systems wird durch einen Erwartungswert der Verschiebungsoperatoren in den Fock-Zuständen dargestellt. Mit steigender Kopplungsstärke zwischen Flussqubit und QFP sinkt der Überlapp bis auf Null.\\
	
	Sobald das QFP-Annealing durchgeführt ist, wird das QFP, in dem das Flussqubit-Signal gespeichert wurde, durch einen Resonator ausgelesen. Dies wurde mit Hilfe der Jaynes-Cummings-Theorie im dispersiven Regime untersucht. Die Gültigkeit der Drehwellennäherung in verschiedenen Basen kann durch das Verhältnis der Werte des Außerdiagonalelements und des Diagonalelements von dem Hamiltonoperator beschrieben werden.\\
	
	Für die Messung eines Flussqubits mittels QFP ist die Fidelity in der Energiebasis immer höher als in der Flussbasis. Die Anwesenheit einer externen Ansteuerung führt zur gleichen Dynamik wie ohne Feld. Daher wurde die externe Ansteuerung für die Simulation vernachlässigt. Die numerischen Ergebnisse zeigen, dass die Energiebasis, auch dem theoretischen Ergebnis entsprechend, immer besser als die Flussbasis ist.\\
	
	\thispagestyle{plain}
	
	Zwei miteinander gekoppelte Flussqubits FQ1 und FQ2, jeweils an ein QFP1 gekoppelt, wurden dann untersucht. Das Ziel war, eines davon (hier FQ2) auszulesen. Dabei wurden zwei Modellierungen erstellt. Der Unterschied dazwischen bestand in der Anwesenheit von QFP1-Annealing. Bei beiden Modellen wurde nur der Messprozess mittels Resonator betrachtet. Das heißt, es wurde angenommen, dass QFP2-Annealing schon adiabatisch durchgeführt und das FQ2-Signal war bereits im QFP2 erfolgreich gespeichert war.\\
	
	Die Bedingungen zum Basis-Crossover wurden in beiden Modellen theoretisch hergeleitet. Der Vorteil der Messung von FQ2 ohne QFP1-Annealing liegt darin, dass die Fidelity in allen Basen bis zum Messzeitpunkt $t_m = \pi/2\chi$, wobei $\chi$ die dispersive Kopplungsstärke des Resonators an QFP2 ist, stabil ist und ein konstantes Maximum erreicht. Die Fidelity ist sowohl in der FQ2-Energiebasis als auch in der FQ1-FQ2-Energiebasis immer höher als in der Flussbasis. Es ist für dieses Modell von Nachteil, wenn die Messzeit zu kurz ist. Im Unterschied dazu bekommt man wegen der Wechselwirkung zwischen den QFPs bei der Messung von FQ2 mit QFP1-Annealing für (spezifische) kurze Messzeiten bessere Ergebnisse als bei langen Messzeiten. Dies ermöglicht uns, die Messung von FQ2 mit QFP1-Annealing bei einer sehr kurzen Messzeit gut zu benutzen.\\	
	
	Für die Messung eines Flussqubits und die Messung von FQ2 ohne QFP1-Annealing wurde übereinstimmend gezeigt, dass für angegebene Parameter die Fidelity in der Energiebasis höher als in der Flussbasis ist. Daraus kann man schließen, dass die Dynamik in beiden Modellierungen vergleichbar ist. Das bedeutet, bei der Messung von FQ2 ohne QFP1-Annealing die Wechselwirkung zwischen beiden Flussqubits nur eine kleine Rolle spielt, während sie bei der Messung von FQ2 mit QFP1-Annealing von großer Bedeutung ist, sodass für eine schnelle Messung die Fidelity darin ziemlich hoch wird.\\
	
	Zusammenfassend lässt sich sagen, dass die Modellierung der Messung eines Flussqubit und die Modellierung von FQ2 ohne QFP1-Annealing ergibt, dass für eine stabile und hohe Fidelity eine festgelegte und relative lange Messzeit von Vorteil ist. Im Vergleich dazu ergibt die Modellierung der Messung mit QFP1-Annealing, dass eine hohe Fidelity mit einer kurzen Messzeit erreicht wird.\\
	
	\vspace{3em}
	
	In zukünftigen Arbeiten könnte die Messung in einem offenen Quantensystem untersucht und die durch Umgebung verursachte Dämpfung berücksichtigt werden. Dies erfordert die Methode zur Lösung der Mastergleichung.\\
	
	\thispagestyle{plain}
	
	Des Weiteren könnte man die optimalen Parameter für die komplexeren Setups finden.
	Es wäre auch interessant, die Ursache für die bei den gekoppelten Qubits auftretenden Oszillationen genauer zu untersuchen.\\
	
	Es wäre möglich, die Messung gekoppelter Flussqubits im Experiment durchzuführen und die entsprechende Fidelity in verschiedenen Basen zu messen.\\
	
	Außerdem könnte man versuchen, das Modell zu verallgemeinern und eine beliebige Anzahl an gekoppelten Flussqubits und QFPs untersuchen.
	
	\clearpage
	\thispagestyle{empty}
	
	\addcontentsline{toc}{chapter}{Literaturverzeichnis}
	
	\bibliographystyle{alpha}

\end{document}